\documentclass[%
reprint,
superscriptaddress,
showpacs,
amsmath,amssymb,
prc,
floatfix, ]%
{revtex4-1}

\usepackage{color}

\usepackage{graphicx}
\usepackage{dcolumn}
\usepackage{bm}
\usepackage[dvipdfm,bookmarks=true,colorlinks,%
            citecolor=blue,linkcolor=blue,hypertex, %
            breaklinks=true]{hyperref}

\newcommand{\br}{\bm{r}}
\newcommand{\bp}{\bm{p}}
\newcommand{\balp}{\bm{\alpha}}

\begin{document}

\title{Deformed relativistic Hartree Bogoliubov theory in continuum}


\author{Lulu Li}
 \affiliation{School of Physics, Peking University,
              Beijing 100871, China}
\author{Jie Meng}
 \affiliation{School of Physics, Peking University,
              Beijing 100871, China}
 \affiliation{State Key Laboratory of Theoretical Physics,
              Institute of Theoretical Physics, Chinese Academy of Sciences,
              Beijing 100190, China}
 \affiliation{Center of Theoretical Nuclear Physics, National Laboratory
              of Heavy Ion Accelerator, Lanzhou 730000, China}
 \affiliation{Department of Physics, University of Stellenbosch, Stellenbosch, South Africa}
\author{P. Ring}
 \affiliation{Physikdepartment, Technische Universit\"at M\"unchen,
              85748 Garching, Germany}
 \affiliation{School of Physics, Peking University,
              Beijing 100871, China}
\author{En-Guang Zhao}
 \affiliation{State Key Laboratory of Theoretical Physics,
              Institute of Theoretical Physics, Chinese Academy of Sciences,
              Beijing 100190, China}
 \affiliation{School of Physics, Peking University,
              Beijing 100871, China}
 \affiliation{Center of Theoretical Nuclear Physics, National Laboratory
              of Heavy Ion Accelerator, Lanzhou 730000, China}
\author{Shan-Gui Zhou}
 \email{sgzhou@itp.ac.cn}
 \affiliation{State Key Laboratory of Theoretical Physics,
              Institute of Theoretical Physics, Chinese Academy of Sciences,
              Beijing 100190, China}
 \affiliation{Center of Theoretical Nuclear Physics, National Laboratory
              of Heavy Ion Accelerator, Lanzhou 730000, China}

\date{\today}

\begin{abstract}
A deformed relativistic Hartree Bogoliubov (RHB) theory in continuum is developed
aiming at a proper description of exotic nuclei, particularly those with a large
spatial extension. In order to give an
adequate consideration of both the contribution of the continuum and
the large spatial distribution in exotic nuclei, the deformed RHB equations
are solved in a Woods-Saxon (WS) basis in which the radial wave
functions have a proper asymptotic behavior at large distance from the
nuclear center. This is crucial for the proper description of a possible halo.
The formalism of deformed RHB theory in continuum is presented. A stable
nucleus, $^{20}$Mg and a weakly-bound nucleus, $^{42}$Mg, are taken
as examples to present numerical details and to carry out necessary numerical
checks. In addition, the ground state properties of even-even magnesium
isotopes are investigated. The generic conditions of the formation of
a halo in weakly bound deformed systems and the shape of the halo in
deformed nuclei are discussed. We show that the existence and the deformation
of a possible neutron halo depend essentially on the quantum numbers
of the main components of the single particle orbitals in the vicinity
of the Fermi surface.
\end{abstract}

\pacs{21.60.-n, 21.10.-k, 21.10.Gv, 21.60.Jz}


\maketitle


\section{\label{sec:intro}Introduction}

The development of radioactive ion beam facilities around
the world~\cite{Xia2002_NIMA488-11, Zhan2010_NPA834-694c,
Sturm2010_NPA834-682c,Gales2010_NPA834-717c, Motobayashi2007_NPN17-5,
*Motobayashi2010_NPA834-707c, Thoennessen2010_NPA834-688c,
Choi2010_ISNPA} stimulates very much the study of nuclei far from the
$\beta$ stability line~\cite{Bertulani2001_PRB, Mueller1993_ARNPS43-529,
Tanihata1995_PPNP35-505, Hansen1995_ARNPS45-591, Casten2000_PPNP45-S171,
Johnson2004_PR389-1, Jensen2004_RMP76-215, Ershov2010_JPG37-064026, Cao2011_SciChinaPAM54S1-1}.
Some new and exotic phenomena have been observed in nuclei close to drip lines
such as neutron or proton halos~\cite{Tanihata1985_PRL55-2676,
Minamisono1992_PRL69-2058,Schwab1995_ZPA350-283},
changes of nuclear magic numbers~\cite{Ozawa2000_PRL84-5493},
pygmy resonances~\cite{Adrich2005_PRL95-132501}, etc.
In halo nuclei, the fact of an extremely weakly binding
leads to many new features, e.g., the coupling between
bound states and the continuum due to pairing correlations
and very extended spatial density distributions. Therefore one must
consider properly the asymptotic behavior of nuclear densities
at large distance $r$ from the center and treat in a self consistent
way the discrete bound states, the continuum and the coupling between
them in order to give a proper theoretical description of the halo
phenomenon~\cite{Cao2002_PRC66-024311, Dobaczewski2007_PPNP59-432,
Pei2011_PRC84-024311}.
This could be achieved by solving the non-relativistic
Hartree-Fock-Bogoliubov (HFB)~\cite{Bulgac1980_nucl-th9907088,
Dobaczewski1984_NPA422-103, Dobaczewski1996_PRC53-2809} or
the relativistic Hartree Bogoliubov
(RHB)~\cite{Meng1996_PRL77-3963, Poschl1997_PRL79-3841,*Lalazissis1998_PLB418-7,
Meng1998_NPA635-3} equations in coordinate ($r$) space which
can fully take into account the mean-field effects of the coupling to the continuum.
The resonant-BCS (rBCS) approach presents an other method to include the
contribution of the resonant continuum which has been used to study halo
phenomena~\cite{Sandulescu2000_PRC61-061301R,*Sandulescu2003_PRC68-054323, 
Zhang2011_arxiv1105.0504}.

The solution of the coupled differential equations of HFB and RHB theories
is particular simple in spherical systems with local potentials, where one-dimensional
Numerov or Runge-Kutta methods~\cite{Press1992} can be applied and this is true even for non-local
problems where Finite Element Methods (FEM)~\cite{Poschl1997_CPC101-75,*Poschl1997_CPC103-217}
have been used. A different
method to solve such equations is the expansion of the single particle wave functions
in an appropriate basis. The oscillator basis has been used for this purpose
with a great success for deformed or non-local systems in the
past~\cite{Vautherin1973_PRC7-296,Gogny1975_NPA237-399,*Decharge1980_PRC21-1568,
Gambhir1990_APNY198-132,*Gambhir1993_MPLA8-787, Stoitsov2000_PRC61-034311, *Stoitsov2003_PRC68-054312}.
The Woods-Saxon basis has been proposed in Ref.~\cite{Zhou2003_PRC68-034323}
as a reconciler between the harmonic oscillator basis and the integration
in coordinate space. Woods-Saxon wave functions have a much more realistic asymptotic
behavior at large $r$ than the harmonic oscillator wave functions do.
A discrete set of Woods-Saxon wave functions is obtained by
using box boundary conditions to discretize the continuum. It has been shown
in Ref.~\cite{Zhou2003_PRC68-034323} for spherical systems that the solution
of the relativistic Hartree equations in a Woods-Saxon basis is almost equivalent
to the solution in coordinate space. The Woods-Saxon basis has also been used
in more complicated situations, e.g., for the description of exotic nuclei
where both deformation and pairing have to be taken into account. Recently, for
spherical systems, both non-relativistic and relativistic Hartree-Fock-Bogoliubov
theories with forces of finite range have been investigated in a Woods-Saxon
basis~\cite{Schunck2008_PRC77-011301R,*Schunck2008_PRC78-064305,Long2010_PRC81-024308}.

Over the past years, lots of efforts have been made to develop a deformed relativistic
Hartree (RH) theory~\cite{Zhou2006_AIPCP865-90} and a deformed relativistic Hartree Bogoliubov
theory in continuum~\cite{Zhou2008_ISPUN2007}. As a first application, halo phenomena in deformed
nuclei have been investigated within the continuum RHB theory and some brief results can be found in
Ref.~\cite{Zhou2010_PRC82-011301R,*Zhou2011_JPCS312-092067}. 
In this paper we present the full version of the
theoretical framework with all the details.

Spherical symmetry facilitates considerably the treatment of the continuum
in non-relativistic 
HFB~\cite{Bulgac1980_nucl-th9907088, Dobaczewski1984_NPA422-103, *Dobaczewski1996_PRC53-2809} 
and in relativistic RHB theory~\cite{Meng1996_PRL77-3963, Poschl1997_PRL79-3841, Meng1998_NPA635-3} 
in $r$-space. Since most of the known nuclei are deformed, interesting questions arise,
whether or not deformed halos exist and what new features can be expected in deformed exotic
nuclei~\cite{Li1996_PRC54-1617,Misu1997_NPA614-44,Guo2003_CTP40-573, Meng2003_NPA722-C366, 
Nunes2005_NPA757-349, Pei2006_NPA765-29}. Such questions can be answered by the deformed
counterparts of the HFB or RHB theories in coordinate space.
From the experimental point of view, $^{31}$Ne is measured to be
a strongly deformed halo nucleus~\cite{Nakamura2009_PRL103-262501},
and for the well deformed magnesium isotopes, $^{35}$Mg is probably
a halo nucleus too~\cite{Kanungo2011_PRC83-021302R}.
Nevertheless for deformed nuclei, to solve the HFB or RHB equations
in $r$ space becomes much more sophisticated and numerically very
time consuming. Many efforts have been made to develop
non-relativistic HFB theories either in (discretized) coordinate space
or in a scaled oscillator basis with improved asymptotic
behavior~\cite{Stoitsov2000_PRC61-034311, *Stoitsov2003_PRC68-054312}.
The HFB equations have been solved in three-dimensional coordinate
space by combining the imaginary time approach and the two basis
method~\cite{Terasaki1996_NPA600-371, *Terasaki1997_NPA621-706} with a
truncated basis composed of discrete localized states and discretized
continuum states up to a few MeV~\cite{Yamagami2001_NPA693-579}.
Alternatively, the HFB equations have been solved on a
two-dimensional basis-spline Galerkin
lattices~\cite{Teran2003_PRC67-064314, Oberacker2003_PRC68-064302,
Pei2008_PRC78-064306} or on a three-dimensional Cartesian
mesh~\cite{Tajima2004_PRC69-034305} using the canonical-basis
approach~\cite{Reinhard1997_ZPA358-277}. Recently, the Gaussian
expansion method is used to solve the HF and HFB equations for
deformed nuclei~\cite{Nakada2008_NPA808-47}
and continuum Skyrme-Hartree-Fock-Bogoliubov approaches have been
developed both for spherical and deformed nuclei~\cite{Oba2009_PRC80-024301,
*Zhang2011_PRC83-054301}.
The deformed relativistic Hartree Bogoliubov (RHB) theory has only been solved
in the conventional harmonic oscillator basis~\cite{Vretenar1999_PRL82-4595,
Lalazissis1999_PRC60-014310,*Lalazissis1999_NPA650-133,
Niksic2002_PRC65-054320,Niksic2004_PRC69-047301} and
neither the above-mentioned approaches nor other methods
which could improve the asymptotic behavior of the nuclear densities
at large $r$ have been implemented in the deformed RHB theory so far.

In this paper we present a method, which allows to take into
account at the same time the coupling to the continuum, deformations,
and pairing correlations in a fully self-consistent way. For this purpose
we expand the deformed Dirac spinors in a basis of spherical Dirac
wave functions obtained by the solution of the Dirac equations
for potentials with spherical Woods-Saxon shape. This idea is
similar to a method proposed in Ref.~\cite{Zhang1988_PLB209-145,*Zhang1991_NPA526-245}
for the solution of the deformed relativistic mean field (RMF) equations in light nuclei,
where the deformed Dirac-spinors were expanded in terms of
the self-consistent solutions of the spherical RMF-equations.
As compared to these early calculations our method is simpler,
because it is based on Woods-Saxon wave functions. On the other side it
is more general, because it allows to include  pairing correlations,
which play an essential role in the formation of halo structures.

The paper is organized as follows. In Sec.~\ref{sec:formalism},
we give the formalism of the deformed RHB theory in continuum.
The numerical details are presented in Sec.~\ref{sec:comparison}
and we discuss applications and detailed results for magnesium isotopes
in Sec.~\ref{sec:results}. A summary is given in Sec.~\ref{sec:summary}.

\section{\label{sec:formalism}Formalism of the deformed relativistic
Hartree Bogoliubov theory in continuum}

The starting point of relativistic mean field theory is a
Lagrangian density where nucleons are described as Dirac spinors
which interact via the exchanges of effective mesons ($\sigma$,
$\omega$, and $\rho$) and the
photon~\cite{Duerr1956_PR0101-494,*Duerr1956_PR0103-469,Walecka1974_APNY83-491,
Serot1986_ANP16-1, Reinhard1989_RPP52-439, Ring1996_PPNP37-193, *Ring2001_PPNP46-165,
Vretenar2005_PR409-101, Meng2006_PPNP57-470},
\begin{eqnarray}
\displaystyle
 {\cal L}
   & = &
     \bar\psi \left( i\rlap{/}\partial -M \right) \psi
    + \frac{1}{2} \partial_\mu \sigma \partial^\mu \sigma
    - U(\sigma)
    - g_{\sigma} \bar\psi \sigma \psi
   \nonumber \\
   &   & \mbox{}
    - \frac{1}{4} \Omega_{\mu\nu} \Omega^{\mu\nu}
    + \frac{1}{2} m_\omega^2 \omega_\mu \omega^\mu
    - g_{\omega} \bar\psi \rlap{/}\omega \psi
   \nonumber \\
   &   & \mbox{}
    - \frac{1}{4} \vec{R}_{\mu\nu} \vec{R}^{\mu\nu}
    + \frac{1}{2} m_{\rho}^{2} \vec{\rho}_\mu \vec{\rho}^\mu
    - g_{\rho} \bar\psi \rlap{/} \vec{\rho} \vec{\tau} \psi
   \nonumber \\
   &   &\mbox{}
    - \frac{1}{4} F_{\mu\nu} F^{\mu\nu}
    - e \bar\psi \frac{1-\tau_3}{2}\rlap{/}A \psi ,
\label{eq:Lagrangian}
\end{eqnarray}
where $M$ is the nucleon mass, and $m_\sigma$, $g_\sigma$, $m_\omega$,
$g_\omega$, $m_\rho$, $g_\rho$ masses and coupling constants of the
respective mesons. The nonlinear self-coupling for the scalar meson
is given by~\cite{Boguta1977_NPA292-413}
\begin{equation}
   U(\sigma) = \displaystyle\frac{1}{2} m^2_\sigma \sigma^2
              +\displaystyle\frac{g_2}{3}\sigma^3
              + \displaystyle\frac{g_3}{4}\sigma^4 ,
\end{equation}
and field tensors for the vector mesons and the photon fields are
defined as
\begin{eqnarray}
 \left\{
  \begin{array}{rcl}
   \Omega_{\mu\nu}  & = & \partial_\mu\omega_\nu
                         -\partial_\nu\omega_\mu, \\
   \vec{R}_{\mu\nu} & = & \partial_\mu\vec{\rho}_\nu
                         -\partial_\nu\vec{\rho}_\mu
                         -g_{\rho} (\vec{\rho}_\mu
                                    \times \vec{\rho}_\nu ), \\
   F_{\mu\nu}       & = & \partial_\mu {A}_\nu
                         - \partial_\nu {A}_\mu.
  \end{array}
 \right.
 \label{eq:tensors}
\end{eqnarray}

Pairing correlations are crucial in the description of open shell nuclei.
For exotic nuclei, the conventional BCS
approach turns out to be only a poor approximation~\cite{Dobaczewski1984_NPA422-103}. 
Starting from the Lagrangian density~(\ref{eq:Lagrangian}), a relativistic theory of
pairing correlations in nuclei has been developed  by Kucharek and
Ring~\cite{Kucharek1991_ZPA339-23}. If we neglect the Fock terms as
it is usually done in the covariant density functional theory, the Dirac Hartree
Bogoliubov (RHB) equation for the nucleons reads,
\begin{eqnarray}
 \int d^3 \br'
 \left(
  \begin{array}{cc}
   h_D
   - \lambda &
   \Delta
   \\
  -\Delta^*
   & -h_D
   + \lambda \\
  \end{array}
 \right)
 \left(
  { U_{k}
  \atop V_{k}
   }
 \right)
 & = &
 E_{k}
  \left(
   { U_{k}
   \atop V_{k}
    }
  \right)
 ,
 \label{eq:RHB0}
\end{eqnarray}
where $E_{k}$ is the quasiparticle energy, $\lambda$ is the chemical potential,
and $h_D$ is the Dirac Hamiltonian,
\begin{equation}
 h_D(\bm{r}, \bm{r}') =
  \bm{\alpha} \cdot \bm{p} + V(\bm{r}) + \beta (M + S(\bm{r})).
\label{eq:Dirac0}
\end{equation}
The scalar and vector potentials
\begin{eqnarray}
S(\bm{r}) & = & g_\sigma \sigma(\bm{r}), \label{eq:vaspot}\\
V(\bm{r}) & = & g_\omega \omega^0(\bm{r}) +g_\rho \tau_3 \rho^0(\bm{r})
                    +e \displaystyle\frac{1-\tau_3}{2} A^0(\bm{r}) ,
\label{eq:vavpot}
\end{eqnarray}
depend on the scalar field $\sigma$ and on the time-like components
$\omega^0$, $\rho^0$, and $A^0$ of the iso-scalar vector field $\omega$,
the 3-component of iso-vector vector field $\rho$ and the photon field.

The equations of motion for the mesons and the photon
\begin{eqnarray}
 \left\{
   \begin{array}{rcl}
    \left( -\Delta + \partial_\sigma U(\sigma) \right )\sigma(\bm{r})
      & = & -g_\sigma \rho_s(\bm{r}) , \\
    \left( -\Delta + m_\omega^2 \right )             \omega^0(\bm{r})
      & = &  g_\omega \rho_v(\bm{r}) , \\
    \left( -\Delta + m_\rho^2 \right)                  \rho^0(\bm{r})
      & = &  g_\rho   \rho_3(\bm{r}) , \\
    -\Delta                                               A^0(\bm{r})
      & = &  e        \rho_p(\bm{r}) ,
   \end{array}
 \right.
 \label{eq:mesonmotion}
\end{eqnarray}
have as sources the various densities
\begin{eqnarray}
 \left\{
  \begin{array}{rcl}
   \rho_s(\bm{r})
   & = &
    \sum\limits_{k>0} V_{k}^\dagger(\bm{r})\gamma_0 V_{k}(\bm{r}) ,\\
   \rho_v(\bm{r})
   & = &
    \sum\limits_{k>0} V_{k}^\dagger(\bm{r}) V_{k}(\bm{r}) ,\\
   \rho_3(\bm{r})
   & = &
    \sum\limits_{k>0} V_{k}^\dagger(\bm{r}) \tau_3 V_{k}(\bm{r}) ,\\
   \rho_c(\bm{r})
   & = &
    \sum\limits_{k>0} V_{k}^\dagger(\bm{r})
                 \displaystyle\frac{1-\tau_3}{2}V_{k}(\bm{r}) ,
  \end{array}
 \right.
 \label{eq:mesonsource}
\end{eqnarray}
where, according to the no-sea approximation, the sum over $k>0$ runs
over the quasi-particle states corresponding to single particle
energies in and above the Fermi sea.

The pairing potential reads,
\begin{eqnarray}
 \Delta(\br_1{s}_1 p_1,\br_2{s}_2 p_2)
 & = &
   \sum^{{s}'_2p'_2}_{{s}'_1p'_1}
   V^\mathrm{pp}(\br_1,\br_2;{s}_1p_1,{s}_2p_2,{s}'_1p'_1,{s}'_2p'_2)
 \nonumber \\
 &   & \mbox{}\times
   \kappa(\br_1{s}'_1 p'_1,\br_2{s}_2' p_2')
 ,
\label{eq:gap12}
\end{eqnarray}
where $p=1, 2$ is used to represent the large and small components
of the Dirac spinors. $V^\mathrm{pp}$ is the effective pairing interaction and
$   \kappa(\br_1{s}'_1 p'_1,\br_2{s}_2' p_2')$ is the pairing tensor~\cite{Ring1980}.

In the particle-particle (pp) channel, we use a density dependent  zero range force,
\begin{equation}
 V^\mathrm{pp}(\br_1,\br_2) =  V_0 \frac{1}{2}(1-P^\sigma)\delta( \mathbf{r}_1 - \mathbf{r}_2 )
   \left(1-\frac{\rho(\br_1)}{\rho_\mathrm{sat}}\right).
 \label{eq:pairing_force}
\end{equation}
$\frac12(1-P^\sigma)$ projects onto spin $S=0$ component in the pairing field.
In this case the gap equation (\ref{eq:gap12}) has the simple form
\begin{equation}
 \Delta(\br)=V_0(1-\rho(\br)/\rho_{\rm sat})\kappa(\br) ,
\end{equation}
and we need only the local part of the pairing tensor
\begin{equation}
 \kappa(\br)= \sum_{k>0} V_{k}^\dagger(\bm{r})U^{}_{k}(\bm{r}) ,
\label{E12}
\end{equation}
Details of the calculation of the pairing interaction and the pairing tensor are given
in Appendices~\ref{appendix:matrix} and \ref{appendix:pair} respectively.

For axially deformed nuclei with spatial reflection symmetry, we
expand the potentials $S(\bm{r})$ and $V(\bm{r})$ in Eqs.~(\ref{eq:vaspot}) and (\ref{eq:vavpot})
and the densities in Eq.~(\ref{eq:mesonsource}) in terms of the Legendre
polynomials~\cite{Price1987_PRC36-354},
\begin{equation}
 f(\bm{r})   = \sum_\lambda f_\lambda({r}) P_\lambda(\cos\theta),\
 \lambda = 0,2,4,\cdots
 ,
 \label{eq:expansion}
\end{equation}
with
\begin{equation}
 f_\lambda(r) = \frac{2\lambda+1}{4\pi}\int d\Omega f({\bf r})P_\lambda(\Omega).
\end{equation}

The quasiparticle wave functions $U_k$ and $V_k$ in Eq.~(\ref{eq:RHB0}) are Dirac spinors.
Each of them is expanded in terms of spherical Dirac spinors $\varphi_{n\kappa m}(\br{s} p)$ 
with the eigenvalues $\epsilon_{n\kappa}$ obtained from the solution of a Dirac equation 
$h^{(0)}_D$ containing spherical potentials $S^{(0)}(r)$ and $V^{(0)}(r)$ of
Woods-Saxon shape~\cite{Zhou2003_PRC68-034323, Koepf1991_ZPA339-81}:
\begin{eqnarray}
 U_{k} (\br{s} p)
 & = & \displaystyle
 \sum_{n\kappa} u^{(m)}_{k,(n\kappa)}     \varphi_{n\kappa m}(\br{s} p),
 \label{eq:Uexpansion0} \\
 V_{k} (\br{s} p)
 & = & \displaystyle
 \sum_{n\kappa} v^{(m)}_{k,(n\kappa)} \bar\varphi_{n\kappa m}(\br{s} p).
\label{eq:Vexpansion0}
\end{eqnarray}
The basis wave function reads
\begin{equation}
 \varphi_{n\kappa m}(\bm{r}{s}) =
   \frac{1}{r}
   \left(
     \begin{array}{c}
       i G_{n\kappa}(r) Y^l _{jm} (\Omega{s})
       \\
       - F_{n\kappa}(r) Y^{\tilde l}_{jm}(\Omega{s})
     \end{array}
   \right) ,
\label{eq:SRHspinor}
\end{equation}
where $G_{n\kappa}(r) / r$ and $F_{n\kappa}(r) / r$ the radial wave
functions for the upper and lower components. 
The spherical spinor $\varphi_{n\kappa m}$ is characterized
by the radial quantum number $n$, angular quantum $j$ and the parity $\pi$. 
$j$ and $\pi$ are combined to the relativistic quantum number  
$\kappa=\pi(-1)^{j+1/2} (j+1/2)$ which runs over positive and negative
integers $\kappa=\pm 1,\pm 2,\cdots$. $Y^l _{jm}$ and
$Y^{\tilde{l}}_{jm}$ are the spinor spherical harmonics where
$l = j + \frac{1}{2}{\rm sign}(\kappa)$ and
$\tilde l = j - \frac{1}{2}{\rm sign}(\kappa)$.

$\bar\varphi_{n\kappa m}(\br{s} p)$ is the time reversal state of $\varphi_{n\kappa
m}(\br{s} p)$. These states form a complete spherical and discrete basis in Dirac space
(see Appendix~\ref{appendix:pre} for details). 
Because of the axial symmetry the $z$-component $m$ of
the angular momentum $j$ is a conserved quantum number and the RHB Hamiltonian can
be decomposed into blocks characterized by $m$ and parity $\pi$. For
each $m\pi$-block, solving the RHB equation (\ref{eq:RHB0}) is
equivalent to the diagonalization of the matrix
\begin{equation}
 \left( \begin{array}{cc}
  {\cal A}-\lambda & {\cal B} \\
  {\cal B^\dag} & -{\cal A}^\ast+\lambda \\
 \end{array} \right)
 \left(
  { {\cal U}_k
    \atop
    {\cal V}_k
  }
 \right)
 = E_k
 \left(
  { {\cal U}_k
    \atop
    {\cal V}_k
  }
 \right),
 \label{eq:RHB1}
\end{equation}
where
\begin{equation}
 {\cal U}_k = \left(u^{(m)}_{k,(n\kappa)}\right),\
 {\cal V}_k = \left(v^{(m)}_{k,(n\kappa)}\right),
\end{equation}
and
\begin{eqnarray}
 {\cal A}
 & = &
 \left( h^{(m)}_{D(n\kappa)(n'\kappa')} \right)
 = 
 \left( \langle n\kappa m|h_D|n'\kappa',m\rangle \right) ,
 \\
 {\cal B}
 & = &
 \left( \Delta^{(m)}_{(n\kappa)(n'\kappa)} \right)~
 = 
 \left( \langle n\kappa m |\Delta| \overline{n'\kappa',m} \rangle \right).
\label{eq:pairing_matrix}
\end{eqnarray}
Further details are given in Appendix~\ref{appendix:matrix}.

Since we use a zero range pairing force we have to introduce a pairing cutoff in
the sums of Eqs.~(\ref{eq:mesonsource}) and  (\ref{E12})
over the quasiparticle space. In the present work, a smooth cut
off is adopted where two parameters, $E^\mathrm{q.p.}_\mathrm{cut}$
and $\Gamma^\mathrm{q.p.}_\mathrm{cut}$, are introduced and the
square root of the factor
\begin{equation}
 s(E_k) =
 \frac{1}{2} \left( 1 - \frac{E_k-E^\mathrm{q.p.}_\mathrm{cut}}
                             {\sqrt{(E_k-E^\mathrm{q.p.}_\mathrm{cut})^2
                             +(\Gamma^\mathrm{q.p.}_\mathrm{cut})^2}} \right)
 \ ,
 \label{eq:smooth}
\end{equation}
is multiplied in the occupation component $V_k(\br)$ of each quasi
particle state with $v^2 < 1/2$. Note that this smooth cutoff is
similar as the soft cutoff proposed in
Ref.~\cite{Bonche1985_NPA443-39}.

The total energy of a nucleus is
\begin{eqnarray}
 E
 & = &
  E_\mathrm{nucleon} + E_{\sigma} + E_{\omega} + E_{\rho}
  + E_c + E_\mathrm{c.m.}
 \nonumber  \\
 & = &
  \sum_{k} 2(\lambda - E_{k}) v^2_{k} - E_\mathrm{pair}
 \nonumber  \\
 &   &
 -\dfrac{1}{2} \int d^3\br \left[ g_\sigma\sigma(\br)\rho_s(\br) + U(\sigma)
                           \right]
 \nonumber  \\
 &   &
 -\dfrac{1}{2} \int d^3\br g_\omega \omega(\br) \rho_v(\br)
 \nonumber  \\
 &   &
 -\dfrac{1}{2} \int d^3\br g_\rho \rho(\br)
                \left[ \rho_v^Z(\br)-\rho_v^N(\br) \right]
 \nonumber  \\
 &   &
 -\dfrac{1}{2} \int d^3\br   A_0 \rho_v^Z(\br)
 + E_{\rm c.m.} .
\end{eqnarray}
where
\begin{equation}
 v^2_k = \int d^3r V_k^\dagger({\bm r})V_k^{}({\bm r})
       = \sum_{n\kappa m} \left( v^{(m)}_{k,(n\kappa)} \right)^2.
\end{equation}

For a zero range force the pairing field $\Delta({\bm r})$ is local and
the pairing energy is calculated as
\begin{equation}
 E_\mathrm{pair} = -\frac{1}{2}\int d^3r \kappa({\bm r})\Delta({\bm r}).
\end{equation}
The center of mass correction energy
\begin{equation}
 E_{\rm c.m.} = - \frac{1}{2Am} \langle \hat{\mathbf{P}}^2 \rangle,
\label{Ecm}
\end{equation}
is calculated after variation with the wave functions of the
self-consistent solution~\cite{Long2004_PRC69-034319, Zhao2009_CPL26-112102}
or in the oscillator approximation
\begin{eqnarray}
 E_{\rm c.m.} = -\dfrac{3}{4} \times 41 \times A^{1/3}\ \mathrm{MeV},
 \label{eq:ECM}
\end{eqnarray}
Details are given in Appendix~\ref{appendix:com}.
The root mean square (rms) radius is calculated as
\begin{eqnarray}
 R_{\tau,\mathrm{rms}} \equiv \langle r^2 \rangle^{1/2}
 & = &
  \left( \int d^3\br \left[ r^2 \rho_\tau(\br) \right] \right)^{1/2}
 \nonumber \\
 & = &
  \left( \int dr \left[ r^4 \rho^\tau_{v,\lambda=0}(r)\right] \right)^{1/2}
 ,
\end{eqnarray}
where $\tau$ represents the proton, the neutron, or the nucleon.
The rms charge radius is calculated simply as $r_{\rm ch}^2 =
r_{\mathrm p}^2 + 0.64$ fm$^2$. The intrinsic multipole moment is
calculated by
\begin{eqnarray}
 Q_{\tau,\lambda}
 & = &
  \sqrt{\frac{16\pi}{2\lambda+1}} \langle r^2 Y_{\lambda0}(\theta,\phi) \rangle
  = 2 \langle r^2 P_{\lambda}(\theta) \rangle
 \nonumber \\
 & = &
  \frac{8\pi}{2\lambda+1} \int dr  \left[ r^4 \rho^\tau_{v,\lambda}(r) \right]
 .
\end{eqnarray}
The quadrupole deformation parameter is obtained from the quadrupole
moment by
\begin{equation}
 \beta_{\tau,2} = \frac{ \sqrt{5 \pi} Q_{\tau,2} } {3 N_\tau \langle r_\tau^2\rangle}
 \ ,
\end{equation}
where $N_\tau$ refers to the number of neutron, proton, or nucleon.

\section{\label{sec:comparison}Numerical details and routine checks}

\subsection{Details on the Woods-Saxon basis}

For numerical reasons several parameters have to be introduced
in the calculations, e.g., the mesh size $\Delta r$, the box size
$R_\mathrm{box}$ for the determination of the basis wave functions
by solving the spherical Dirac equations with the Hamiltonian $h^{(0)}_D$,
the maximal $\lambda$-value $\lambda_\mathrm{max}$ in the expansion
Eq.~(\ref{eq:expansion}) of the deformed fields and densities, the cutoff
parameters for the radial and angular quantum numbers $n$ and $\kappa$ in the
expansion of Eqs.~(\ref{eq:Uexpansion0}) and (\ref{eq:Vexpansion0}),
$n_\mathrm{max}$ and $\kappa_\mathrm{max}$. Instead of
$n_\mathrm{max}$, we introduced an energy cutoff parameter
$E^+_\mathrm{cut}$ for positive energy states in the Woods-Saxon
basis and in each $\kappa$-block, the number of negative energy
states in the Dirac sea is the same as that of positive energy states
above the Dirac gap~\cite{Zhou2003_PRC68-034323}. We have investigated
the dependence of our results on these parameters in
spherical and deformed relativistic Hartree
models~\cite{Zhou2003_PRC68-034323, Zhou2006_AIPCP865-90}. It is found
that a box of the size $R_\mathrm{box} = 4r_0 A^{1/3}$ with $r_0 =
1.2$ fm, a step size $\Delta r = 0.1$ fm, $\lambda_\mathrm{max} = 4$,
and $|\kappa_\mathrm{max}| = 15$ leads in light nuclei to an acceptable accuracy
of less than 0.1 \% for the binding energies, the rms radii, and the quadrupole moments .

In the present work we use the determination of the Woods-Saxon basis
a box size $R_\mathrm{box} = 20$ fm, a mesh size
$\Delta r = 0.1$ fm and a cutoff energy $E^+_\mathrm{cut} = 100$ MeV.
In each $\kappa$-block in the Woods-Saxon basis, the number of negative
energy states in the Dirac sea is the same as that of positive energy
states above the Dirac gap. In Sec.~\ref{sec:conv} we investigate the convergence
of our results with respect to these three parameters.

In order to reduce the computational time, $\lambda_\mathrm{max} = 4$ and
$|\kappa_\mathrm{max}| = 10$ are used in this work.
The parameter sets NL3~\cite{Lalazissis1997_PRC55-540} and
PK1~\cite{Long2004_PRC69-034319} are used for the Lagrangian density.
Note that the center of mass correction energy is calculated differently
with these two parameter sets. For NL3, the empirical formula in Eq.~(\ref{eq:ECM})
is used and for PK1, the center of mass correction energy is calculated
microscopically (see Appendix~\ref{appendix:com}).

\subsection{Parameters for the pairing force}

There are two parameters in the phenomenological pairing force
Eq.~(\ref{eq:pairing_force}), namely, $V_0$ and $\rho_\mathrm{sat}$, and
two additional ones in the smooth cutoff Eq.~(\ref{eq:smooth}). We take
the empirical value 0.152 fm$^{-3}$ for the saturation density
$\rho_\mathrm{sat}$. The pairing strength $V_0$, the cutoff
$E^\mathrm{q.p.}_\mathrm{cut}$ is fixed by reproducing the
proton pairing energy of the Gogny force D1S in the spherical
nucleus $^{20}$Mg. We first
calculate the ground state properties of $^{20}$Mg by using the
spherical relativistic Hartree Bogoliubov theory in a harmonic
oscillator basis (SRHBHO)~\cite{Serra2002_PRC65-064324} in which the
Gogny-D1S~\cite{Berger1984_NPA428-23} force is used in the pp
channel. The pairing energy for protons is obtained as $-9.2382$ MeV.
In Table~\ref{tab:pairing} the proton pairing energy
$E^{\mathrm{p}}_{\mathrm{pair}}$ from the SRHBHO and deformed RHB
calculations for $^{20}$Mg are given.  
The deformed RHB calculation using the parameter set NL3 with $V_0 = 380$ MeV fm$^3$,
$E^\mathrm{q.p.}_\mathrm{cut} = 60$ MeV and the smooth parameter
$\Gamma^\mathrm{q.p.}_\mathrm{cut} = 5.65$ MeV reproduces the proton
pairing energy from the SRHBHO calculation for $^{20}$Mg.
These parameters for the pairing are used in all the following
calculations regardless of whether NL3 or PK1 is used for the
RMF Lagrangian density.

\begin{table}
\caption{\label{tab:pairing}%
Determination of the parameters for the pairing force used in the deformed RHB
calculations presented in this work. In the last column is given the
proton pairing energy $E^{\mathrm{p}}_{\mathrm{pair}}$ from the
SRHBHO and deformed RHB calculations for the spherical nucleus $^{20}$Mg.
}
\begin{tabular}{l | l | l | l}
\hline\hline
 Model  & Pairing force    & Parameters        & $E^{\mathrm{p}}_{\mathrm{pair}}$ (MeV) \\
\hline
 SRHBHO & Gogny            & D1S~\cite{Berger1984_NPA428-23}                & $-$9.2382 \\
\hline
  RHB
        & Surface $\delta$ & $V_0$ = 380 MeV fm$^3$                         & $-$9.2382 \\
        & with             & $\rho_{\rm sat}$ = 0.152 fm$^{-3}$             &           \\
        & smooth cutoff    & $E^\mathrm{q.p.}_\mathrm{cut}$ = 60 MeV        &           \\
        &                  & $\Gamma^\mathrm{q.p.}_\mathrm{cut}$ = 5.65 MeV &           \\
\hline\hline
\end{tabular}
\end{table}

\subsection{\label{sec:conv}Completeness of the Woods-Saxon basis}

\begin{table}
\caption{\label{tab:comp_RCHB}%
Ground state properties of $^{20}$Mg from deformed RHB calculations with
different cutoff parameters in the Woods-Saxon basis compared with
the results of spherical RCHB~\cite{Meng1998_NPA635-3} theory.
}
\begin{tabular}{l || rrr | r}
\hline\hline
                                    & \multicolumn{3}{c}{deformed RHB}           & RCHB         \\
\hline
 $E^+_\mathrm{cut}$ (MeV)           & 100          & 200          & 300          & ---          \\
\hline
 $\lambda_\mathrm{p}$ (MeV)         &    $-$0.8992 &    $-$0.9072 &    $-$0.9063 &    $-$0.9061 \\
 $\Delta_\mathrm{p}$ (MeV)          &       2.3823 &       2.3866 &       2.3871 &       2.3876 \\
 $R_\mathrm{n}$ (fm)                &       2.5910 &       2.5902 &       2.5900 &       2.5900 \\
 $R_\mathrm{p}$ (fm)                &       3.0073 &       3.0052 &       3.0049 &       3.0049 \\
 $E^\mathrm{p}_\mathrm{pair}$ (MeV) &    $-$9.1165 &    $-$9.2294 &    $-$9.2381 &    $-$9.2387 \\
 $E$ (MeV)                          &  $-$136.6728 &  $-$136.7608 &  $-$136.7701 &  $-$136.7668 \\
\hline\hline
\end{tabular}
\end{table}

The spherical nucleus $^{20}$Mg has been investigated as the first
test of the deformed RHB theory and some results were given in
Fig.~1 in Ref.~\cite{Zhou2008_ISPUN2007}.
A comparison was made between results obtained for ground state properties
of the spherical nucleus $^{20}$Mg with the spherical RCHB code~\cite{Meng1998_NPA635-3}
based on the Runge-Kutta method in the radial coordinate $r$ and the new deformed RHB
code discussed in this manuscript.
We summarize these comparison in Table~\ref{tab:comp_RCHB}.
In these calculations, the parameter set NL3, a box of
the size $R_\mathrm{box} = 4r_0 A^{1/3} = 13.0$ fm and a step size
$\Delta r = 0.1$ fm are used.
The surface $\delta$ pairing force is used with the strength $V_0 = -374$
MeV fm$^{3}$ and $\rho_\mathrm{sat}$ = 0.152 fm$^{-3}$.
A sharp cutoff is applied on the quasiparticle states with
$E^\mathrm{q.p.}_\mathrm{cut} = $ 60 MeV.
It is shown that when the basis size increases, the total binding
energy $E$, the proton pairing energy $E^\mathrm{p}_\mathrm{pair}$,
and the rms radius $R$ all converge to the corresponding exact values.
In practical calculations, $E^+_\text{cut}$ may be chosen according to
the balance between the desired accuracy and the computational cost.
It is concluded~\cite{Zhou2008_ISPUN2007} that for light nuclei,
one can safely use $E^+_\text{cut}$ = 100 MeV which results in
accuracies in the total binding energy and the proton pairing energy
of about a hundred keV and in the rms radius of around 0.002 fm.

\begin{figure}
\begin{center}
\includegraphics[width=0.23\textwidth]{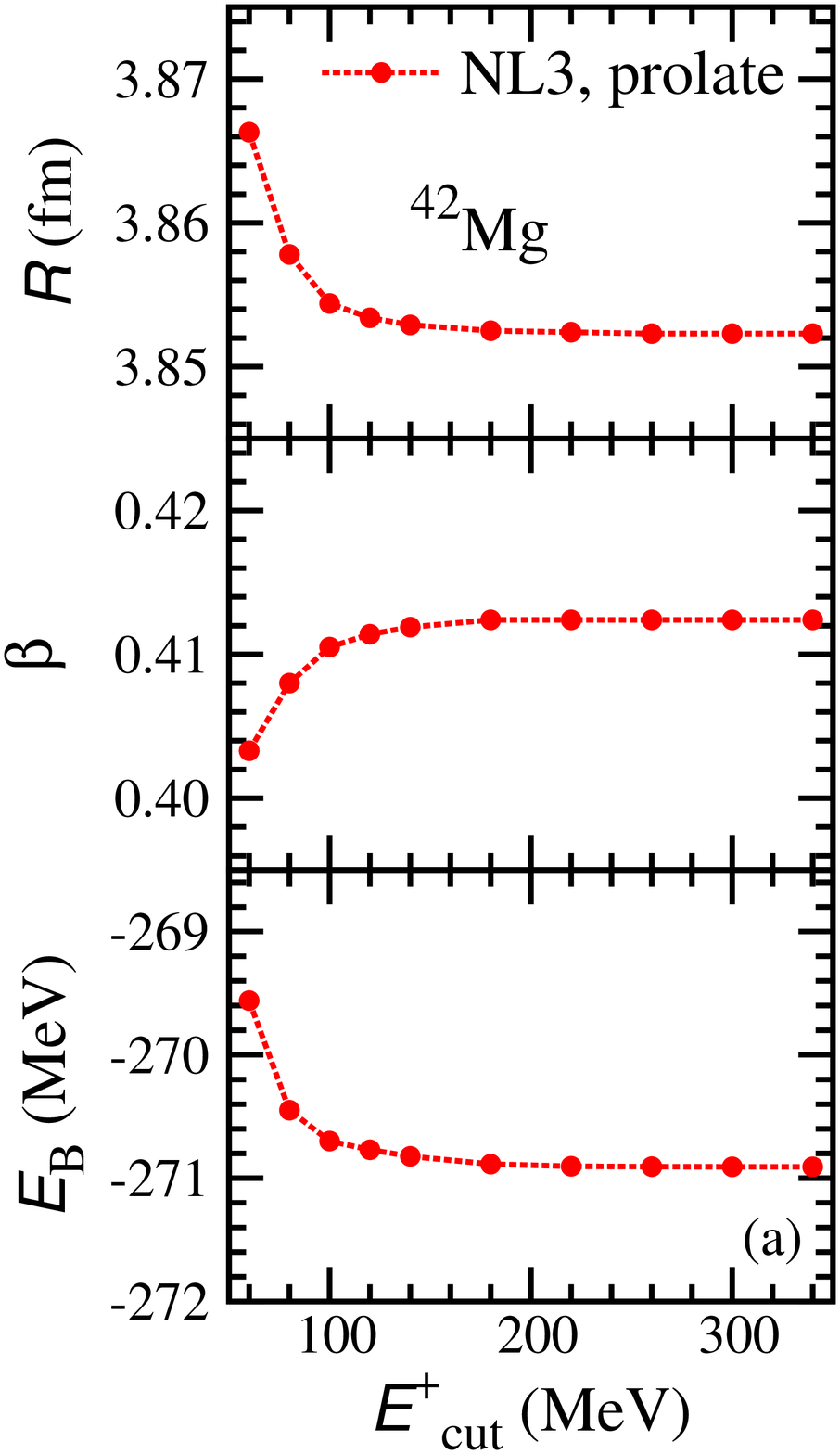}
\includegraphics[width=0.23\textwidth]{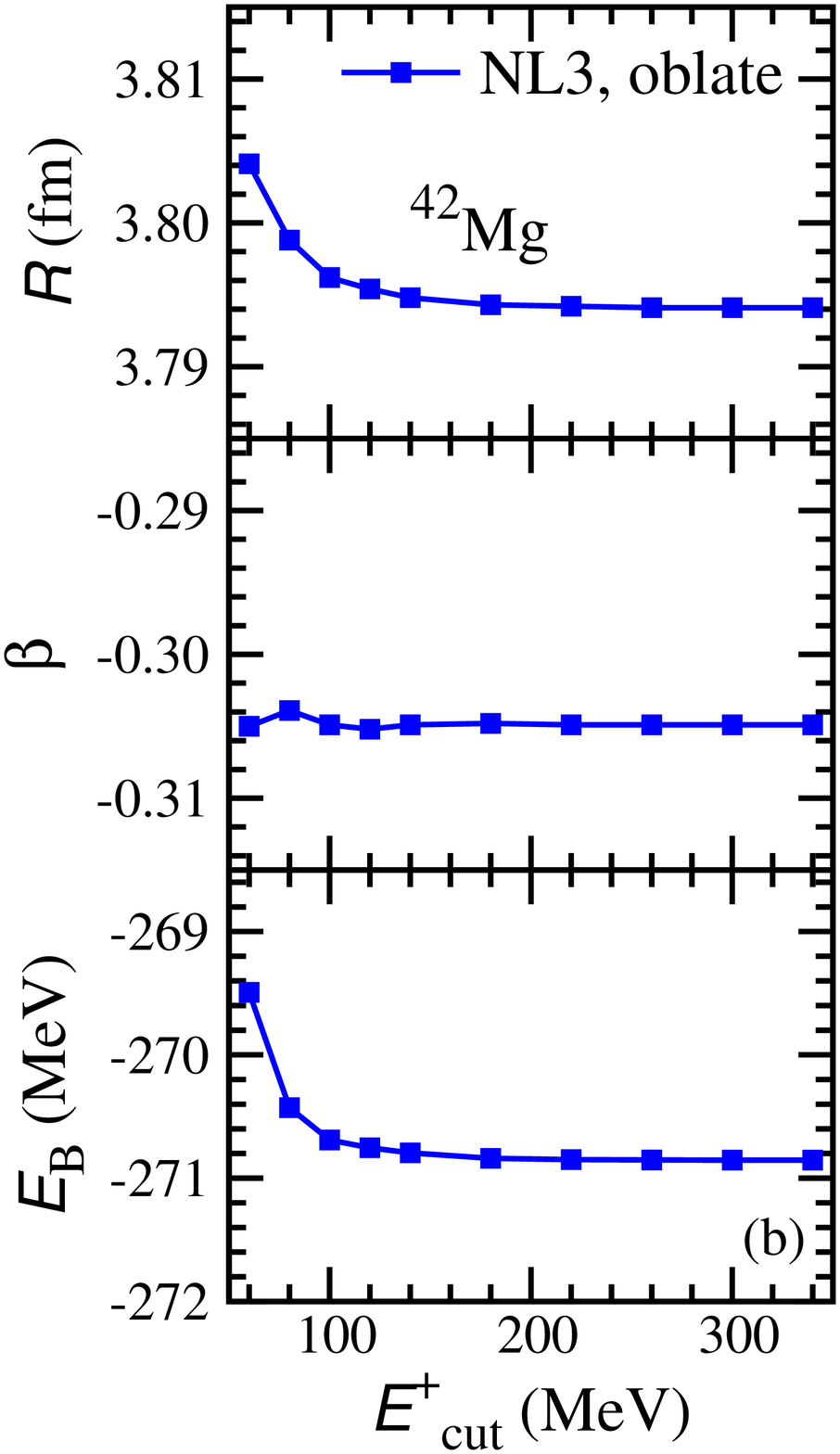}
\end{center}
\caption{(Color online) %
Bulk properties of ground state (a) and the oblate minimum
(b) of $^{42}$Mg as functions of the cutoff energy $E_{\rm cut}^+$.
From the lowest to the top panels, the total binding energy $E_{\rm B}$,
the quadrupole deformation $\beta$, and the rms radius $R$ are plotted.
The parameter set is NL3, the box size is $R_{\rm box} = 20$ fm and
the step size is $\Delta r = 0.1$ fm.}
\label{fig:mg42_ecut}
\end{figure}

Since we are also interested in drip-line nuclei, next we study the
dependence of the deformed RHB results on the completeness of the Woods-Saxon
basis for a very neutron rich nucleus.
In this subsection we study the results with different values of $E^+_\text{cut}$.
For the calculation with a Woods-Saxon basis~\cite{Zhou2003_PRC68-034323},
a box of the size $R_{\rm box} = 4 r_0 A^{1/3}$
with $r_0 = 1.2$ fm is used. In this case $R_{\rm box}$ is different for
different magnesium isotopes, e.g., 13.0 fm for $^{20}$Mg and 16.7 fm for $^{42}$Mg.
In the present work, we prefer to use a fixed box size $R_{\rm box} = 20$ fm which is
large enough for all magnesium isotopes. The mesh size for the radial wave function
of each Woods-Saxon state is taken as 0.1 fm.

For $^{42}$Mg both prolate and oblate minima in the potential energy
surface are searched for and it is found that the ground state is prolate.
In Fig.~\ref{fig:mg42_ecut} the total binding energy $E_{\rm B}$, the quadrupole
deformation $\beta$, and the rms radius $R$ are plotted as functions of $E^+_\text{cut}$
for the prolate ground state and for the oblate minimum of $^{42}$Mg,
respectively. Apparently, when we increase $E^+_\text{cut}$, these quantities all converge
well. Similar as in the case of the spherical nucleus $^{20}$Mg, for light deformed
nuclei, the cutoff $E^+_\text{cut}$ = 100 MeV results in relative accuracies
of 0.5\% for the quadrupole deformation, 0.05\% for the rms radius, and 0.1 \%
for the total binding energy.

\begin{figure}
\begin{center}
\includegraphics[width=0.23\textwidth]{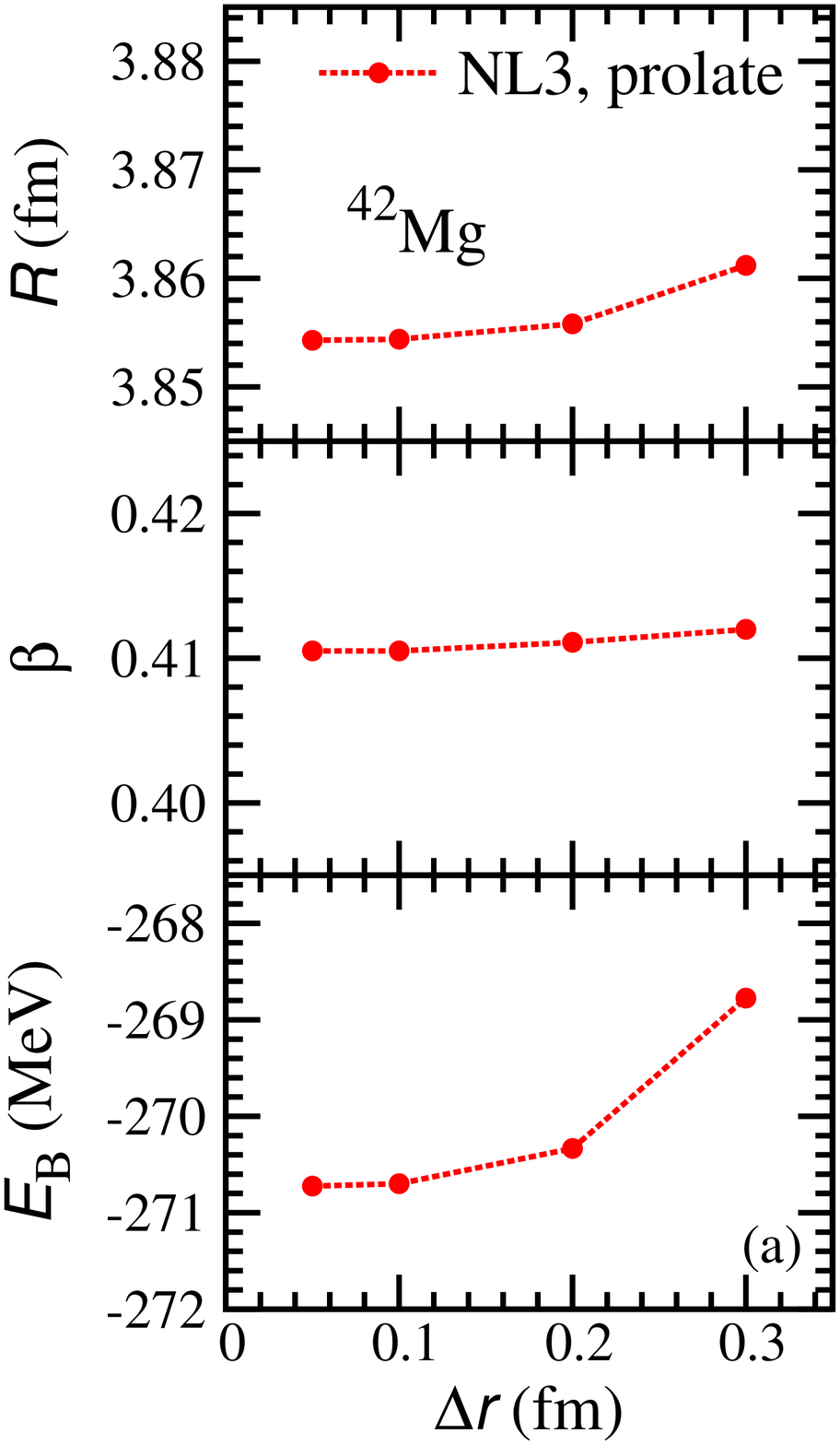}
\includegraphics[width=0.23\textwidth]{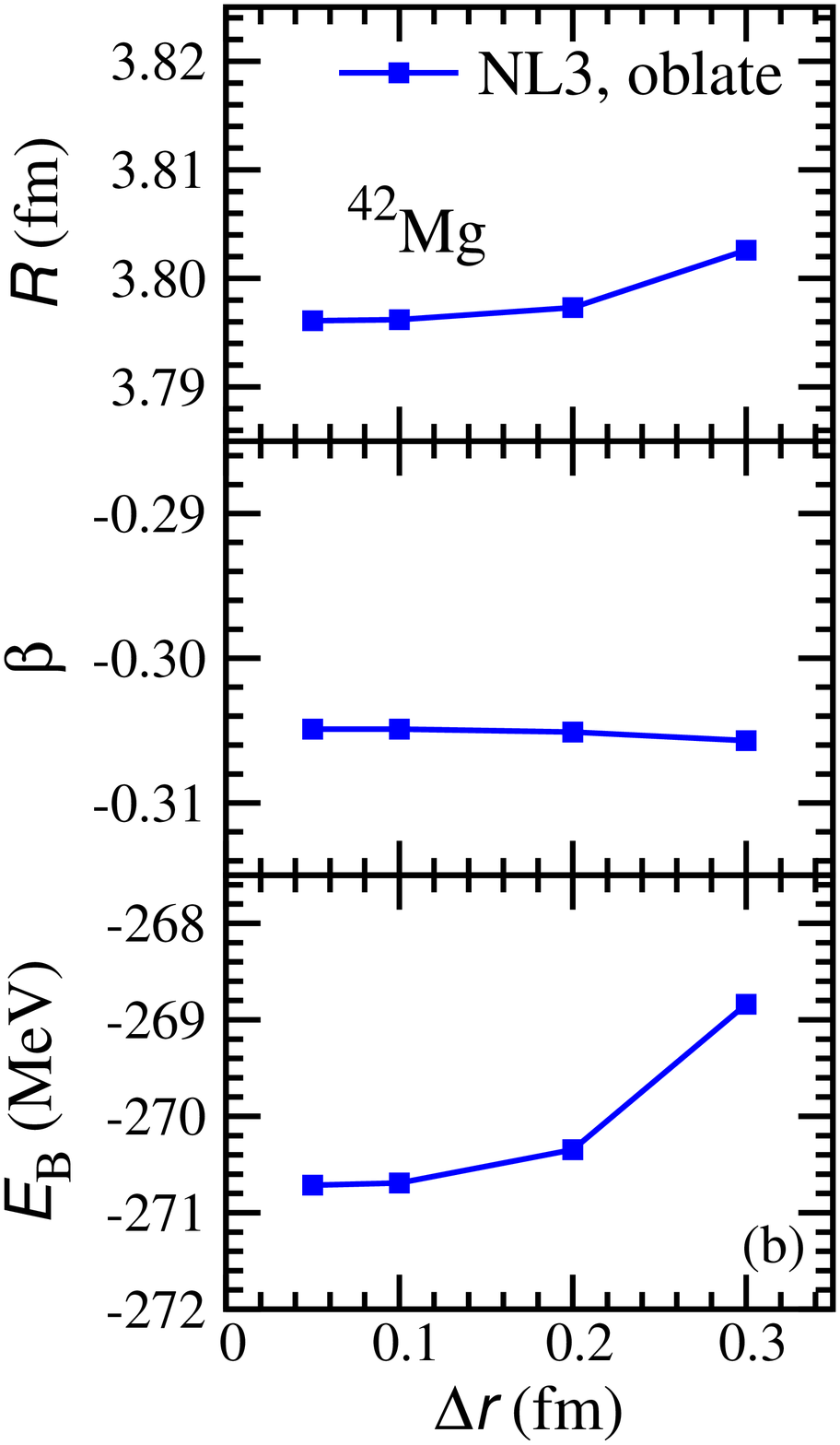}
\end{center}
\caption{(Color online) %
Bulk properties for $^{42}$Mg as in Fig.~\ref{fig:mg42_ecut} but now
as functions of the step size $\Delta r$. The box size is $R_{\rm box} = 20$ fm
and an energy cutoff is $E_{\rm cut}^+ = 100$ MeV.
}
\label{fig:Mg42_dr}
\end{figure}

The box size $R_\mathrm{box} = 20$ fm and the cutoff energy
$E_{\rm cut}^+ = 100$ MeV are fixed when we investigate the convergence
of the deformed RHB results with respect to the mesh size $\Delta r$.
In Fig.~{\ref{fig:Mg42_dr}} it is shown that when the mesh size decreases,
the total binding energy $E_{\rm B}$, the quadrupole deformation $\beta$, and the rms
radius $R$ all converge well. The difference of the binding energy between calculations with
$\Delta r = 0.1$ fm and $\Delta r = 0.05$ fm is smaller
than 0.025 MeV for both minima, which is about 0.008\% of the total binding energy.
When $\Delta r$ is decreased from 0.1 fm to 0.05 fm, the relative changes of
the quadrupole deformation $\beta$ and the radius $R$ are both
smaller than 0.01\%.

\begin{figure}
\begin{center}
 \includegraphics[width=0.23\textwidth]{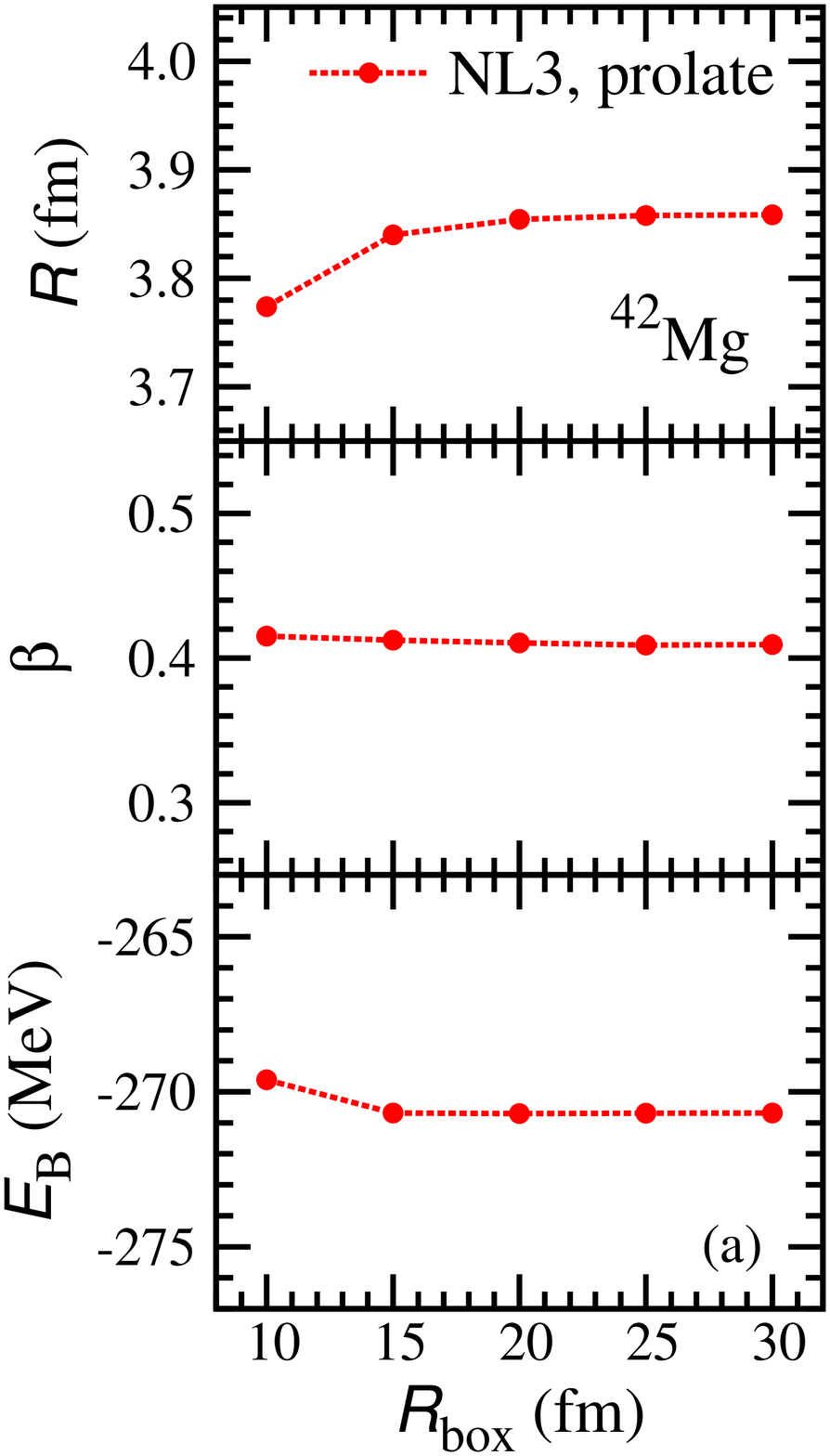}
 \includegraphics[width=0.23\textwidth]{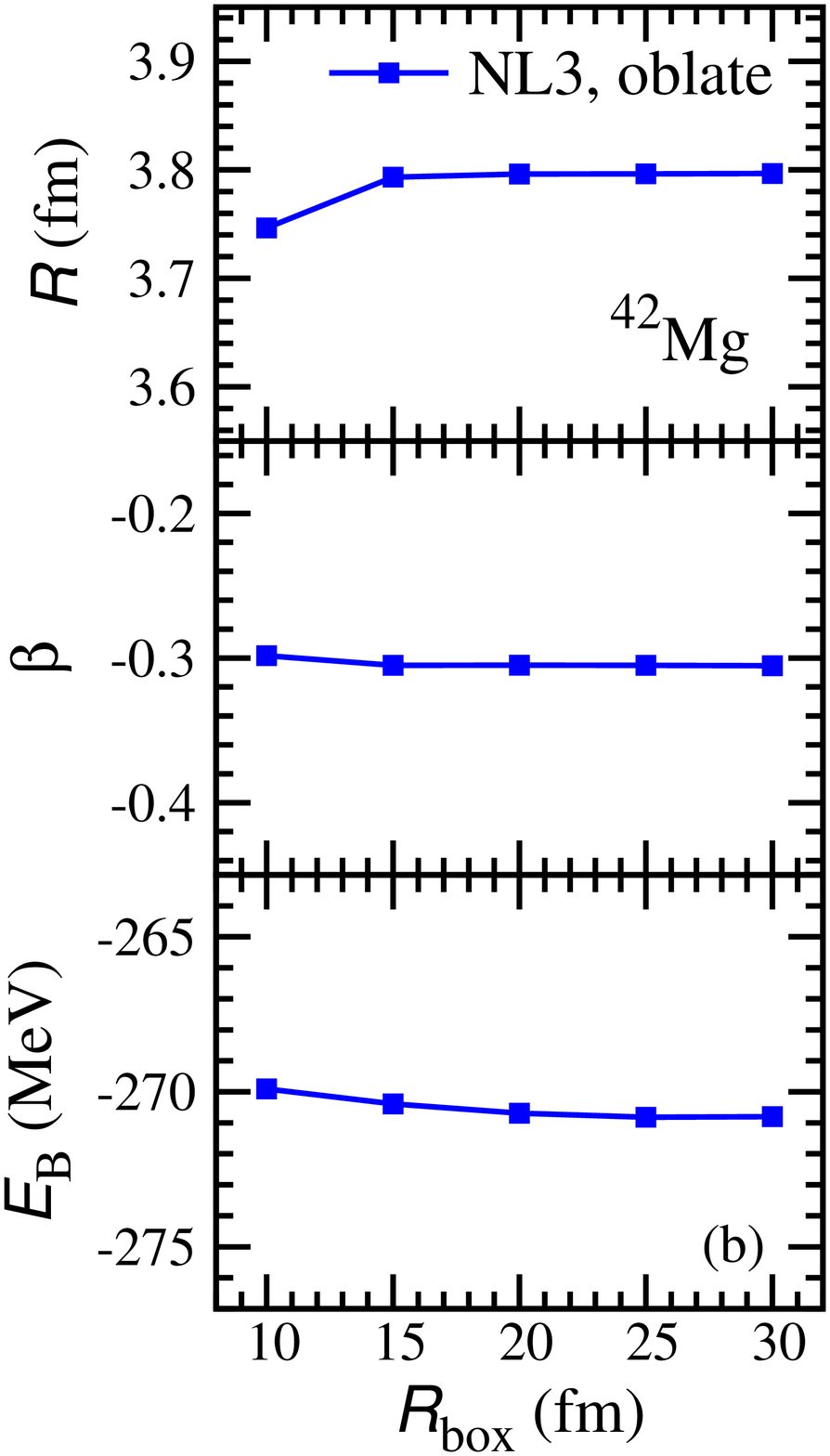}
\end{center}
\caption{(Color online) %
Bulk properties for $^{42}$Mg as in Fig.~\ref{fig:mg42_ecut} but now
as functions of the box size $R_{\rm box}$. The step size is
$\Delta r  = 0.1$ fm and the energy cutoff is $E_{\rm cut}^+ = 100$ MeV.}
\label{fig:Mg42_rbox}
\end{figure}

Figure~\ref{fig:Mg42_rbox} shows the same quantities as a function of the box
size $R_{\rm box}$. The relative deviations between the rms radius $R$ at $R_{\rm box} = 20$ fm
and $R_{\rm box} = 30$ fm are about 0.1\% for the prolate ground state and
0.01\% for the oblate minimum. The box size $R_{\rm box} = 20$ fm gives also
a good accuracy For the quadrupole deformation $\beta$ and the
binding energy.

In conclusion, in the following calculations, we fix the box size at $R_\mathrm{box}
= 20$ fm, the mesh size at $\Delta r = 0.1$ fm, and the cutoff energy for positive
energy states in the Woods-Saxon basis at $E^+_\mathrm{cut} = 100$ MeV.
In each $\kappa$-block, the number of negative energy states in the Dirac sea
is the same as that of positive energy states above the Dirac gap.
The cutoff parameter for $\lambda$ in the expansion Eq.~(\ref{eq:expansion}),
$\lambda_\mathrm{max} = 4$ and the cutoff parameter for the angular quantum number
$\kappa$ in the expansion Eq.~(\ref{eq:Vexpansion0}) is $|\kappa_\mathrm{max}| = 10$.
With these values we do not introduce sizable errors.

\section{\label{sec:results}Results and discussions}

In this section, we present results from the deformed RHB theory in continuum.
We choose magnesium isotopes as examples. After discussing the bulk properties
of magnesium isotopes, we will focus on the neutron rich nucleus $^{42}$Mg.

\subsection{Bulk properties of magnesium isotopes}

\begin{figure}
\begin{center}
 \includegraphics[width=0.45\textwidth]{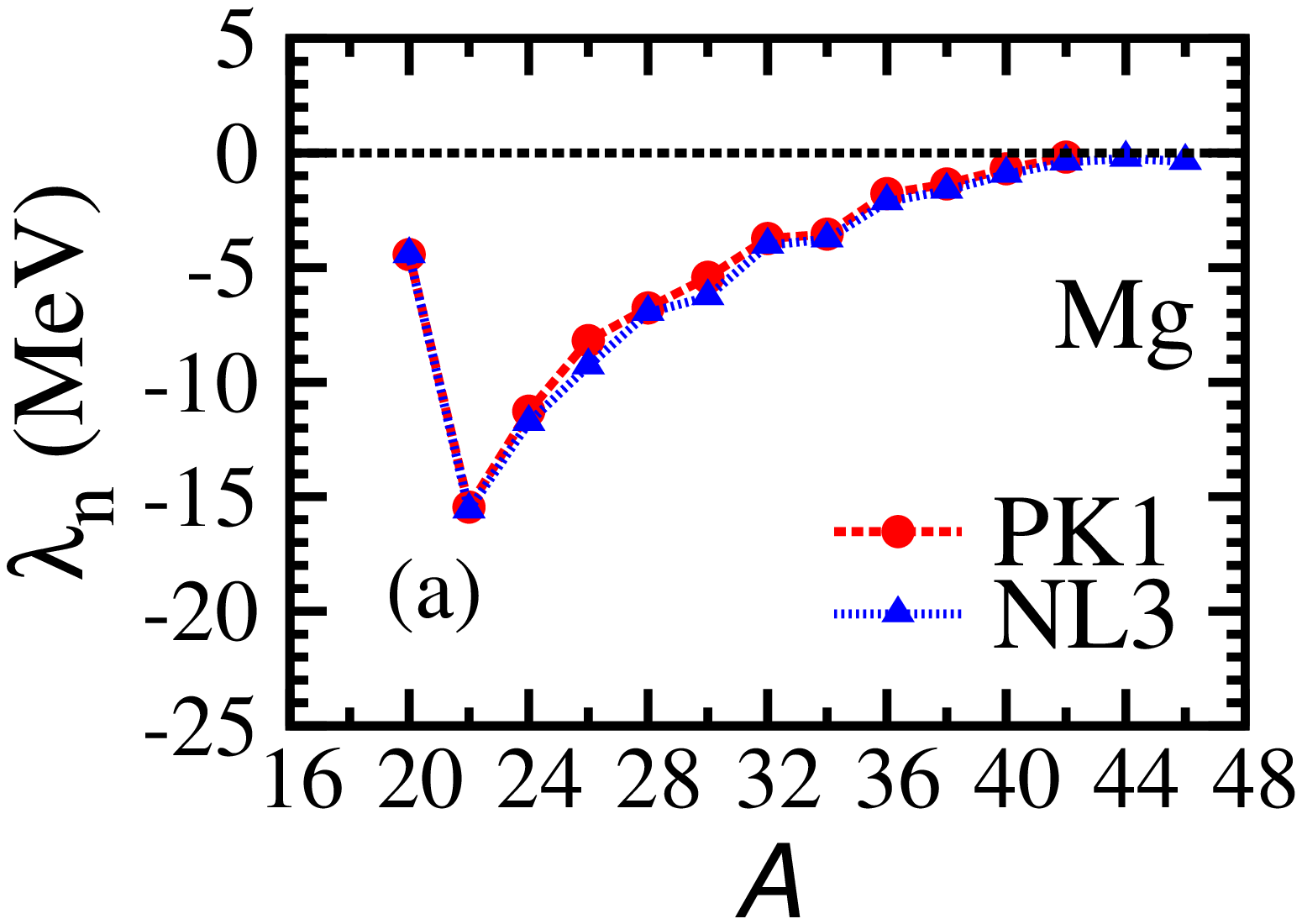}
~~~~~~~~~~~~~~~~~~~~~~~~~~~~~~~~~~~
 \includegraphics[width=0.45\textwidth]{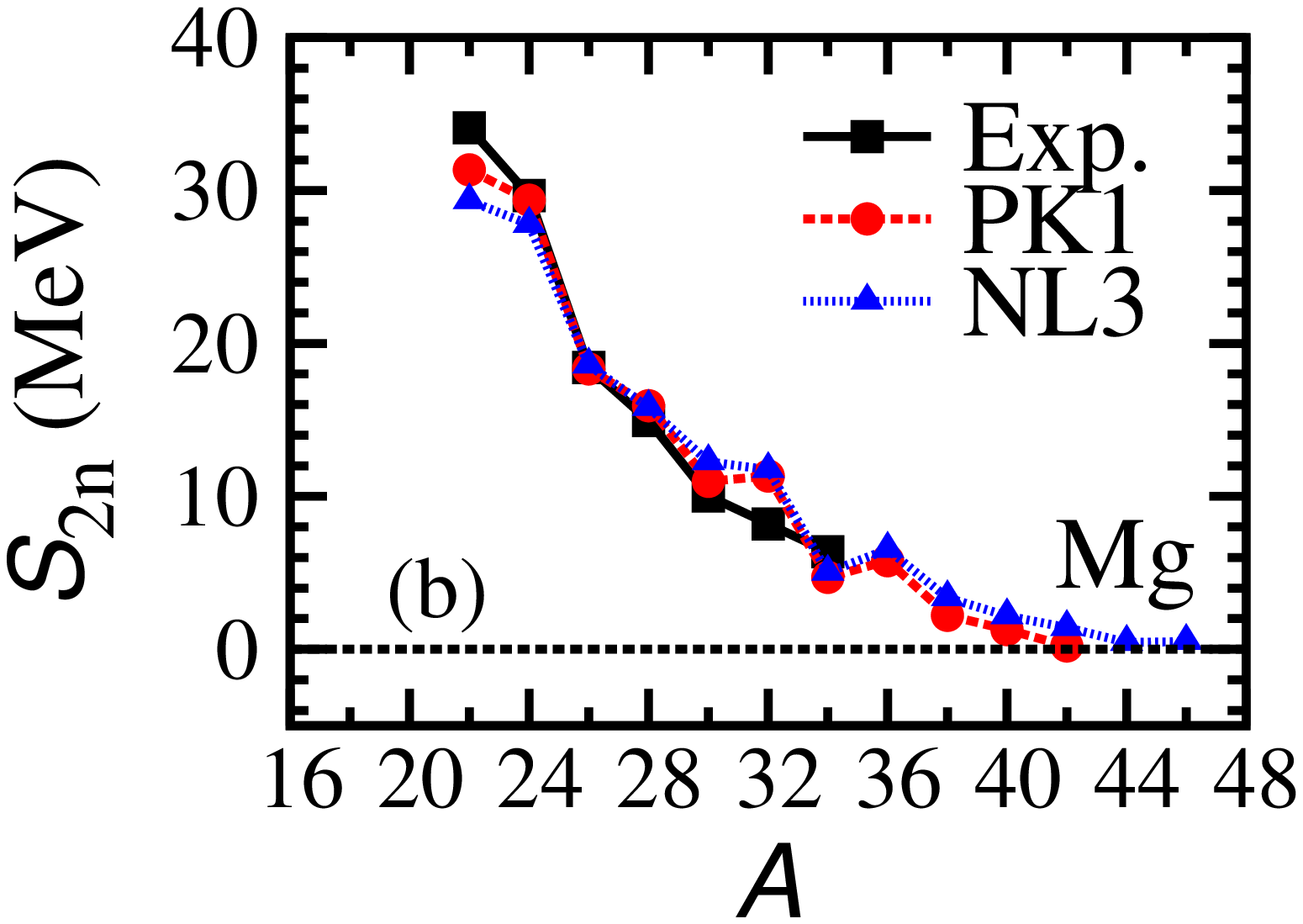}
\end{center}
\caption{(Color online) %
The neutron Fermi energy $\lambda_\mathrm{n}$ (a) and two neutron
separation energy $S_\mathrm{2n}$ (b) of magnesium isotopes
calculated with the parameter sets NL3 and PK1.
The data of $S_\mathrm{2n}$ (labeled as ``Exp.'') are taken
from Ref.~\cite{Audi2003_NPA729-337}.
}
\label{fig:mg-lam}
\end{figure}

Figure~\ref{fig:mg-lam} shows
the neutron Fermi energy $\lambda_\mathrm{n}$ and two neutron separation
energy $S_\mathrm{2n}$ of magnesium isotopes calculated with the parameter
sets NL3~\cite{Lalazissis1997_PRC55-540} and PK1~\cite{Long2004_PRC69-034319}.
The separation energies are compared with data taken
from Ref.~\cite{Audi2003_NPA729-337}.
Except the different prediction of the two-neutron drip line nucleus,
the results of the neutron Fermi surfaces and two neutron separation
energies are very similar for both parameter sets.
The calculated two neutron separation energies $S_{2n}$ of magnesium
isotopes agree reasonably well with the available experimental values
except for $^{32}$Mg. The large discrepancy in $^{32}$Mg is connected
to the shape and the shell structure at $N=20$ and will be discussed later.

Experimentally the nucleus $^{40}$Mg has been observed~\cite{Baumann2007_Nature449-1022}.
Theoretically there are several predictions on the last bound nucleus
in Mg isotopes, e.g., $^{44}$Mg in the phenomenological finite range
droplet model~\cite{Moller1995_ADNDT59-185}, $^{40}$Mg in a macroscopic-
microscopic model~\cite{Zhi2006_PLB638-166}, a RMF model with the parameter
set NLSH~\cite{Ren1996_PLB380-241}, and the Skyrme HFB
model with the parameter set SLy4 and solved in a 3-dimensional Cartesian 
mesh~\cite{Terasaki1997_NPA621-706},
and $^{42}$Mg from the Skyrme HFB model with SLy4 but solved in a transformed harmonic
oscillator basis~\cite{Stoitsov2000_PRC61-034311, *Stoitsov2003_PRC68-054312}
and the HFB21 mass table~\cite{Goriely2010_PRC82-035804}.
Therefore the prediction of the two-neutron drip line nucleus in Mg isotopes
is both model and parametrization dependent.
In our deformed RHB calculations with the parameter set NL3, $^{46}$Mg is
the last nucleus of which the neutron Fermi surface is negative and
the two neutron separation energy is positive.
However, with the parameter set PK1, $^{42}$Mg is predicted to be
the last nucleus within the two-neutron drip line.

\begin{figure}
\begin{center}
 \includegraphics[width=0.45\textwidth]{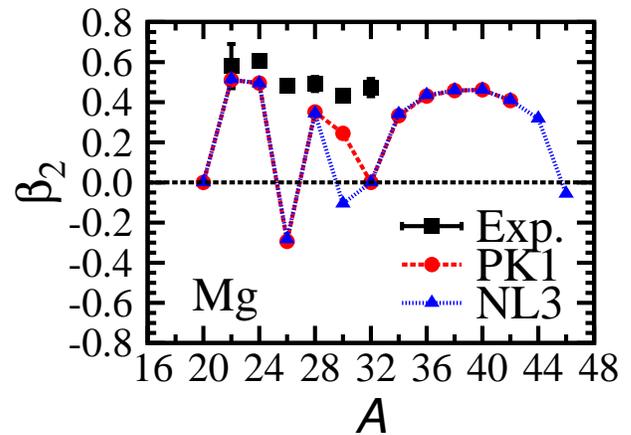}
\end{center}
\caption{(Color online) %
The quadrupole deformation parameter $\beta$ of magnesium isotopes calculated
with the parameter sets NL3 and PK1.
The experimental vales (``Exp.'') are taken from Ref.~\cite{RAMAN2001_ADNDT78-1}.
}
\label{fig:mg-beta}
\end{figure}

The comparison of the quadrupole deformation $\beta$ between the theory
and the experiment are given in Fig.~\ref{fig:mg-beta}.
The experimental values of $\beta$ is extracted from the measured
$B(E2:0^+_1\rightarrow2^+_1)$ values and therefore only absolute values are
available~\cite{RAMAN2001_ADNDT78-1}.
Generally speaking, the ground state quadrupole deformations $\beta$ calculated with both parameter
sets reproduce the data rather well. Exceptions are the nuclei $^{32}$Mg,
which turns out to be spherical in both models and $^{30}$Mg, which is prolate and
slightly less deformed than the experiment for PK1 and slightly oblate for NL3.
In $^{32}$Mg, the gap between the neutron levels $1d_{3/2}$ and $1f_{7/2}$
is almost 7 MeV which results in a strong closed shell at $N = 20$. Therefore the deformed RHB
calculations with both parameter sets predict spherical shapes for this nucleus.
This also results in a large discrepancy from the experiment for 
the two neutron separation energy $S_\mathrm{2n}$
of $^{32}$Mg as it is seen in Fig.~\ref{fig:mg-lam}.
Other mean field models predict spherical or almost spherical shapes for
$^{32}$Mg too~\cite{Patra1991_PLB273-13, Ren1996_PLB380-241, Terasaki1997_NPA621-706,
Lalazissis1999_ADNDT71-1,Rodriguez-Guzman2000_PRC62-054319,
Rodriguez-Guzman2002_NPA709-201,Niksic2006_PRC73-034308,
Yao2010_PRC81-044311, *Yao2011_PRC83-014308}. For the isotopes beyond this nucleus
with $32<A<46$ we observe large deformations, 
the so called ``island of inversion''~\cite{Sorlin2008_PPNP61-602,
Tripathi2008_PRL101-142504, Doornenbal2009_PRL103-032501, Wimmer2010_PRL105-252501}
which is related to the quenching of the $N=20$ shell closure. On the mean field
level the nucleus $^{32}$Mg does not belong to this island yet. In fact, going beyond
mean field and calculating the energy surface as a function of the
deformation parameters one finds that this nucleus is a transitional nucleus
with an extended shoulder reaching to large deformations. This leads in
GCM calculations with the Gogny force~\cite{Rodriguez-Guzman2000_PRC62-054319}
to wavefunctions with large fluctuations in deformation space and to a
large $B(E2:0^+_1\rightarrow2^+_1)$ value as it is observed in the
experiment~\cite{Church2005_PRC72-054320}. So far it is an open question,
why other GCM calculations based on Skyrme forces~\cite{Heenen2000_RIKENRev26-31}
or on the relativistic point coupling model PC-F1~\cite{Yao2011_PRC83-014308} cannot
reproduce this fact.

\begin{figure}
\begin{center}
\includegraphics[width=0.40\textwidth]{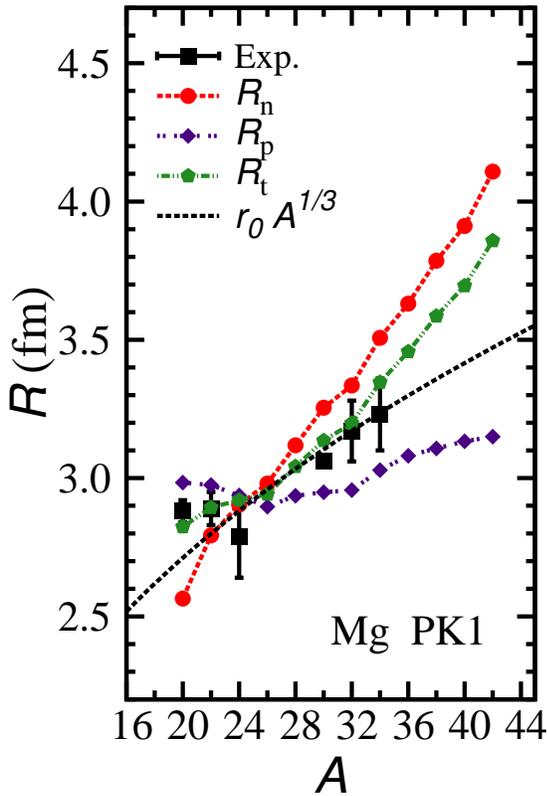}
\end{center}
\caption{(Color online) %
The root mean square radii for
magnesium isotopes are plotted as functions of the neutron number.
We display the neutron radius $R_{\rm n}$, the proton radius $R_{\rm p}$,
the matter radius $R_{\rm t}$, and the available data for
$R_{\rm t}$~\cite{Suzuki1998_NPA630-661, Kanungo2011_PRC83-021302R}.
The $r_0 A^{1/3}$ curve is included to guide the eye.}
\label{fig:mg-radii}
\end{figure}

Up to $^{42}$Mg, the deformed RHB results from the parameter set NL3 are very
similar to those from the parameter set PK1. Therefore in the following we
will mainly focus our discussion on the results from PK1.

In Fig.~\ref{fig:mg-radii}, the root mean square radii for
magnesium isotopes are plotted as functions of the neutron number.
We display neutron radii $R_{\rm n}$, proton radii $R_{\rm p}$, matter radii 
$R_{\rm t}$, the $r_0 A^{1/3}$ curve with $r_0 = 1$~fm, and experimental
matter radii~\cite{Suzuki1998_NPA630-661, Kanungo2011_PRC83-021302R}.
The proton radius are almost a constant with a very slow increase with increasing
$N$ due to the neutron-proton coupling included in the mean field.
With the neutron number increasing, the neutron radius $R_{\rm n}$ increases 
monotonically with an exception at $^{32}$Mg. 
The neutron radius of $^{32}$Mg is relatively small,
which is again due to the strong shell effect at $N=20$ in the mean field calculations.
It is shown that the deformed RHB results agree well with the experiment for the matter radius.
The calculated matter radius follows roughly the $r_0 A^{1/3}$ curve up to $A=34$. 
From $^{36}$Mg on, the matter radius lies much high above the $r_0 A^{1/3}$ curve.
This may indicate some exotic structure in these nuclei.

\begin{figure}
\begin{center}
\includegraphics[width=0.40\textwidth]{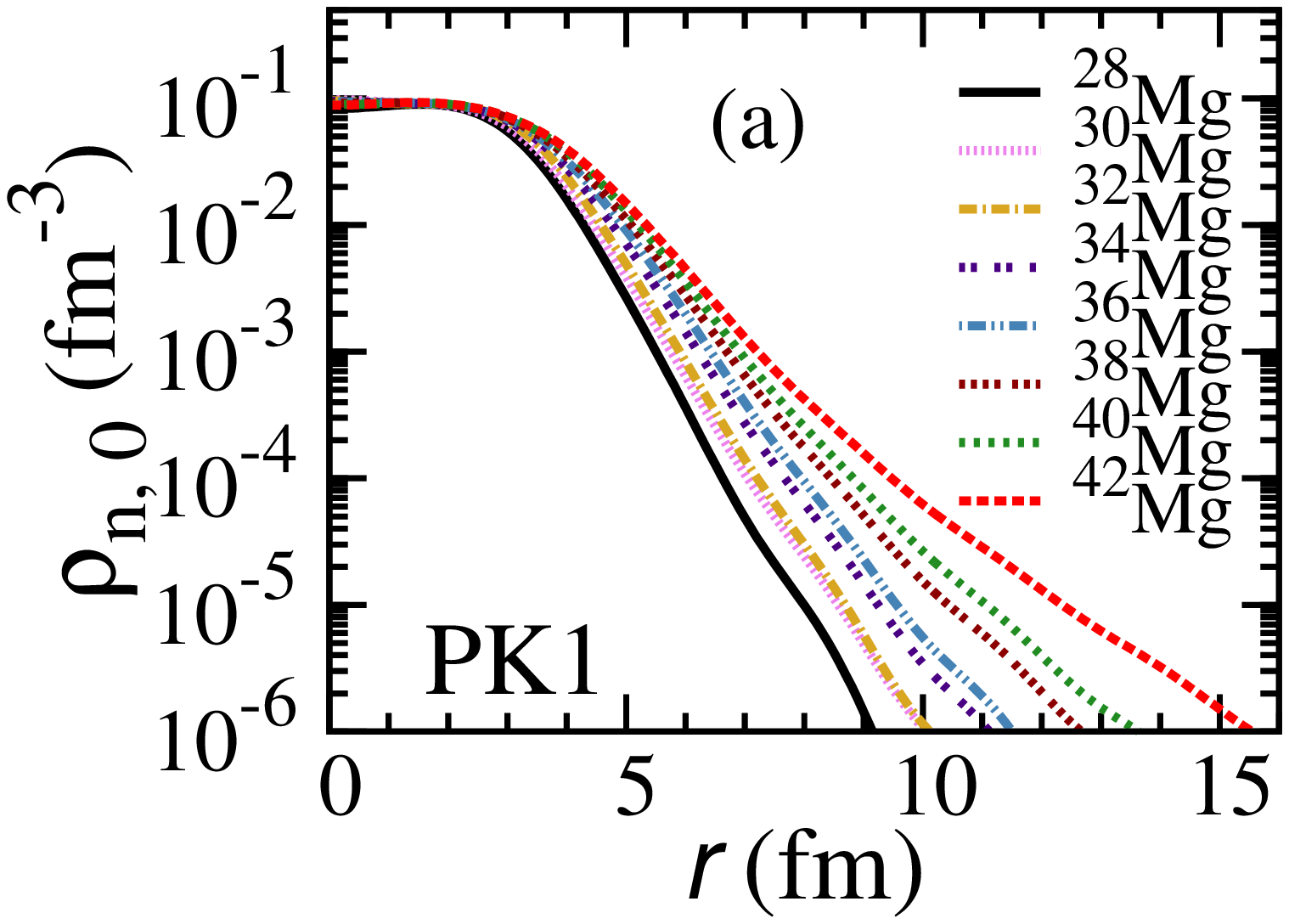}
~~~~~~~~~~~~~~~~~~~~~~~~~~~~~~~~~~~~~~~~~~~~~~~~~~~~~
\includegraphics[width=0.40\textwidth]{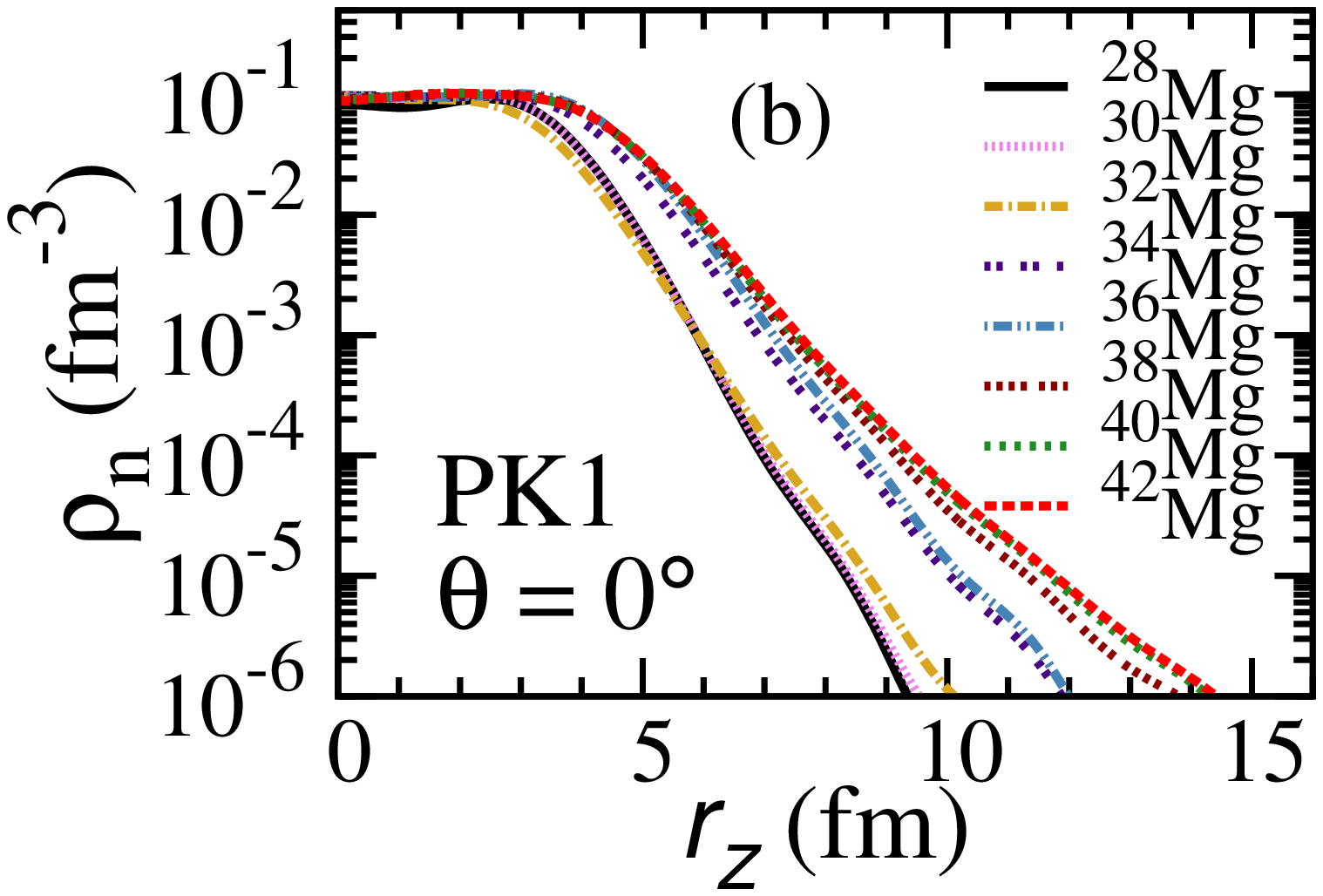}
~~~~~~~~~~~~~~~~~~~~~~~~~~~~~~~~~~~~~~~~~~~~~~~~~~~~~
\includegraphics[width=0.40\textwidth]{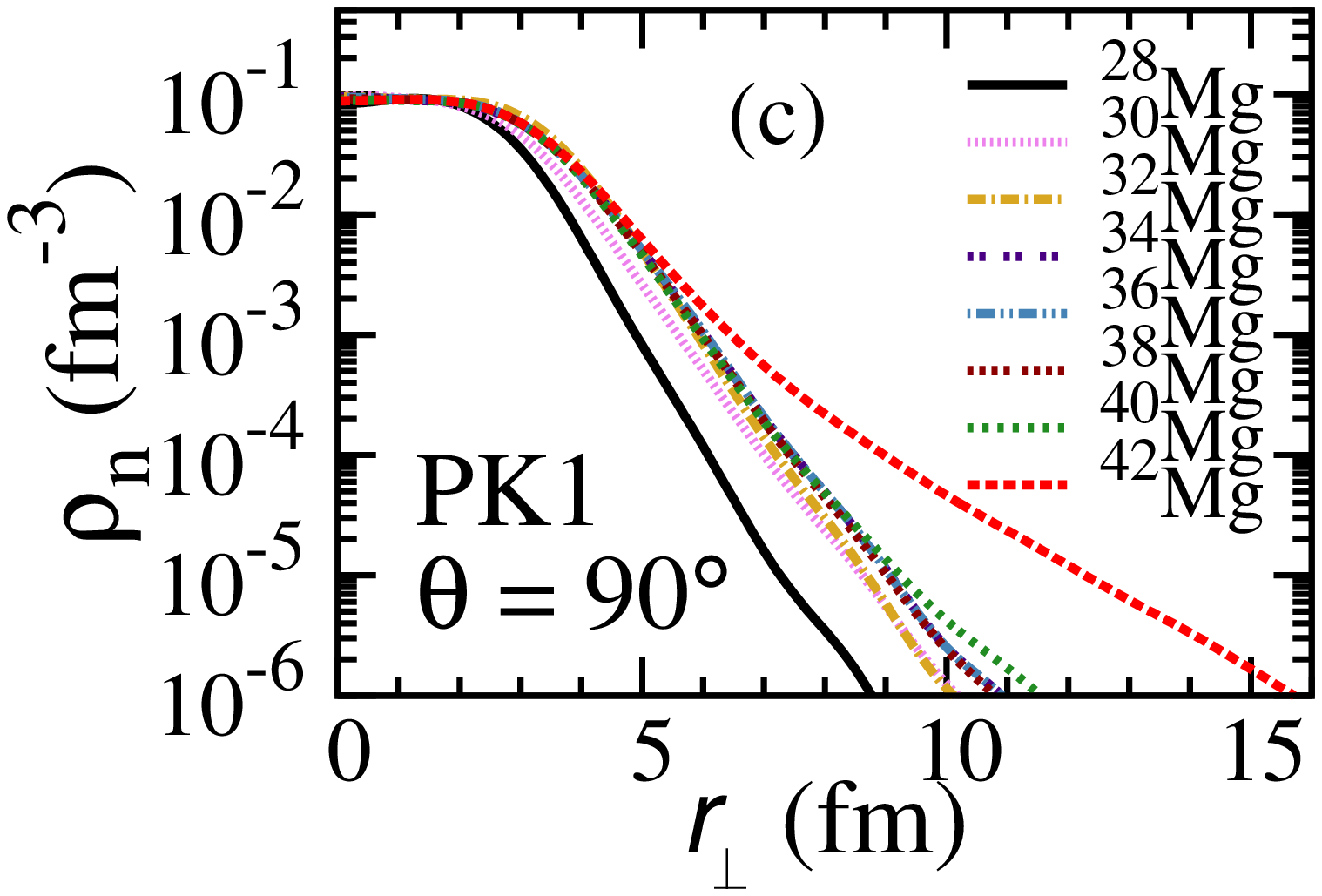}
\end{center}
\caption{(Color online) %
Neutron density profiles of even-even magnesium isotopes with $A \ge 28$
calculated with the parameter set PK1. Details are given in the text.}
\label{fig:mg-dens-profile}
\end{figure}

Figure~\ref{fig:mg-dens-profile} shows
neutron density profiles of even-even magnesium isotopes with $A \ge 28$
calculated with the parameter set PK1.
$\rho_{\mathrm{n},\lambda=0}(r)$ represents the spherical component of the neutron
density distribution (cf. Eq.~\ref{eq:expansion}).
$\rho_\mathrm{n}(z,r_\perp=0)$ with $r_\perp = \sqrt{x^2+y^2}$ refers to
the density distribution along the symmetry axis $z$ ($\theta = 0^\circ$)
and $\rho_\mathrm{n}(z=0,r_\perp)$ refers to that perpendicular to the
symmetry axis $z$ ($\theta = 90^\circ$).
With increasing $A$, the spherical component of the neutron density distribution
$\rho_{\mathrm{n},\lambda=0}(r)$ changes rapidly at
$^{42}$Mg.
The density distribution along the symmetry axis $\rho_\mathrm{n}(z,r_\perp=0)$
changes abruptly from $^{32}$Mg to $^{34}$Mg. This can be understood easily
by the change in shape in going from the spherical $^{32}$Mg to the prolate
$^{34}$Mg where the density is elongated along the $z$ axis.
In the direction perpendicular to the symmetry axis, the neutron density
$\rho_\mathrm{n}(z=0,r_\perp)$ of $^{42}$Mg extends very far away from
the center of the nucleus and a long tail emerges, revealing the formation
of a halo.

By comparing $\rho_\mathrm{n}(z,r_\perp=0)$ and $\rho_\mathrm{n}(z=0,r_\perp)$
for $^{42}$Mg, it is found that in the tail part, the neutron density
extends more along the direction perpendicular to the symmetry axis.
Since this nucleus as a whole is prolate, it indicates that the neutron tail
has a different shape as the core. This fact is similar to
the decoupling of the shape of the halo from the shape of the core found
for $^{44}$Mg in Ref.~\cite{Zhou2010_PRC82-011301R,*Zhou2011_JPCS312-092067}.
Next we will concentrate on $^{42}$Mg and discuss in details the structure
of its ground state.

\subsection{Ground state of $^{42}$Mg}

\begin{table}[htb!]
\begin{center}
\caption{
Properties of $^{42}$Mg at the ground
state and at the oblate minimum derived from deformed RHB calculations
with the parameter sets NL3 and PK1. The neutron and proton Fermi
surface $\lambda_\mathrm{n}$ and $\lambda_\mathrm{p}$, neutron, proton and
total quadrupole deformation $\beta_\mathrm{n}$, $\beta_\mathrm{p}$,
$\beta_\mathrm{t}$, neutron, proton and total radii $R_\mathrm{n}$,
$R_\mathrm{p}$, $R_\mathrm{t}$, neutron and proton pairing energies
$E_{\rm Pair}^\mathrm{n}, E_{\rm Pair}^\mathrm{p}$, and total binding energy
$E_{\rm B}$ are listed.
}
\label{tab:mg42-bulk}
\begin{tabular}{l |rr |rr}
\hline\hline
                  & \multicolumn{2}{c}{PK1}      & \multicolumn{2}{c}{NL3}      \\
\hline
 $\lambda_\mathrm{n}$      & $    -0.6147$ & $   -0.1753$ & $    -0.8805$ & $   -0.3989$ \\
 $\lambda_\mathrm{p}$      & $   -24.6731$ & $  -23.9050$ & $   -24.2695$ & $  -22.8118$ \\
 $\beta_\mathrm{n}$        & $    -0.3282$ & $    0.4155$ & $    -0.3299$ & $    0.4181$ \\
 $\beta_\mathrm{p}$        & $    -0.2426$ & $    0.3911$ & $    -0.2426$ & $    0.3917$ \\
 $\beta_\mathrm{t}$        & $    -0.3038$ & $    0.4085$ & $    -0.3049$ & $    0.4105$ \\
 $R_\mathrm{n}$            & $     4.0250$ & $    4.1077$ & $     4.0291$ & $    4.0971$ \\
 $R_\mathrm{p}$            & $     3.1208$ & $    3.1499$ & $     3.1393$ & $    3.1673$ \\
 $R_\mathrm{t}$            & $     3.7888$ & $    3.8584$ & $     3.7962$ & $    3.8544$ \\
 $E_{\rm Pair}^\mathrm{n}$ & $   -18.2511$ & $   -6.2620$ & $   -17.1509$ & $   -6.1595$ \\
 $E_{\rm Pair}^\mathrm{p}$ & $    -7.0405$ & $    0.0000$ & $    -6.7639$ & $    0.0000$ \\
 $E_{\rm B}$               & $  -265.4629$ & $ -266.4505$ & $  -270.6907$ & $ -270.6993$ \\
\hline
\end{tabular}
\end{center}
\end{table}

\begin{figure}
\begin{center}
\includegraphics[width=0.45\textwidth]{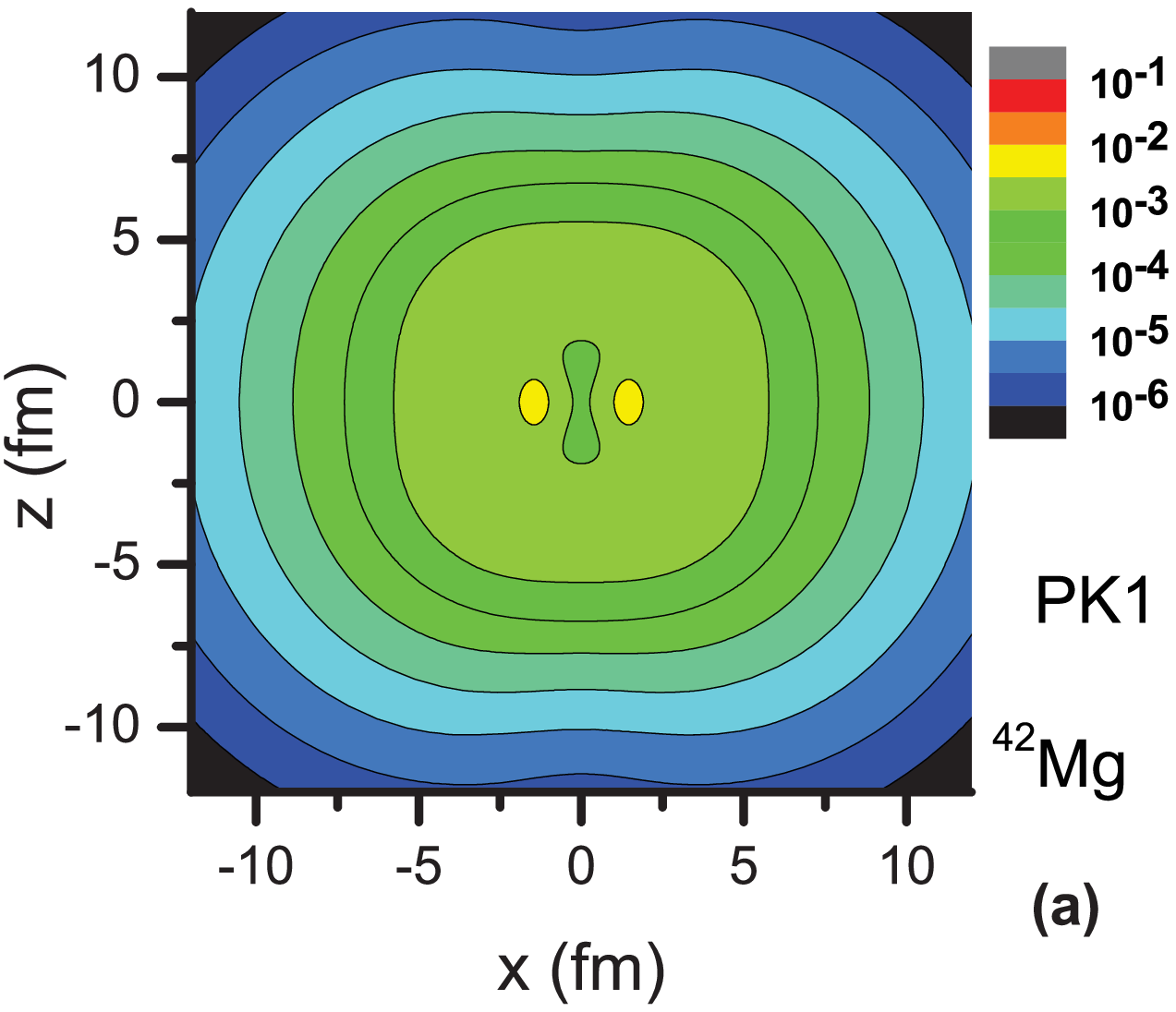}
~~~~~~~~~~~~~~~~~~~~~~~~~~~~~~~~~~~~~~~~~~~~~~~~~~~~~~~
\includegraphics[width=0.45\textwidth]{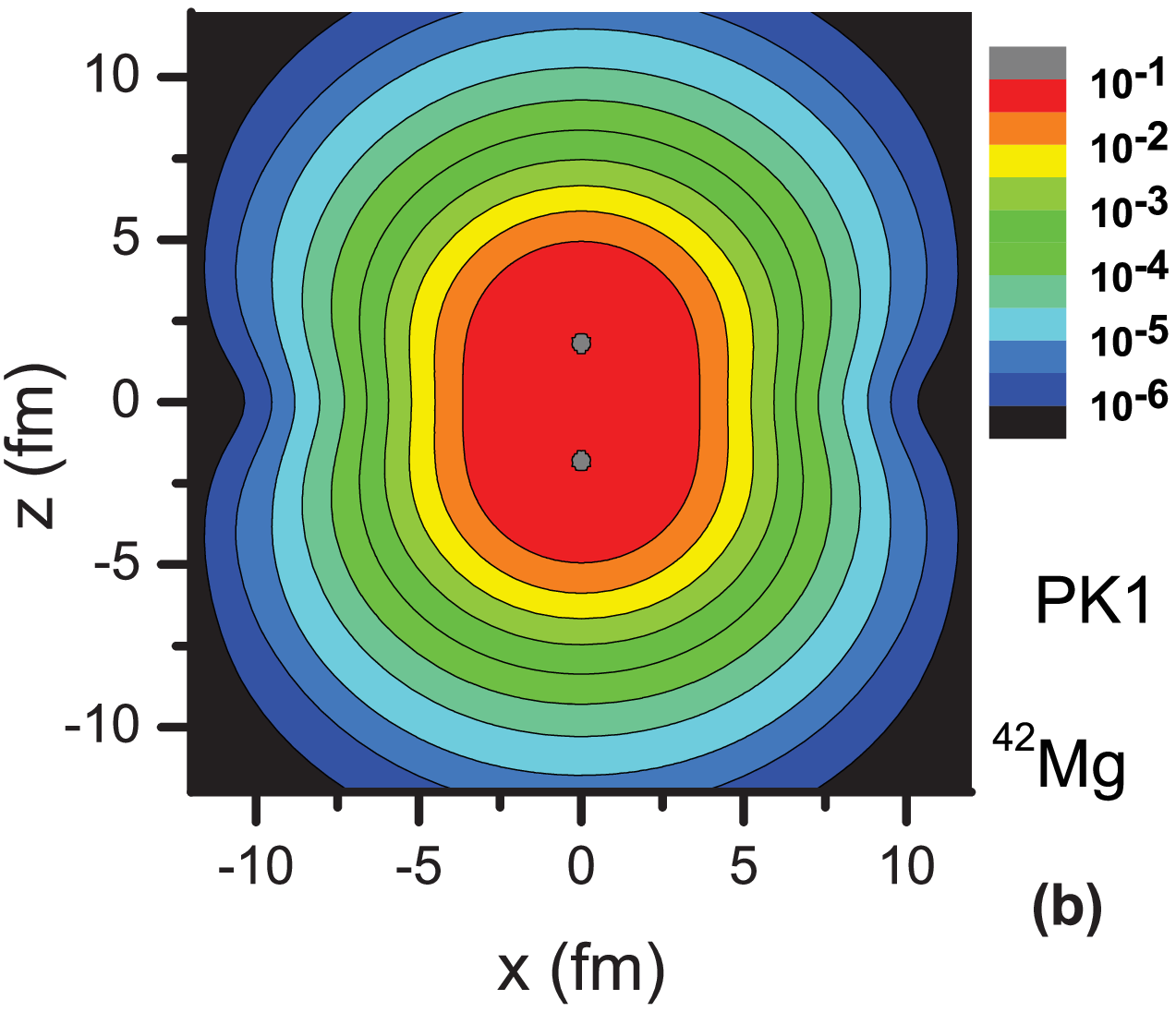}
\caption{(Color online) Density distributions of the ground state of $^{42}$Mg with the
$z$ axis as the symmetry axis: (a) the neutron halo, and (b) the neutron core.}
\label{fig:Mg42_pro_2d}
\end{center}
\end{figure}

In the calculations based on the parameter set PK1, the chain of Mg isotopes
reaches the two-neutron drip line at the nucleus $^{42}$Mg. Its properties are summarized
in Table~\ref{tab:mg42-bulk}. For $^{42}$Mg we find two minima in the energy
surface as a function of the deformation parameter $\beta$. The lower one has a
prolate shape and corresponds to the ground state of $^{42}$Mg. 
The second minimum has an oblate shape. 
From RMF calculations allowing for triaxial deformations~\cite{Lu2011_PRC84-014328,*Lu2012_PRC85-011301R}
we know, however, that the oblate minimum is not stable. It forms a saddle point
in the ($\beta$-$\gamma$) plane and therefore it does not correspond to an isomeric state.
The ground state is well deformed with a quadrupole deformation $\beta \approx 0.41$,
and a very small two neutron separation energy $S_{2n} \approx 0.22$ MeV.
The density distribution of this weakly bound nucleus has a very long tail
in the direction perpendicular to the symmetry axis
(cf. Fig.~\ref{fig:mg-dens-profile}), which indicates the prolate nucleus $^{42}$Mg
has an oblate halo.

The density distribution in Fig.~\ref{fig:Mg42_pro_2d} is decomposed into
contributions of the oblate ``Halo'' and of the prolate ``Core''.
Details of this decomposition will be given further down.
This indicates the decoupling between the deformations of the core and the halo.

\begin{figure}
\begin{center}
\includegraphics[width=0.40\textwidth]{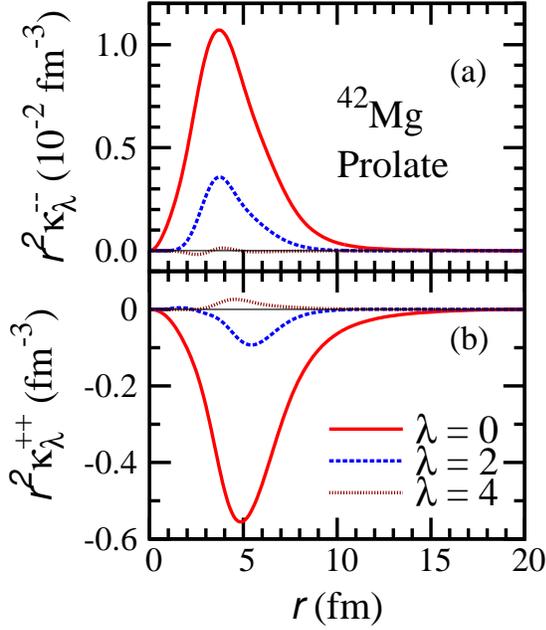}
\caption{(Color online)
The neutron pairing tensor $r^2\kappa^{--}_\lambda(r)$ (a) and 
$r^2\kappa^{++}_\lambda(r)$ (b) with $\lambda = 0$, 2, and 4 of the ground state
of $^{42}$Mg from the deformed RHB theory in continuum with the parameter set PK1. 
 }
\label{fig:Mg42_pro_kappa}
\end{center}
\end{figure}

Pairing correlations play a very important role in the formation of the
halo~\cite{Meng1996_PRL77-3963}. For the parameter set PK1 we find
in Table~\ref{tab:mg42-bulk} in the ground state of $^{42}$Mg a vanishing
pairing energy for protons and a paring energy $E_{\rm Pair}^\mathrm{n} = -6.26$ MeV
for the neutrons. For the zero range pairing interaction in Eq.~(\ref{eq:pairing_force})
only spin singlet ($S=0$) states and elements diagonal in the quantum number $p$
are taken into account in the pairing tensor. See appendix \ref{appendix:reldel} 
for more details concerning this assumption.
In Fig.~\ref{fig:Mg42_pro_kappa} we show the components $\kappa^{++}_\lambda(r)$
in Eq. (\ref{eq:E5})and $\kappa^{--}_\lambda(r)$ in Eq. (\ref{eq:E6}) of the pairing tensor
in the ground state of $^{42}$Mg for the parameter set PK1.
Figure~\ref{fig:Mg42_pro_kappa}(b) shows the main component $\kappa^{++}_{\lambda}(r)$
corresponding to the large components of the Dirac spinor.
Comparing Fig.~\ref{fig:Mg42_pro_kappa}(a) and Fig.~\ref{fig:Mg42_pro_kappa}(b) 
one finds that $\kappa^{--}_{\lambda}(r)$ is smaller by two orders of
magnitude than $\kappa^{++}_{\lambda}(r)$. 
The same sign for the quadrupole ($\lambda=2$) and the spherical ($\lambda=0$)
components can be understood by the fact that the ground state of $^{42}$Mg is prolate
in the present calculation.
The maximum of $\kappa^{++}_\lambda(r)$ appears at about 4.8 fm indicating
that paring in nuclei is a surface effect. The hexadecapole components ($\lambda=4$)
are much smaller than the spherical components ($\lambda=0$).

\begin{figure}
\begin{center}
\includegraphics[width=0.48\textwidth]{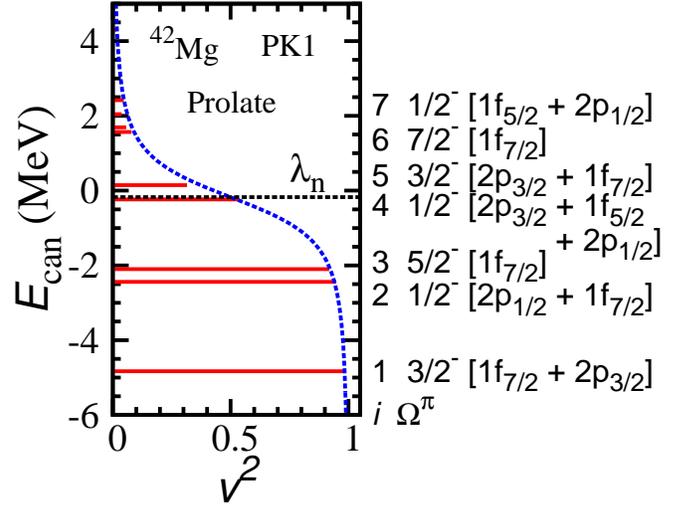}
\caption{(Color online) 
Single neutron levels of ground state of $^{42}$Mg in the canonical basis
as a function of the occupation probability $v^2$.
The order $i$, good quantum numbers $\Omega^\pi$, and the main spherical
components for orbitals close to the threshold are also given.
The blue dashed line corresponds to the BCS-formula with an average pairing gap.
}
\label{fig:Mg42_pro_lev}
\end{center}
\end{figure}

Weakly bound orbitals or those embedded in the continuum play a crucial role
in the formation of a nuclear halo~\cite{Meng1996_PRL77-3963,Meng1997_ZPA358-123,Meng1998_PRL80-460}.
In order to have an intuitive understanding of the single particle structure,
the canonical basis is constructed by the method given in Ref.~\cite{Meng1998_NPA635-3}.
The single particle spectrum around the Fermi level for the ground state of
$^{42}$Mg is shown in Fig.~\ref{fig:Mg42_pro_lev}.
For an axially deformed nucleus with spatial reflection symmetry,
the good quantum numbers of each single particle state include the parity
$\pi$ and the third component of the angular momentum $m$ (labeled by the
Nilsson quantum number $\Omega$ in the figures).
The occupation probabilities $v^2$ in the canonical basis have BCS-form~\cite{Ring1980}
and are given by the length of the horizontal lines in Fig.~\ref{fig:Mg42_pro_lev}.
To guide the eye we also show by a blue dashed line the BCS-formula calculated
with an average gap parameter.
The levels close to the threshold are labeled by the number $i$ according
to their energies, and their conserved quantum number
$\Omega^\pi$ as well as the main spherical components are given at the right hand side.
The neutron Fermi level is within the $pf$ shell and most of the single
particle levels have negative parities.
Since the chemical potential $\lambda_n \approx -175$ keV is negative,
the corresponding density $\rho(r)$ is localized and the particles occupying
the levels in the continuum are bound~\cite{Dobaczewski1984_NPA422-103}.
Since the chemical potential $\lambda_n$ is close to the continuum, orbitals
above the threshold have noticeable occupations due to the pairing correlations.
For instance, the occupation probability of the fifth level
($\Omega^\pi = 3/2^-$) is 31.5\%.
The fourth level $\Omega^\pi = 1/2^-$ is just below the threshold with
a single particle energy in the canonical basis $\varepsilon_{\rm can} = -0.234$ MeV
and an occupation probability of 53.0\%.
All the other levels below that orbital are well bound with $\varepsilon_{\rm can} < -2 $ MeV.
Similar to those of $^{44}$Mg in Ref.~\cite{Zhou2010_PRC82-011301R,*Zhou2011_JPCS312-092067},
the single neutron levels of $^{42}$Mg can be divided into two parts, the
deeply bound levels ($\varepsilon_{\rm can} < -2$ MeV) corresponding
to the ``core'', and the remaining weakly bound levels close to the
threshold ($\varepsilon_{\rm can} > -0.3$ MeV) and in the continuum
corresponding to the ``halo''.

\begin{figure}
\begin{center}
\includegraphics[width=0.40\textwidth]{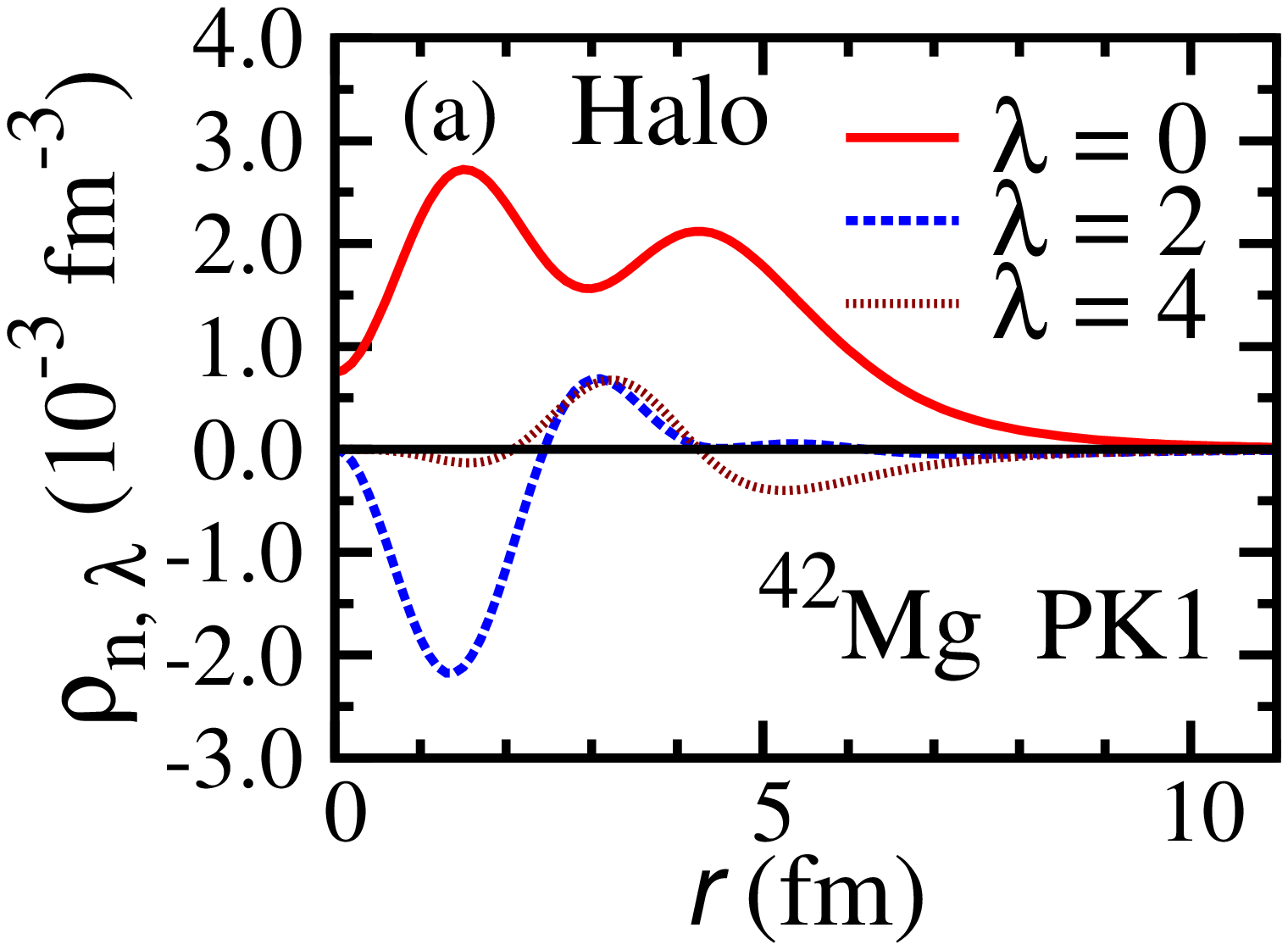}
~~~~~~~~~~~~~~~~~~~~~~~~~~~~~~~~~~~~~~~~~~~~~~~~~~~~
\includegraphics[width=0.40\textwidth]{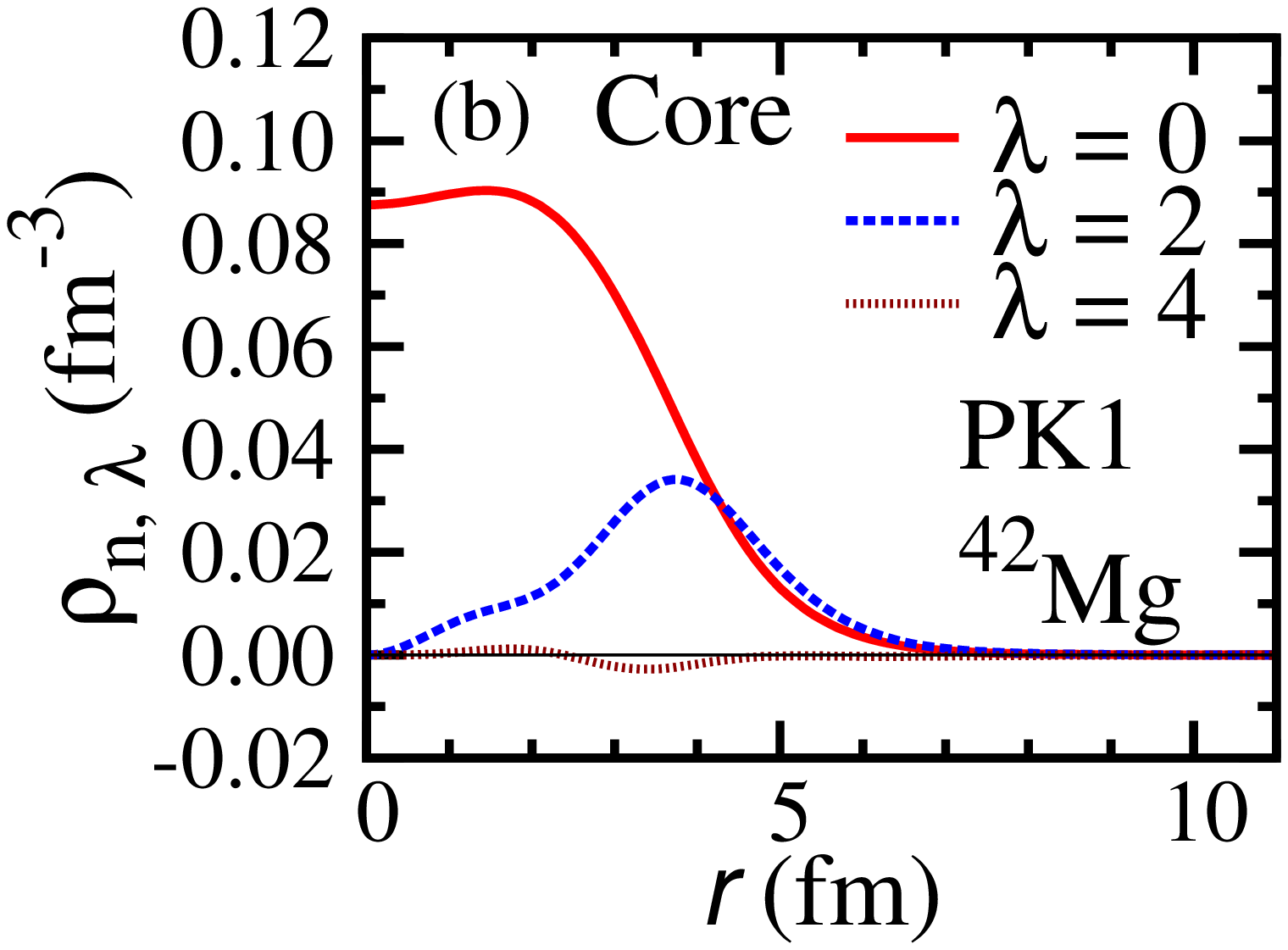}
\caption{(Color online)
Decomposition of the neutron density of the ground state of $^{42}$Mg into
spherical ($\lambda$ = 0), quadrupole ($\lambda$ = 2), and hexadecapole
($\lambda$ = 4) components for the halo (a) and the core (b).
}
\label{fig:Mg42_decom}
\end{center}
\end{figure}

We have already seen in Fig.~\ref{fig:Mg42_pro_2d} that the core is prolate
and the halo is oblate.
According to Eq.~(\ref{eq:expansion}) the density distributions of the core
and of the halo are decomposed into spherical ($\lambda=0$),
quadrupole ($\lambda=2$), and hexadecapole ($\lambda=4$) components
in Fig.~\ref{fig:Mg42_decom}.
The quadrupole component of the core turns out to be positive,
which is consistent with the prolate shape of $^{42}$Mg in the ground state.
However, for the halo, the quadrupole component is mainly negative,
which means the halo has an oblate shape.
This explains the decoupling between the quadrupole deformations
of the core and the halo. We also find in Fig.~\ref{fig:Mg42_decom}
that the spherical component is absolutely the main part of the density
distribution for both the core and the halo, and that the hexadecapole component in the
density distribution of the neutron halo is also noticeable.

\begin{figure}[htb!]
\begin{center}
\includegraphics[width=0.45\textwidth]{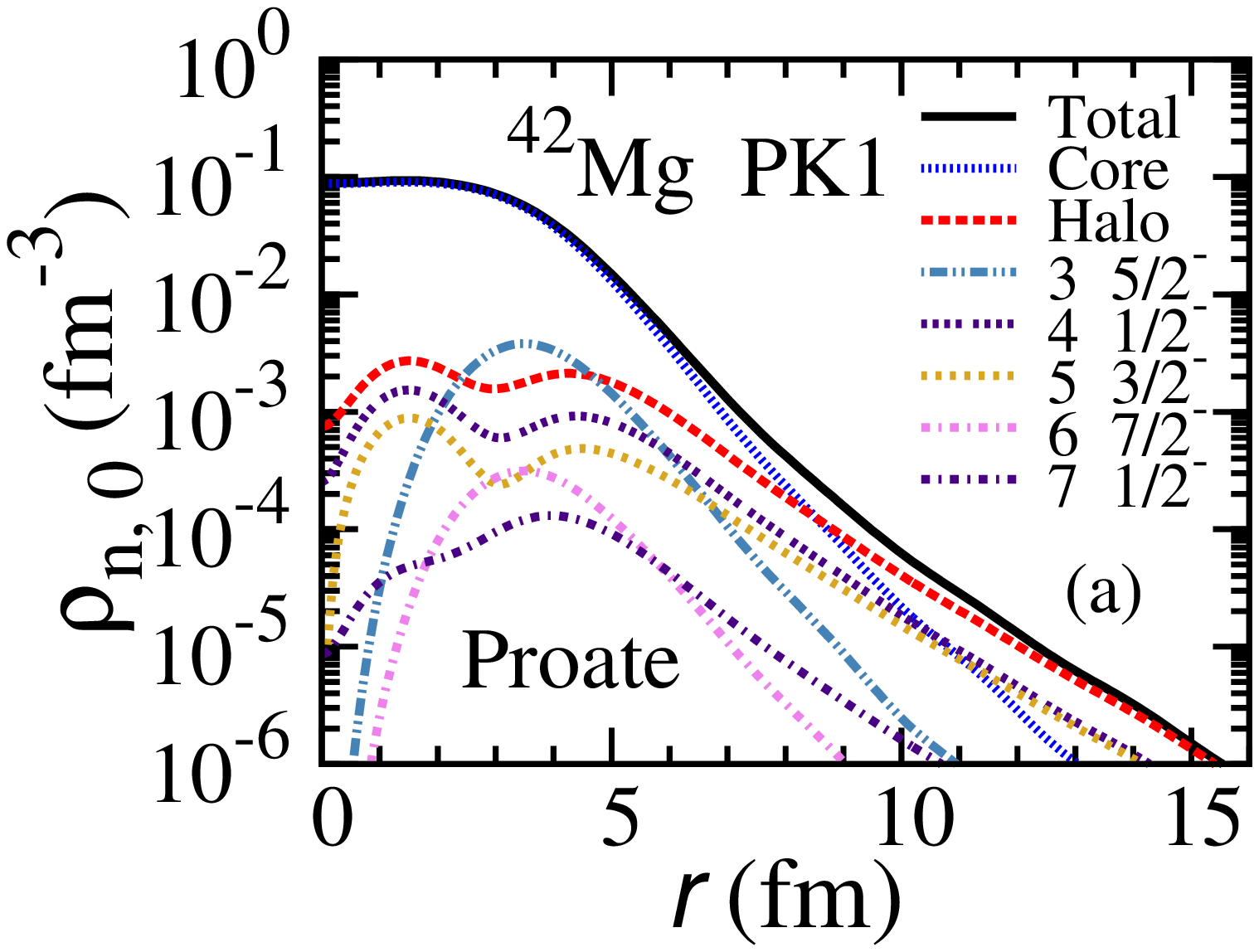}
~~~~~~~~~~~~~~~~~~~~~~~~~~~~~~~~~~~~~~~~~~~~~~~~~~~~
\includegraphics[width=0.45\textwidth]{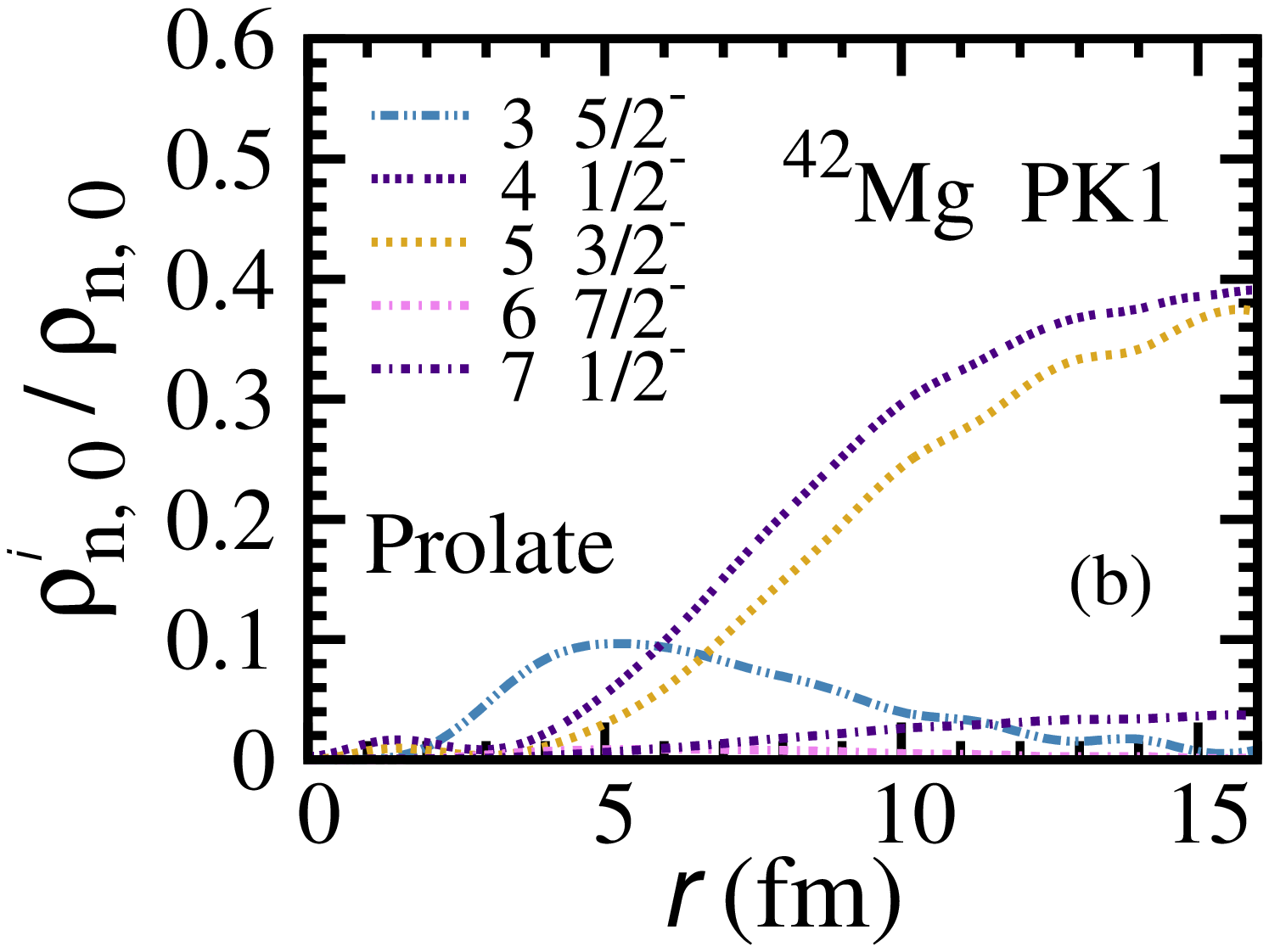}
\caption{(Color online)
Spherical components of neutron density distributions of ground state of $^{42}$Mg:
(a) the total density and its decomposition into the core and the halo 
and contributions from several neutron orbitals around the Fermi level;
(b) relative contributions of these neutron orbitals to
the total neutron density.
}
\label{fig:mg42_den_spl}
\end{center}
\end{figure}

In order to study the formation mechanism of the halo in more detail, we show in 
Fig.~\ref{fig:mg42_den_spl}(a) the main (spherical) components 
$\rho^i_{{\mathrm n},\lambda=0}$
of the density distribution for the weakly bound neutron orbitals $i$. 
Figure~\ref{fig:mg42_den_spl}(b) gives the ratio of these
spherical components $\rho^i_{{\mathrm n},\lambda=0}$ to the
spherical component of the total neutron density $\rho_{{\mathrm n},\lambda=0}$.
One can clearly see that far away from the center, the main contribution
comes from the 4th and 5th levels. Almost 80\% of
the total density distribution in the tail part comes from these two levels which are close
to the Fermi surface. Level 7 is embedded in the continuum and gives also some
contribution to the tail of the total density distribution.
However, the occupation probability of this level is just 5.7\%, so its
contribution is very small. The occupation probability of level 6 is
7.9\%, a bit lager than that of level 7. But there is almost no contribution
to the tail of total density from this level. By examining the spherical Woods-Saxon
components, it is found that the main component of level 6 is 1$f_{7/2}$.
The large centrifugal barrier of $f$ states with $l=3$ hinders strongly its
spatial extension. For level 7, about 31.3\% contribution comes from 2$p_{1/2}$
with a small centrifugal barrier and therefore the density can extend far
away from the center of the nucleus.

As it is shown in Fig.~\ref{fig:mg42_den_spl}, the halo is mainly
formed by level 4 and level 5 with occupation probabilities
of 53.0\% and 31.5\% respectively.
Having in mind the degeneracy 2 for each single particle level,
the occupation number of these two orbitals is about 1.7.
If we decompose the deformed wave functions of these
two orbitals in the spherical Woods-Saxon basis, it turns out that
in both cases the major part comes from $p$ waves, as indicated
on the right-hand side of Fig.~\ref{fig:Mg42_pro_lev}.
For level 4 ($\Omega^\pi = 1/2^-$), the probability of 2$p_{3/2}$,  1$f_{5/2}$,
and  2$p_{1/2}$ are 37.0\%, 32.3\%, and 21.2\% respectively.
For level 5 ($\Omega^\pi = 3/2^-$), 2$p_{3/2}$ is the dominant component with
a probability of 78.6\%.
The low centrifugal barrier for $p$ waves gives rise to the formation of
the halo.

The shape of the halo originates from the intrinsic structure of the weakly
bound or continuum orbitals~\cite{Misu1997_NPA614-44, Zhou2010_PRC82-011301R,*Zhou2011_JPCS312-092067}.
As discussed before, for the ground state of $^{42}$Mg, the halo is mainly
formed by level 4 and level 5.
We know that the angular distribution of $ |Y_{10} (\theta, \phi)|^2 \propto
\cos^2 \theta$ with a projection of the orbital angular momentum on the
symmetry axis $\Lambda = 0$ is prolate and that of
$ |Y_{1\pm 1} (\theta, \phi)|^2 \propto \sin^2 \theta$ with
$\Lambda = 1$ is oblate~\cite{Misu1997_NPA614-44}.
For level 4 ($\Omega^\pi = 1/2^-$), $\Lambda$ could be 0 or 1 since
the third component of total spin is $1/2$.
However, it turns out that the $\Lambda=0$ component dominates which results in
an oblate shape.
For level 5, since the third component of the total spin is $3/2$, $\Lambda$
can only be 1, which corresponds to an oblate shape too.
Therefore in $^{42}$Mg the shape of the halo is oblate and decouples from the prolate core.

\section{\label{sec:summary}Summary}

A deformed relativistic Hartree Bogoliubov theory in continuum
is developed in order to describe deformation effects in exotic nuclei
allowing for halo structures. The deformed RHB equations
are solved in a Woods-Saxon basis where the radial wave
functions have a proper asymptotic behavior at large distance from the
nuclear center. This is crucial for the formation of a halo.
The formalism and the numerical details of the deformed RHB theory are presented. 
Routine checks are made including convergence studies of the deformed RHB results 
concerning the mesh size, the box size and the size of the Woods-Saxon basis. 
The results are compared for spherical nuclei with solutions of the 1D continuum RHB
equations in the radial coordinate $r$ based on the Runge-Kutta method.

The deformed RHB theory in continuum is applied to study the chain of
magnesium isotopes with the parameter sets NL3 and PK1 of the Lagrangian.
Except for the different prediction of the two-neutron drip line nucleus,
the results of neutron Fermi surfaces and two neutron separation
energies are very similar for both parameter sets.
The calculated two neutron separation energies $S_{2n}$ of magnesium
isotopes agree reasonably well with the available experimental values
except for $^{32}$Mg, a well known problem connected with the shape and the shell
structure at $N=20$. For $^{32}$Mg, the gap between the neutron levels $1d_{3/2}$ and $1f_{7/2}$
is almost 7 MeV which results in a strong shell closure at $N = 20$.
The nuclear radii are also investigated, the deformed RHB results agree well with
the experiment for matter radii.
The proton radius is almost a constant with a very slow increase with increasing
$N$ due to the neutron-proton coupling included in the mean field.
A sharp increase in the neutron radius is
observed at $^{42}$Mg.

Detailed results are shown for the two-neutron drip line nucleus
$^{42}$Mg with the parameter set PK1, which is well deformed.
The ground state of $^{42}$Mg is prolate, however, it has an oblate neutron halo.
By examining in detail the density distributions, the pairing tensor,
and the single particle levels in the canonical basis in the deformed nucleus $^{42}$Mg,
it can be understood, why the shape of the neutron halo decouples from that of the core.
It is shown that the existence and the deformation of a possible neutron halo depends
essentially on the quantum numbers of the main components of the
single-particle orbits in the vicinity of the Fermi surface and
the shape of their single-particle density distributions.

In stable nuclei, there are situations that the levels of valence nucleons are sometimes also well
separated from the core. It is, however, a difficult question, whether there
exists cases of such a decoupling of shapes as we have seen here in the case
of loosely bound valence orbits close to the continuum limit, because in
stable nuclei even the valence nucleons are well bound in the average potential.

We can conclude that spherical and deformed relativistic Hartree Bogoliubov
theory in continuum is a very powerful tool providing a proper description
of exotic nuclei including halo phenomena, because it takes into account in a
self-consistent and microscopic way polarization effects, shape changes of
individual orbitals, pairing correlations and the coupling to the continuum
with proper boundary conditions.

\begin{acknowledgments}
This work has been supported in part by 
the Natural Science Foundation of China 
(10875157, 10975100, 10979066, 11105005, 11175002, and 11175252), 
by the Major State Basic Research Development Program of China (2007CB815000), 
by the Knowledge Innovation Project of Chinese Academy of Sciences 
(KJCX2-EW-N01 and KJCX2-YW-N32), 
and by the DFG cluster of excellence \textquotedblleft Origin and Structure of the
Universe\textquotedblright\ (www.universe-cluster.de). 
Part of the results described in this paper is obtained on the ScGrid of 
Supercomputing Center, Computer Network Information Center of Chinese Academy
of Sciences.
One of the authors (P.R.) would like to express his gratitude to J. Meng
for the kind hospitality extended to him at the Peking University.
Helpful discussions with N. V. Giai, B. N. Lu, Z. Y. Ma, N. Sandulescu, J. Terasaki, 
D. Vretenar, S. J. Wang, and S. Yamaji are gratefully acknowledged.
\end{acknowledgments}


\begin{thebibliography}{127}%
\makeatletter
\providecommand \@ifxundefined [1]{%
 \@ifx{#1\undefined}
}%
\providecommand \@ifnum [1]{%
 \ifnum #1\expandafter \@firstoftwo
 \else \expandafter \@secondoftwo
 \fi
}%
\providecommand \@ifx [1]{%
 \ifx #1\expandafter \@firstoftwo
 \else \expandafter \@secondoftwo
 \fi
}%
\providecommand \natexlab [1]{#1}%
\providecommand \enquote  [1]{``#1''}%
\providecommand \bibnamefont  [1]{#1}%
\providecommand \bibfnamefont [1]{#1}%
\providecommand \citenamefont [1]{#1}%
\providecommand \href@noop [0]{\@secondoftwo}%
\providecommand \href [0]{\begingroup \@sanitize@url \@href}%
\providecommand \@href[1]{\@@startlink{#1}\@@href}%
\providecommand \@@href[1]{\endgroup#1\@@endlink}%
\providecommand \@sanitize@url [0]{\catcode `\\12\catcode `\$12\catcode
  `\&12\catcode `\#12\catcode `\^12\catcode `\_12\catcode `\%12\relax}%
\providecommand \@@startlink[1]{}%
\providecommand \@@endlink[0]{}%
\providecommand \url  [0]{\begingroup\@sanitize@url \@url }%
\providecommand \@url [1]{\endgroup\@href {#1}{\urlprefix }}%
\providecommand \urlprefix  [0]{URL }%
\providecommand \Eprint [0]{\href }%
\providecommand \doibase [0]{http://dx.doi.org/}%
\providecommand \selectlanguage [0]{\@gobble}%
\providecommand \bibinfo  [0]{\@secondoftwo}%
\providecommand \bibfield  [0]{\@secondoftwo}%
\providecommand \translation [1]{[#1]}%
\providecommand \BibitemOpen [0]{}%
\providecommand \bibitemStop [0]{}%
\providecommand \bibitemNoStop [0]{.\EOS\space}%
\providecommand \EOS [0]{\spacefactor3000\relax}%
\providecommand \BibitemShut  [1]{\csname bibitem#1\endcsname}%
\let\auto@bib@innerbib\@empty
\bibitem [{\citenamefont {Xia}\ \emph {et~al.}(2002)\citenamefont {Xia},
  \citenamefont {Zhan}, \citenamefont {Wei}, \citenamefont {Yuan},
  \citenamefont {Song}, \citenamefont {Zhang}, \citenamefont {Yang},
  \citenamefont {Yuan}, \citenamefont {Gao}, \citenamefont {Zhao},
  \citenamefont {Yang}, \citenamefont {Xiao}, \citenamefont {Man},
  \citenamefont {Dang}, \citenamefont {Cai}, \citenamefont {Wang},
  \citenamefont {Tang}, \citenamefont {Qiao}, \citenamefont {Rao},
  \citenamefont {He}, \citenamefont {Mao},\ and\ \citenamefont
  {Zhou}}]{Xia2002_NIMA488-11}%
  \BibitemOpen
  \bibfield  {author} {\bibinfo {author} {\bibfnamefont {J.~W.}\ \bibnamefont
  {Xia}}, \bibinfo {author} {\bibfnamefont {W.~L.}\ \bibnamefont {Zhan}},
  \bibinfo {author} {\bibfnamefont {B.~W.}\ \bibnamefont {Wei}}, \bibinfo
  {author} {\bibfnamefont {Y.~J.}\ \bibnamefont {Yuan}}, \bibinfo {author}
  {\bibfnamefont {M.~T.}\ \bibnamefont {Song}}, \bibinfo {author}
  {\bibfnamefont {W.~Z.}\ \bibnamefont {Zhang}}, \bibinfo {author}
  {\bibfnamefont {X.~D.}\ \bibnamefont {Yang}}, \bibinfo {author}
  {\bibfnamefont {P.}~\bibnamefont {Yuan}}, \bibinfo {author} {\bibfnamefont
  {D.~Q.}\ \bibnamefont {Gao}}, \bibinfo {author} {\bibfnamefont {H.~W.}\
  \bibnamefont {Zhao}}, \bibinfo {author} {\bibfnamefont {X.~T.}\ \bibnamefont
  {Yang}}, \bibinfo {author} {\bibfnamefont {G.~Q.}\ \bibnamefont {Xiao}},
  \bibinfo {author} {\bibfnamefont {K.~T.}\ \bibnamefont {Man}}, \bibinfo
  {author} {\bibfnamefont {J.~R.}\ \bibnamefont {Dang}}, \bibinfo {author}
  {\bibfnamefont {X.~H.}\ \bibnamefont {Cai}}, \bibinfo {author} {\bibfnamefont
  {Y.~F.}\ \bibnamefont {Wang}}, \bibinfo {author} {\bibfnamefont {J.~Y.}\
  \bibnamefont {Tang}}, \bibinfo {author} {\bibfnamefont {W.~M.}\ \bibnamefont
  {Qiao}}, \bibinfo {author} {\bibfnamefont {Y.~N.}\ \bibnamefont {Rao}},
  \bibinfo {author} {\bibfnamefont {Y.}~\bibnamefont {He}}, \bibinfo {author}
  {\bibfnamefont {L.~Z.}\ \bibnamefont {Mao}}, \ and\ \bibinfo {author}
  {\bibfnamefont {Z.~Z.}\ \bibnamefont {Zhou}},\ }\href {\doibase
  10.1016/S0168-9002(02)00475-8} {\bibfield  {journal} {\bibinfo  {journal}
  {Nucl. Instrum. Methods Phys. Res. A}\ }\textbf {\bibinfo {volume} {488}},\
  \bibinfo {pages} {11} (\bibinfo {year} {2002})}\BibitemShut {NoStop}%
\bibitem [{\citenamefont {Zhan}\ \emph {et~al.}(2010)\citenamefont {Zhan},
  \citenamefont {Xu}, \citenamefont {Xiao}, \citenamefont {Xia}, \citenamefont
  {Zhao},\ and\ \citenamefont {Yuan}}]{Zhan2010_NPA834-694c}%
  \BibitemOpen
  \bibfield  {author} {\bibinfo {author} {\bibfnamefont {W.}~\bibnamefont
  {Zhan}}, \bibinfo {author} {\bibfnamefont {H.}~\bibnamefont {Xu}}, \bibinfo
  {author} {\bibfnamefont {G.}~\bibnamefont {Xiao}}, \bibinfo {author}
  {\bibfnamefont {J.}~\bibnamefont {Xia}}, \bibinfo {author} {\bibfnamefont
  {H.}~\bibnamefont {Zhao}}, \ and\ \bibinfo {author} {\bibfnamefont
  {Y.}~\bibnamefont {Yuan}},\ }\href {\doibase 10.1016/j.nuclphysa.2010.01.126}
  {\bibfield  {journal} {\bibinfo  {journal} {Nucl. Phys. A}\ }\textbf
  {\bibinfo {volume} {834}},\ \bibinfo {pages} {694c} (\bibinfo {year}
  {2010})}\BibitemShut {NoStop}%
\bibitem [{\citenamefont {Sturm}\ \emph {et~al.}(2010)\citenamefont {Sturm},
  \citenamefont {Sharkov},\ and\ \citenamefont
  {St\"ocker}}]{Sturm2010_NPA834-682c}%
  \BibitemOpen
  \bibfield  {author} {\bibinfo {author} {\bibfnamefont {C.}~\bibnamefont
  {Sturm}}, \bibinfo {author} {\bibfnamefont {B.}~\bibnamefont {Sharkov}}, \
  and\ \bibinfo {author} {\bibfnamefont {H.}~\bibnamefont {St\"ocker}},\ }\href
  {\doibase 10.1016/j.nuclphysa.2010.01.124} {\bibfield  {journal} {\bibinfo
  {journal} {Nucl. Phys. A}\ }\textbf {\bibinfo {volume} {834}},\ \bibinfo
  {pages} {682c} (\bibinfo {year} {2010})}\BibitemShut {NoStop}%
\bibitem [{\citenamefont {Gales}(2010)}]{Gales2010_NPA834-717c}%
  \BibitemOpen
  \bibfield  {author} {\bibinfo {author} {\bibfnamefont {S.}~\bibnamefont
  {Gales}},\ }\href {\doibase 10.1016/j.nuclphysa.2010.01.130} {\bibfield
  {journal} {\bibinfo  {journal} {Nucl. Phys. A}\ }\textbf {\bibinfo {volume}
  {834}},\ \bibinfo {pages} {717c} (\bibinfo {year} {2010})}\BibitemShut
  {NoStop}%
\bibitem [{\citenamefont {Motobayashi}\ and\ \citenamefont
  {Yano}(2007)}]{Motobayashi2007_NPN17-5}%
  \BibitemOpen
  \bibfield  {author} {\bibinfo {author} {\bibfnamefont {T.}~\bibnamefont
  {Motobayashi}}\ and\ \bibinfo {author} {\bibfnamefont {Y.}~\bibnamefont
  {Yano}},\ }\href {www.nupecc.org/npn/npn174.pdf} {\bibfield  {journal}
  {\bibinfo  {journal} {Nucl. Phys. News}\ }\textbf {\bibinfo {volume} {17}},\
  \bibinfo {pages} {5} (\bibinfo {year} {2007})}\BibitemShut {NoStop}%
\bibitem [{\citenamefont {Motobayashi}(2010)}]{Motobayashi2010_NPA834-707c}%
  \BibitemOpen
  \bibfield  {author} {\bibinfo {author} {\bibfnamefont {T.}~\bibnamefont
  {Motobayashi}},\ }\href {\doibase 10.1016/j.nuclphysa.2010.01.128} {\bibfield
   {journal} {\bibinfo  {journal} {Nucl. Phys. A}\ }\textbf {\bibinfo {volume}
  {834}},\ \bibinfo {pages} {707c} (\bibinfo {year} {2010})}\BibitemShut
  {NoStop}%
\bibitem [{\citenamefont {Thoennessen}(2010)}]{Thoennessen2010_NPA834-688c}%
  \BibitemOpen
  \bibfield  {author} {\bibinfo {author} {\bibfnamefont {M.}~\bibnamefont
  {Thoennessen}},\ }\href {\doibase 10.1016/j.nuclphysa.2010.01.125} {\bibfield
   {journal} {\bibinfo  {journal} {Nucl. Phys. A}\ }\textbf {\bibinfo {volume}
  {834}},\ \bibinfo {pages} {688c} (\bibinfo {year} {2010})}\BibitemShut
  {NoStop}%
\bibitem [{\citenamefont {Choi}()}]{Choi2010_ISNPA}%
  \BibitemOpen
  \bibfield  {author} {\bibinfo {author} {\bibfnamefont {S.}~\bibnamefont
  {Choi}},\ }\href
  {http://indico.ihep.ac.cn/conferenceTimeTable.py?confId=1714} {\enquote
  {\bibinfo {title} {{KoRIA} project - {RI} accelerator in {Korea}},}\
  }\bibinfo {note} {Invited talk given in International Symposium on Nuclear
  Physics in Asia, 14-15 October, 2010, Beihang University,
  Beijing}\BibitemShut {NoStop}%
\bibitem [{\citenamefont {Bertulani}\ \emph {et~al.}(2001)\citenamefont
  {Bertulani}, \citenamefont {Hussein},\ and\ \citenamefont
  {M\"unzenberg}}]{Bertulani2001_PRB}%
  \BibitemOpen
  \bibfield  {author} {\bibinfo {author} {\bibfnamefont {C.~A.}\ \bibnamefont
  {Bertulani}}, \bibinfo {author} {\bibfnamefont {M.~S.}\ \bibnamefont
  {Hussein}}, \ and\ \bibinfo {author} {\bibfnamefont {G.}~\bibnamefont
  {M\"unzenberg}},\ }\href {http://books.google.com/books?id=sxWMQAAACAAJ}
  {\emph {\bibinfo {title} {Physics of Radioactive Beams}}}\ (\bibinfo
  {publisher} {Nova Science Publishers, Inc.},\ \bibinfo {year}
  {2001})\BibitemShut {NoStop}%
\bibitem [{\citenamefont {M\"uller}\ and\ \citenamefont
  {Sherrill}(1993)}]{Mueller1993_ARNPS43-529}%
  \BibitemOpen
  \bibfield  {author} {\bibinfo {author} {\bibfnamefont {A.~C.}\ \bibnamefont
  {M\"uller}}\ and\ \bibinfo {author} {\bibfnamefont {B.~M.}\ \bibnamefont
  {Sherrill}},\ }\href {\doibase 10.1146/annurev.ns.43.120193.002525}
  {\bibfield  {journal} {\bibinfo  {journal} {Annu. Rev. Nucl. Part. Sci.}\
  }\textbf {\bibinfo {volume} {43}},\ \bibinfo {pages} {529} (\bibinfo {year}
  {1993})}\BibitemShut {NoStop}%
\bibitem [{\citenamefont {Tanihata}(1995)}]{Tanihata1995_PPNP35-505}%
  \BibitemOpen
  \bibfield  {author} {\bibinfo {author} {\bibfnamefont {I.}~\bibnamefont
  {Tanihata}},\ }\href {\doibase 10.1016/0146-6410(95)00046-L} {\bibfield
  {journal} {\bibinfo  {journal} {Prog. Part. Nucl. Phys.}\ }\textbf {\bibinfo
  {volume} {35}},\ \bibinfo {pages} {505} (\bibinfo {year} {1995})}\BibitemShut
  {NoStop}%
\bibitem [{\citenamefont {Hansen}\ \emph {et~al.}(1995)\citenamefont {Hansen},
  \citenamefont {Jensen},\ and\ \citenamefont
  {Jonson}}]{Hansen1995_ARNPS45-591}%
  \BibitemOpen
  \bibfield  {author} {\bibinfo {author} {\bibfnamefont {P.~G.}\ \bibnamefont
  {Hansen}}, \bibinfo {author} {\bibfnamefont {A.~S.}\ \bibnamefont {Jensen}},
  \ and\ \bibinfo {author} {\bibfnamefont {B.}~\bibnamefont {Jonson}},\ }\href
  {\doibase 10.1146/annurev.ns.45.120195.003111} {\bibfield  {journal}
  {\bibinfo  {journal} {Annu. Rev. Nucl. Part. Sci.}\ }\textbf {\bibinfo
  {volume} {45}},\ \bibinfo {pages} {591} (\bibinfo {year} {1995})}\BibitemShut
  {NoStop}%
\bibitem [{\citenamefont {Casten}\ and\ \citenamefont
  {Sherrill}(2000)}]{Casten2000_PPNP45-S171}%
  \BibitemOpen
  \bibfield  {author} {\bibinfo {author} {\bibfnamefont {R.~F.}\ \bibnamefont
  {Casten}}\ and\ \bibinfo {author} {\bibfnamefont {B.~M.}\ \bibnamefont
  {Sherrill}},\ }\href {\doibase 10.1016/S0146-6410(00)90013-9} {\bibfield
  {journal} {\bibinfo  {journal} {Prog. Part. Nucl. Phys.}\ }\textbf {\bibinfo
  {volume} {45}},\ \bibinfo {pages} {S171} (\bibinfo {year}
  {2000})}\BibitemShut {NoStop}%
\bibitem [{\citenamefont {Johnson}(2004)}]{Johnson2004_PR389-1}%
  \BibitemOpen
  \bibfield  {author} {\bibinfo {author} {\bibfnamefont {B.}~\bibnamefont
  {Johnson}},\ }\href {\doibase 10.1016/j.physrep.2003.07.004} {\bibfield
  {journal} {\bibinfo  {journal} {Phys. Rep.}\ }\textbf {\bibinfo {volume}
  {389}},\ \bibinfo {pages} {1} (\bibinfo {year} {2004})}\BibitemShut {NoStop}%
\bibitem [{\citenamefont {Jensen}\ \emph {et~al.}(2004)\citenamefont {Jensen},
  \citenamefont {Riisager}, \citenamefont {Fedorov},\ and\ \citenamefont
  {Garrido}}]{Jensen2004_RMP76-215}%
  \BibitemOpen
  \bibfield  {author} {\bibinfo {author} {\bibfnamefont {A.~S.}\ \bibnamefont
  {Jensen}}, \bibinfo {author} {\bibfnamefont {K.}~\bibnamefont {Riisager}},
  \bibinfo {author} {\bibfnamefont {D.~V.}\ \bibnamefont {Fedorov}}, \ and\
  \bibinfo {author} {\bibfnamefont {E.}~\bibnamefont {Garrido}},\ }\href
  {\doibase 10.1103/RevModPhys.76.215} {\bibfield  {journal} {\bibinfo
  {journal} {Rev. Mod. Phys.}\ }\textbf {\bibinfo {volume} {76}},\ \bibinfo
  {pages} {215} (\bibinfo {year} {2004})}\BibitemShut {NoStop}%
\bibitem [{\citenamefont {Ershov}\ \emph {et~al.}(2010)\citenamefont {Ershov},
  \citenamefont {Grigorenko}, \citenamefont {Vaagen},\ and\ \citenamefont
  {Zhukov}}]{Ershov2010_JPG37-064026}%
  \BibitemOpen
  \bibfield  {author} {\bibinfo {author} {\bibfnamefont {S.~N.}\ \bibnamefont
  {Ershov}}, \bibinfo {author} {\bibfnamefont {L.~V.}\ \bibnamefont
  {Grigorenko}}, \bibinfo {author} {\bibfnamefont {J.~S.}\ \bibnamefont
  {Vaagen}}, \ and\ \bibinfo {author} {\bibfnamefont {M.~V.}\ \bibnamefont
  {Zhukov}},\ }\href {\doibase 10.1088/0954-3899/37/6/064026} {\bibfield
  {journal} {\bibinfo  {journal} {J. Phys. G: Nucl. Phys.}\ }\textbf {\bibinfo
  {volume} {37}},\ \bibinfo {pages} {064026} (\bibinfo {year}
  {2010})}\BibitemShut {NoStop}%
\bibitem [{\citenamefont {Cao}\ and\ \citenamefont
  {Ye}(2011)}]{Cao2011_SciChinaPAM54S1-1}%
  \BibitemOpen
  \bibfield  {author} {\bibinfo {author} {\bibfnamefont {Z.-X.}\ \bibnamefont
  {Cao}}\ and\ \bibinfo {author} {\bibfnamefont {Y.-L.}\ \bibnamefont {Ye}},\
  }\href {\doibase 10.1007/s11433-011-4423-9} {\bibfield  {journal} {\bibinfo
  {journal} {Sci. China-Phys. Mech. Astron.}\ }\textbf {\bibinfo {volume} {54
  (Suppl. 1)}},\ \bibinfo {pages} {s1} (\bibinfo {year} {2011})}\BibitemShut
  {NoStop}%
\bibitem [{\citenamefont {Tanihata}\ \emph {et~al.}(1985)\citenamefont
  {Tanihata}, \citenamefont {Hamagaki}, \citenamefont {Hashimoto},
  \citenamefont {Shida}, \citenamefont {Yoshikawa}, \citenamefont {Sugimoto},
  \citenamefont {Yamakawa}, \citenamefont {Kobayashi},\ and\ \citenamefont
  {Takahashi}}]{Tanihata1985_PRL55-2676}%
  \BibitemOpen
  \bibfield  {author} {\bibinfo {author} {\bibfnamefont {I.}~\bibnamefont
  {Tanihata}}, \bibinfo {author} {\bibfnamefont {H.}~\bibnamefont {Hamagaki}},
  \bibinfo {author} {\bibfnamefont {O.}~\bibnamefont {Hashimoto}}, \bibinfo
  {author} {\bibfnamefont {Y.}~\bibnamefont {Shida}}, \bibinfo {author}
  {\bibfnamefont {N.}~\bibnamefont {Yoshikawa}}, \bibinfo {author}
  {\bibfnamefont {K.}~\bibnamefont {Sugimoto}}, \bibinfo {author}
  {\bibfnamefont {O.}~\bibnamefont {Yamakawa}}, \bibinfo {author}
  {\bibfnamefont {T.}~\bibnamefont {Kobayashi}}, \ and\ \bibinfo {author}
  {\bibfnamefont {N.}~\bibnamefont {Takahashi}},\ }\href {\doibase
  10.1103/PhysRevLett.55.2676} {\bibfield  {journal} {\bibinfo  {journal}
  {Phys. Rev. Lett.}\ }\textbf {\bibinfo {volume} {55}},\ \bibinfo {pages}
  {2676} (\bibinfo {year} {1985})}\BibitemShut {NoStop}%
\bibitem [{\citenamefont {Minamisono}\ \emph {et~al.}(1992)\citenamefont
  {Minamisono}, \citenamefont {Ohtsubo}, \citenamefont {Minami}, \citenamefont
  {Fukuda}, \citenamefont {Kitagawa}, \citenamefont {Fukuda}, \citenamefont
  {Matsuta}, \citenamefont {Nojiri}, \citenamefont {Takeda}, \citenamefont
  {Sagawa},\ and\ \citenamefont {Kitagawa}}]{Minamisono1992_PRL69-2058}%
  \BibitemOpen
  \bibfield  {author} {\bibinfo {author} {\bibfnamefont {T.}~\bibnamefont
  {Minamisono}}, \bibinfo {author} {\bibfnamefont {T.}~\bibnamefont {Ohtsubo}},
  \bibinfo {author} {\bibfnamefont {I.}~\bibnamefont {Minami}}, \bibinfo
  {author} {\bibfnamefont {S.}~\bibnamefont {Fukuda}}, \bibinfo {author}
  {\bibfnamefont {A.}~\bibnamefont {Kitagawa}}, \bibinfo {author}
  {\bibfnamefont {M.}~\bibnamefont {Fukuda}}, \bibinfo {author} {\bibfnamefont
  {K.}~\bibnamefont {Matsuta}}, \bibinfo {author} {\bibfnamefont
  {Y.}~\bibnamefont {Nojiri}}, \bibinfo {author} {\bibfnamefont
  {S.}~\bibnamefont {Takeda}}, \bibinfo {author} {\bibfnamefont
  {H.}~\bibnamefont {Sagawa}}, \ and\ \bibinfo {author} {\bibfnamefont
  {H.}~\bibnamefont {Kitagawa}},\ }\href {\doibase 10.1103/PhysRevLett.69.2058}
  {\bibfield  {journal} {\bibinfo  {journal} {Phys. Rev. Lett.}\ }\textbf
  {\bibinfo {volume} {69}},\ \bibinfo {pages} {2058} (\bibinfo {year}
  {1992})}\BibitemShut {NoStop}%
\bibitem [{\citenamefont {Schwab}\ \emph {et~al.}(1995)\citenamefont {Schwab},
  \citenamefont {Geissel}, \citenamefont {Lenske}, \citenamefont {Behr},
  \citenamefont {Br\"unle}, \citenamefont {Burkard}, \citenamefont {Irnich},
  \citenamefont {Kobayashi}, \citenamefont {Kraus}, \citenamefont {Magel},
  \citenamefont {M\"unzenberg}, \citenamefont {Nickel}, \citenamefont
  {Riisager}, \citenamefont {Scheidenberger}, \citenamefont {Sherrill},
  \citenamefont {Suzuki},\ and\ \citenamefont {Voss}}]{Schwab1995_ZPA350-283}%
  \BibitemOpen
  \bibfield  {author} {\bibinfo {author} {\bibfnamefont {W.}~\bibnamefont
  {Schwab}}, \bibinfo {author} {\bibfnamefont {H.}~\bibnamefont {Geissel}},
  \bibinfo {author} {\bibfnamefont {H.}~\bibnamefont {Lenske}}, \bibinfo
  {author} {\bibfnamefont {K.~H.}\ \bibnamefont {Behr}}, \bibinfo {author}
  {\bibfnamefont {A.}~\bibnamefont {Br\"unle}}, \bibinfo {author}
  {\bibfnamefont {K.}~\bibnamefont {Burkard}}, \bibinfo {author} {\bibfnamefont
  {H.}~\bibnamefont {Irnich}}, \bibinfo {author} {\bibfnamefont
  {T.}~\bibnamefont {Kobayashi}}, \bibinfo {author} {\bibfnamefont
  {G.}~\bibnamefont {Kraus}}, \bibinfo {author} {\bibfnamefont
  {A.}~\bibnamefont {Magel}}, \bibinfo {author} {\bibfnamefont
  {G.}~\bibnamefont {M\"unzenberg}}, \bibinfo {author} {\bibfnamefont
  {F.}~\bibnamefont {Nickel}}, \bibinfo {author} {\bibfnamefont
  {K.}~\bibnamefont {Riisager}}, \bibinfo {author} {\bibfnamefont
  {C.}~\bibnamefont {Scheidenberger}}, \bibinfo {author} {\bibfnamefont
  {B.~M.}\ \bibnamefont {Sherrill}}, \bibinfo {author} {\bibfnamefont
  {T.}~\bibnamefont {Suzuki}}, \ and\ \bibinfo {author} {\bibfnamefont
  {B.}~\bibnamefont {Voss}},\ }\href {\doibase 10.1007/BF01291183} {\bibfield
  {journal} {\bibinfo  {journal} {Z. Phys. A}\ }\textbf {\bibinfo {volume}
  {350}},\ \bibinfo {pages} {283} (\bibinfo {year} {1995})}\BibitemShut
  {NoStop}%
\bibitem [{\citenamefont {Ozawa}\ \emph {et~al.}(2000)\citenamefont {Ozawa},
  \citenamefont {Kobayashi}, \citenamefont {Suzuki}, \citenamefont {Yoshida},\
  and\ \citenamefont {Tanihata}}]{Ozawa2000_PRL84-5493}%
  \BibitemOpen
  \bibfield  {author} {\bibinfo {author} {\bibfnamefont {A.}~\bibnamefont
  {Ozawa}}, \bibinfo {author} {\bibfnamefont {T.}~\bibnamefont {Kobayashi}},
  \bibinfo {author} {\bibfnamefont {T.}~\bibnamefont {Suzuki}}, \bibinfo
  {author} {\bibfnamefont {K.}~\bibnamefont {Yoshida}}, \ and\ \bibinfo
  {author} {\bibfnamefont {I.}~\bibnamefont {Tanihata}},\ }\href {\doibase
  10.1103/PhysRevLett.84.5493} {\bibfield  {journal} {\bibinfo  {journal}
  {Phys. Rev. Lett.}\ }\textbf {\bibinfo {volume} {84}},\ \bibinfo {pages}
  {5493} (\bibinfo {year} {2000})}\BibitemShut {NoStop}%
\bibitem [{\citenamefont {Adrich}\ \emph {et~al.}(2005)\citenamefont {Adrich},
  \citenamefont {Klimkiewicz}, \citenamefont {Fallot}, \citenamefont
  {Boretzky}, \citenamefont {Aumann}, \citenamefont {Cortina-Gil},
  \citenamefont {Pramanik}, \citenamefont {Elze}, \citenamefont {Emling},
  \citenamefont {Geissel}, \citenamefont {Hellstrom}, \citenamefont {Jones},
  \citenamefont {Kratz}, \citenamefont {Kulessa}, \citenamefont {Leifels},
  \citenamefont {Nociforo}, \citenamefont {Palit}, \citenamefont {Simon},
  \citenamefont {Surowka}, \citenamefont {Summerer},\ and\ \citenamefont
  {Walus}}]{Adrich2005_PRL95-132501}%
  \BibitemOpen
  \bibfield  {author} {\bibinfo {author} {\bibfnamefont {P.}~\bibnamefont
  {Adrich}}, \bibinfo {author} {\bibfnamefont {A.}~\bibnamefont {Klimkiewicz}},
  \bibinfo {author} {\bibfnamefont {M.}~\bibnamefont {Fallot}}, \bibinfo
  {author} {\bibfnamefont {K.}~\bibnamefont {Boretzky}}, \bibinfo {author}
  {\bibfnamefont {T.}~\bibnamefont {Aumann}}, \bibinfo {author} {\bibfnamefont
  {D.}~\bibnamefont {Cortina-Gil}}, \bibinfo {author} {\bibfnamefont {U.~D.}\
  \bibnamefont {Pramanik}}, \bibinfo {author} {\bibfnamefont {T.~W.}\
  \bibnamefont {Elze}}, \bibinfo {author} {\bibfnamefont {H.}~\bibnamefont
  {Emling}}, \bibinfo {author} {\bibfnamefont {H.}~\bibnamefont {Geissel}},
  \bibinfo {author} {\bibfnamefont {M.}~\bibnamefont {Hellstrom}}, \bibinfo
  {author} {\bibfnamefont {K.~L.}\ \bibnamefont {Jones}}, \bibinfo {author}
  {\bibfnamefont {J.~V.}\ \bibnamefont {Kratz}}, \bibinfo {author}
  {\bibfnamefont {R.}~\bibnamefont {Kulessa}}, \bibinfo {author} {\bibfnamefont
  {Y.}~\bibnamefont {Leifels}}, \bibinfo {author} {\bibfnamefont
  {C.}~\bibnamefont {Nociforo}}, \bibinfo {author} {\bibfnamefont
  {R.}~\bibnamefont {Palit}}, \bibinfo {author} {\bibfnamefont
  {H.}~\bibnamefont {Simon}}, \bibinfo {author} {\bibfnamefont
  {G.}~\bibnamefont {Surowka}}, \bibinfo {author} {\bibfnamefont
  {K.}~\bibnamefont {Summerer}}, \ and\ \bibinfo {author} {\bibfnamefont
  {W.}~\bibnamefont {Walus}},\ }\href {\doibase 10.1103/PhysRevLett.95.132501}
  {\bibfield  {journal} {\bibinfo  {journal} {Phys. Rev. Lett.}\ }\textbf
  {\bibinfo {volume} {95}},\ \bibinfo {pages} {132501} (\bibinfo {year}
  {2005})}\BibitemShut {NoStop}%
\bibitem [{\citenamefont {Cao}\ and\ \citenamefont
  {Ma}(2002)}]{Cao2002_PRC66-024311}%
  \BibitemOpen
  \bibfield  {author} {\bibinfo {author} {\bibfnamefont {L.-g.}\ \bibnamefont
  {Cao}}\ and\ \bibinfo {author} {\bibfnamefont {Z.-y.}\ \bibnamefont {Ma}},\
  }\href {\doibase 10.1103/PhysRevC.66.024311} {\bibfield  {journal} {\bibinfo
  {journal} {Phys. Rev. C}\ }\textbf {\bibinfo {volume} {66}},\ \bibinfo
  {pages} {024311} (\bibinfo {year} {2002})}\BibitemShut {NoStop}%
\bibitem [{\citenamefont {Dobaczewski}\ \emph {et~al.}(2007)\citenamefont
  {Dobaczewski}, \citenamefont {Michel}, \citenamefont {Nazarewicz},
  \citenamefont {Ploszajczak},\ and\ \citenamefont
  {Rotureau}}]{Dobaczewski2007_PPNP59-432}%
  \BibitemOpen
  \bibfield  {author} {\bibinfo {author} {\bibfnamefont {J.}~\bibnamefont
  {Dobaczewski}}, \bibinfo {author} {\bibfnamefont {N.}~\bibnamefont {Michel}},
  \bibinfo {author} {\bibfnamefont {W.}~\bibnamefont {Nazarewicz}}, \bibinfo
  {author} {\bibfnamefont {M.}~\bibnamefont {Ploszajczak}}, \ and\ \bibinfo
  {author} {\bibfnamefont {J.}~\bibnamefont {Rotureau}},\ }\href {\doibase
  10.1016/j.ppnp.2007.01.022} {\bibfield  {journal} {\bibinfo  {journal} {Prog.
  Part. Nucl. Phys.}\ }\textbf {\bibinfo {volume} {59}},\ \bibinfo {pages}
  {432} (\bibinfo {year} {2007})}\BibitemShut {NoStop}%
\bibitem [{\citenamefont {Pei}\ \emph {et~al.}(2011)\citenamefont {Pei},
  \citenamefont {Kruppa},\ and\ \citenamefont
  {Nazarewicz}}]{Pei2011_PRC84-024311}%
  \BibitemOpen
  \bibfield  {author} {\bibinfo {author} {\bibfnamefont {J.~C.}\ \bibnamefont
  {Pei}}, \bibinfo {author} {\bibfnamefont {A.~T.}\ \bibnamefont {Kruppa}}, \
  and\ \bibinfo {author} {\bibfnamefont {W.}~\bibnamefont {Nazarewicz}},\
  }\href {\doibase 10.1103/PhysRevC.84.024311} {\bibfield  {journal} {\bibinfo
  {journal} {Phys. Rev. C}\ }\textbf {\bibinfo {volume} {84}},\ \bibinfo
  {pages} {024311} (\bibinfo {year} {2011})}\BibitemShut {NoStop}%
\bibitem [{\citenamefont {Bulgac}(1980)}]{Bulgac1980_nucl-th9907088}%
  \BibitemOpen
  \bibfield  {author} {\bibinfo {author} {\bibfnamefont {A.}~\bibnamefont
  {Bulgac}},\ }\href {http://arxiv.org/abs/nucl-th/9907088} {\enquote {\bibinfo
  {title} {{Hartree}-{Fock}-{Bogoliubov} approximation for finite systems},}\
  }\bibinfo {howpublished} {IPNE FT-194-1980, Bucharest (arXiv:
  nucl-th/9907088)} (\bibinfo {year} {1980})\BibitemShut {NoStop}%
\bibitem [{\citenamefont {Dobaczewski}\ \emph {et~al.}(1984)\citenamefont
  {Dobaczewski}, \citenamefont {Flocard},\ and\ \citenamefont
  {Treiner}}]{Dobaczewski1984_NPA422-103}%
  \BibitemOpen
  \bibfield  {author} {\bibinfo {author} {\bibfnamefont {J.}~\bibnamefont
  {Dobaczewski}}, \bibinfo {author} {\bibfnamefont {H.}~\bibnamefont
  {Flocard}}, \ and\ \bibinfo {author} {\bibfnamefont {J.}~\bibnamefont
  {Treiner}},\ }\href {\doibase 10.1016/0375-9474(84)90433-0} {\bibfield
  {journal} {\bibinfo  {journal} {Nucl. Phys. A}\ }\textbf {\bibinfo {volume}
  {422}},\ \bibinfo {pages} {103} (\bibinfo {year} {1984})}\BibitemShut
  {NoStop}%
\bibitem [{\citenamefont {Dobaczewski}\ \emph {et~al.}(1996)\citenamefont
  {Dobaczewski}, \citenamefont {Nazarewicz}, \citenamefont {Werner},
  \citenamefont {Berger}, \citenamefont {Chinn},\ and\ \citenamefont
  {Decharg\'{e}}}]{Dobaczewski1996_PRC53-2809}%
  \BibitemOpen
  \bibfield  {author} {\bibinfo {author} {\bibfnamefont {J.}~\bibnamefont
  {Dobaczewski}}, \bibinfo {author} {\bibfnamefont {W.}~\bibnamefont
  {Nazarewicz}}, \bibinfo {author} {\bibfnamefont {T.~R.}\ \bibnamefont
  {Werner}}, \bibinfo {author} {\bibfnamefont {J.~F.}\ \bibnamefont {Berger}},
  \bibinfo {author} {\bibfnamefont {C.~R.}\ \bibnamefont {Chinn}}, \ and\
  \bibinfo {author} {\bibfnamefont {J.}~\bibnamefont {Decharg\'{e}}},\ }\href
  {\doibase 10.1103/PhysRevC.53.2809} {\bibfield  {journal} {\bibinfo
  {journal} {Phys. Rev. C}\ }\textbf {\bibinfo {volume} {53}},\ \bibinfo
  {pages} {2809} (\bibinfo {year} {1996})}\BibitemShut {NoStop}%
\bibitem [{\citenamefont {Meng}\ and\ \citenamefont
  {Ring}(1996)}]{Meng1996_PRL77-3963}%
  \BibitemOpen
  \bibfield  {author} {\bibinfo {author} {\bibfnamefont {J.}~\bibnamefont
  {Meng}}\ and\ \bibinfo {author} {\bibfnamefont {P.}~\bibnamefont {Ring}},\
  }\href {\doibase 10.1103/PhysRevLett.77.3963} {\bibfield  {journal} {\bibinfo
   {journal} {Phys. Rev. Lett.}\ }\textbf {\bibinfo {volume} {77}},\ \bibinfo
  {pages} {3963} (\bibinfo {year} {1996})}\BibitemShut {NoStop}%
\bibitem [{\citenamefont {P\"{o}schl}\ \emph {et~al.}(1997)\citenamefont
  {P\"{o}schl}, \citenamefont {Vretenar}, \citenamefont {Lalazissis},\ and\
  \citenamefont {Ring}}]{Poschl1997_PRL79-3841}%
  \BibitemOpen
  \bibfield  {author} {\bibinfo {author} {\bibfnamefont {W.}~\bibnamefont
  {P\"{o}schl}}, \bibinfo {author} {\bibfnamefont {D.}~\bibnamefont
  {Vretenar}}, \bibinfo {author} {\bibfnamefont {G.~A.}\ \bibnamefont
  {Lalazissis}}, \ and\ \bibinfo {author} {\bibfnamefont {P.}~\bibnamefont
  {Ring}},\ }\href {\doibase 10.1103/PhysRevLett.79.3841} {\bibfield  {journal}
  {\bibinfo  {journal} {Phys. Rev. Lett.}\ }\textbf {\bibinfo {volume} {79}},\
  \bibinfo {pages} {3841} (\bibinfo {year} {1997})}\BibitemShut {NoStop}%
\bibitem [{\citenamefont {Lalazissis}\ \emph {et~al.}(1998)\citenamefont
  {Lalazissis}, \citenamefont {Vretenar}, \citenamefont {P\"oschl},\ and\
  \citenamefont {Ring}}]{Lalazissis1998_PLB418-7}%
  \BibitemOpen
  \bibfield  {author} {\bibinfo {author} {\bibfnamefont {G.}~\bibnamefont
  {Lalazissis}}, \bibinfo {author} {\bibfnamefont {D.}~\bibnamefont
  {Vretenar}}, \bibinfo {author} {\bibfnamefont {W.}~\bibnamefont {P\"oschl}},
  \ and\ \bibinfo {author} {\bibfnamefont {P.}~\bibnamefont {Ring}},\ }\href
  {\doibase 10.1016/S0370-2693(97)01473-1} {\bibfield  {journal} {\bibinfo
  {journal} {Phys. Lett. B}\ }\textbf {\bibinfo {volume} {418}},\ \bibinfo
  {pages} {7} (\bibinfo {year} {1998})}\BibitemShut {NoStop}%
\bibitem [{\citenamefont {Meng}(1998)}]{Meng1998_NPA635-3}%
  \BibitemOpen
  \bibfield  {author} {\bibinfo {author} {\bibfnamefont {J.}~\bibnamefont
  {Meng}},\ }\href {\doibase 10.1016/S0375-9474(98)00178-X} {\bibfield
  {journal} {\bibinfo  {journal} {Nucl. Phys. A}\ }\textbf {\bibinfo {volume}
  {635}},\ \bibinfo {pages} {3} (\bibinfo {year} {1998})}\BibitemShut {NoStop}%
\bibitem [{\citenamefont {Sandulescu}\ \emph {et~al.}(2000)\citenamefont
  {Sandulescu}, \citenamefont {Van~Giai},\ and\ \citenamefont
  {Liotta}}]{Sandulescu2000_PRC61-061301R}%
  \BibitemOpen
  \bibfield  {author} {\bibinfo {author} {\bibfnamefont {N.}~\bibnamefont
  {Sandulescu}}, \bibinfo {author} {\bibfnamefont {N.}~\bibnamefont
  {Van~Giai}}, \ and\ \bibinfo {author} {\bibfnamefont {R.~J.}\ \bibnamefont
  {Liotta}},\ }\href {\doibase 10.1103/PhysRevC.61.061301} {\bibfield
  {journal} {\bibinfo  {journal} {Phys. Rev. C}\ }\textbf {\bibinfo {volume}
  {61}},\ \bibinfo {pages} {061301(R)} (\bibinfo {year} {2000})}\BibitemShut
  {NoStop}%
\bibitem [{\citenamefont {Sandulescu}\ \emph {et~al.}(2003)\citenamefont
  {Sandulescu}, \citenamefont {Geng}, \citenamefont {Toki},\ and\ \citenamefont
  {Hillhouse}}]{Sandulescu2003_PRC68-054323}%
  \BibitemOpen
  \bibfield  {author} {\bibinfo {author} {\bibfnamefont {N.}~\bibnamefont
  {Sandulescu}}, \bibinfo {author} {\bibfnamefont {L.~S.}\ \bibnamefont
  {Geng}}, \bibinfo {author} {\bibfnamefont {H.}~\bibnamefont {Toki}}, \ and\
  \bibinfo {author} {\bibfnamefont {G.~C.}\ \bibnamefont {Hillhouse}},\ }\href
  {\doibase 10.1103/PhysRevC.68.054323} {\bibfield  {journal} {\bibinfo
  {journal} {Phys. Rev. C}\ }\textbf {\bibinfo {volume} {68}},\ \bibinfo
  {pages} {054323} (\bibinfo {year} {2003})}\BibitemShut {NoStop}%
\bibitem [{\citenamefont {Zhang}\ \emph
  {et~al.}(2011{\natexlab{a}})\citenamefont {Zhang}, \citenamefont {Zhao},\
  and\ \citenamefont {Zhou}}]{Zhang2011_arxiv1105.0504}%
  \BibitemOpen
  \bibfield  {author} {\bibinfo {author} {\bibfnamefont {S.-S.}\ \bibnamefont
  {Zhang}}, \bibinfo {author} {\bibfnamefont {E.-G.}\ \bibnamefont {Zhao}}, \
  and\ \bibinfo {author} {\bibfnamefont {S.-G.}\ \bibnamefont {Zhou}},\ }\href
  {http://arxiv.org/abs/1105.0504} {\enquote {\bibinfo {title} {Theoretical
  study of the two-proton halo candidate $^{17}${Ne} including contributions
  from a resonant continuum and pairing correlations},}\ }\bibinfo
  {howpublished} {arXiv: 1105.0504 [nucl-th]} (\bibinfo {year}
  {2011}{\natexlab{a}}),\ \bibinfo {note} {{submitted to Nucl. Phys.
  A}}\BibitemShut {NoStop}%
\bibitem [{\citenamefont {Press}\ \emph {et~al.}(1992)\citenamefont {Press},
  \citenamefont {Teukolsky}, \citenamefont {Vetterling},\ and\ \citenamefont
  {Flannery}}]{Press1992}%
  \BibitemOpen
  \bibfield  {author} {\bibinfo {author} {\bibfnamefont {W.~H.}\ \bibnamefont
  {Press}}, \bibinfo {author} {\bibfnamefont {S.~A.}\ \bibnamefont
  {Teukolsky}}, \bibinfo {author} {\bibfnamefont {W.~T.}\ \bibnamefont
  {Vetterling}}, \ and\ \bibinfo {author} {\bibfnamefont {B.~P.}\ \bibnamefont
  {Flannery}},\ }\href {http://www.nrbook.com/a/bookfpdf.php} {\emph {\bibinfo
  {title} {Numerical Recipes in Fortran 77: the Art of Scientific Computing
  (Vol. 1 of Fortran Numerical Recipes)}}},\ \bibinfo {edition} {2nd}\ ed.\
  (\bibinfo  {publisher} {Cambridge University Press},\ \bibinfo {year}
  {1992})\BibitemShut {NoStop}%
\bibitem [{\citenamefont {P\"oschl}\ \emph
  {et~al.}(1997{\natexlab{a}})\citenamefont {P\"oschl}, \citenamefont
  {Vretenar}, \citenamefont {Rummel},\ and\ \citenamefont
  {Ring}}]{Poschl1997_CPC101-75}%
  \BibitemOpen
  \bibfield  {author} {\bibinfo {author} {\bibfnamefont {W.}~\bibnamefont
  {P\"oschl}}, \bibinfo {author} {\bibfnamefont {D.}~\bibnamefont {Vretenar}},
  \bibinfo {author} {\bibfnamefont {A.}~\bibnamefont {Rummel}}, \ and\ \bibinfo
  {author} {\bibfnamefont {P.}~\bibnamefont {Ring}},\ }\href {\doibase
  10.1016/S0010-4655(97)84583-3} {\bibfield  {journal} {\bibinfo  {journal}
  {Comput. Phys. Commun.}\ }\textbf {\bibinfo {volume} {101}},\ \bibinfo
  {pages} {75} (\bibinfo {year} {1997}{\natexlab{a}})}\BibitemShut {NoStop}%
\bibitem [{\citenamefont {P\"oschl}\ \emph
  {et~al.}(1997{\natexlab{b}})\citenamefont {P\"oschl}, \citenamefont
  {Vretenar},\ and\ \citenamefont {Ring}}]{Poschl1997_CPC103-217}%
  \BibitemOpen
  \bibfield  {author} {\bibinfo {author} {\bibfnamefont {W.}~\bibnamefont
  {P\"oschl}}, \bibinfo {author} {\bibfnamefont {D.}~\bibnamefont {Vretenar}},
  \ and\ \bibinfo {author} {\bibfnamefont {P.}~\bibnamefont {Ring}},\ }\href
  {\doibase 10.1016/S0010-4655(97)00042-8} {\bibfield  {journal} {\bibinfo
  {journal} {Comput. Phys. Commun.}\ }\textbf {\bibinfo {volume} {103}},\
  \bibinfo {pages} {217} (\bibinfo {year} {1997}{\natexlab{b}})}\BibitemShut
  {NoStop}%
\bibitem [{\citenamefont {Vautherin}(1973)}]{Vautherin1973_PRC7-296}%
  \BibitemOpen
  \bibfield  {author} {\bibinfo {author} {\bibfnamefont {D.}~\bibnamefont
  {Vautherin}},\ }\href {\doibase 10.1103/PhysRevC.7.296} {\bibfield  {journal}
  {\bibinfo  {journal} {Phys. Rev. C}\ }\textbf {\bibinfo {volume} {7}},\
  \bibinfo {pages} {296} (\bibinfo {year} {1973})}\BibitemShut {NoStop}%
\bibitem [{\citenamefont {Gogny}(1975)}]{Gogny1975_NPA237-399}%
  \BibitemOpen
  \bibfield  {author} {\bibinfo {author} {\bibfnamefont {D.}~\bibnamefont
  {Gogny}},\ }\href {\doibase 10.1016/0375-9474(75)90407-8} {\bibfield
  {journal} {\bibinfo  {journal} {Nucl. Phys. A}\ }\textbf {\bibinfo {volume}
  {237}},\ \bibinfo {pages} {399} (\bibinfo {year} {1975})}\BibitemShut
  {NoStop}%
\bibitem [{\citenamefont {Decharge}\ and\ \citenamefont
  {Gogny}(1980)}]{Decharge1980_PRC21-1568}%
  \BibitemOpen
  \bibfield  {author} {\bibinfo {author} {\bibfnamefont {J.}~\bibnamefont
  {Decharge}}\ and\ \bibinfo {author} {\bibfnamefont {D.}~\bibnamefont
  {Gogny}},\ }\href {\doibase 10.1103/PhysRevC.21.1568} {\bibfield  {journal}
  {\bibinfo  {journal} {Phys. Rev. C}\ }\textbf {\bibinfo {volume} {21}},\
  \bibinfo {pages} {1568} (\bibinfo {year} {1980})}\BibitemShut {NoStop}%
\bibitem [{\citenamefont {Gambhir}\ \emph {et~al.}(1990)\citenamefont
  {Gambhir}, \citenamefont {Ring},\ and\ \citenamefont
  {Thimet}}]{Gambhir1990_APNY198-132}%
  \BibitemOpen
  \bibfield  {author} {\bibinfo {author} {\bibfnamefont {Y.~K.}\ \bibnamefont
  {Gambhir}}, \bibinfo {author} {\bibfnamefont {P.}~\bibnamefont {Ring}}, \
  and\ \bibinfo {author} {\bibfnamefont {A.}~\bibnamefont {Thimet}},\ }\href
  {\doibase 10.1016/0003-4916(90)90330-Q} {\bibfield  {journal} {\bibinfo
  {journal} {Ann. Phys.}\ }\textbf {\bibinfo {volume} {198}},\ \bibinfo {pages}
  {132} (\bibinfo {year} {1990})}\BibitemShut {NoStop}%
\bibitem [{\citenamefont {Gambhir}\ and\ \citenamefont
  {Ring}(1993)}]{Gambhir1993_MPLA8-787}%
  \BibitemOpen
  \bibfield  {author} {\bibinfo {author} {\bibfnamefont {Y.~K.}\ \bibnamefont
  {Gambhir}}\ and\ \bibinfo {author} {\bibfnamefont {P.}~\bibnamefont {Ring}},\
  }\href {\doibase http://dx.doi.org/10.1142/S0217732393000817} {\bibfield
  {journal} {\bibinfo  {journal} {Mod. Phys. Lett. A}\ }\textbf {\bibinfo
  {volume} {8}},\ \bibinfo {pages} {787} (\bibinfo {year} {1993})}\BibitemShut
  {NoStop}%
\bibitem [{\citenamefont {Stoitsov}\ \emph {et~al.}(2000)\citenamefont
  {Stoitsov}, \citenamefont {Dobaczewski}, \citenamefont {Ring},\ and\
  \citenamefont {Pittel}}]{Stoitsov2000_PRC61-034311}%
  \BibitemOpen
  \bibfield  {author} {\bibinfo {author} {\bibfnamefont {M.~V.}\ \bibnamefont
  {Stoitsov}}, \bibinfo {author} {\bibfnamefont {J.}~\bibnamefont
  {Dobaczewski}}, \bibinfo {author} {\bibfnamefont {P.}~\bibnamefont {Ring}}, \
  and\ \bibinfo {author} {\bibfnamefont {S.}~\bibnamefont {Pittel}},\ }\href
  {\doibase 10.1103/PhysRevC.61.034311} {\bibfield  {journal} {\bibinfo
  {journal} {Phys. Rev. C}\ }\textbf {\bibinfo {volume} {61}},\ \bibinfo
  {pages} {034311} (\bibinfo {year} {2000})}\BibitemShut {NoStop}%
\bibitem [{\citenamefont {Stoitsov}\ \emph {et~al.}(2003)\citenamefont
  {Stoitsov}, \citenamefont {Dobaczewski}, \citenamefont {Nazarewicz},
  \citenamefont {Pittel},\ and\ \citenamefont
  {Dean}}]{Stoitsov2003_PRC68-054312}%
  \BibitemOpen
  \bibfield  {author} {\bibinfo {author} {\bibfnamefont {M.~V.}\ \bibnamefont
  {Stoitsov}}, \bibinfo {author} {\bibfnamefont {J.}~\bibnamefont
  {Dobaczewski}}, \bibinfo {author} {\bibfnamefont {W.}~\bibnamefont
  {Nazarewicz}}, \bibinfo {author} {\bibfnamefont {S.}~\bibnamefont {Pittel}},
  \ and\ \bibinfo {author} {\bibfnamefont {D.~J.}\ \bibnamefont {Dean}},\
  }\href {\doibase 10.1103/PhysRevC.68.054312} {\bibfield  {journal} {\bibinfo
  {journal} {Phys. Rev. C}\ }\textbf {\bibinfo {volume} {68}},\ \bibinfo
  {pages} {054312} (\bibinfo {year} {2003})}\BibitemShut {NoStop}%
\bibitem [{\citenamefont {Zhou}\ \emph {et~al.}(2003)\citenamefont {Zhou},
  \citenamefont {Meng},\ and\ \citenamefont {Ring}}]{Zhou2003_PRC68-034323}%
  \BibitemOpen
  \bibfield  {author} {\bibinfo {author} {\bibfnamefont {S.-G.}\ \bibnamefont
  {Zhou}}, \bibinfo {author} {\bibfnamefont {J.}~\bibnamefont {Meng}}, \ and\
  \bibinfo {author} {\bibfnamefont {P.}~\bibnamefont {Ring}},\ }\href {\doibase
  10.1103/PhysRevC.68.034323} {\bibfield  {journal} {\bibinfo  {journal} {Phys.
  Rev. C}\ }\textbf {\bibinfo {volume} {68}},\ \bibinfo {pages} {034323}
  (\bibinfo {year} {2003})}\BibitemShut {NoStop}%
\bibitem [{\citenamefont {Schunck}\ and\ \citenamefont
  {Egido}(2008{\natexlab{a}})}]{Schunck2008_PRC77-011301R}%
  \BibitemOpen
  \bibfield  {author} {\bibinfo {author} {\bibfnamefont {N.}~\bibnamefont
  {Schunck}}\ and\ \bibinfo {author} {\bibfnamefont {J.~L.}\ \bibnamefont
  {Egido}},\ }\href {\doibase 10.1103/PhysRevC.77.011301} {\bibfield  {journal}
  {\bibinfo  {journal} {Phys. Rev. C}\ }\textbf {\bibinfo {volume} {77}},\
  \bibinfo {pages} {011301(R)} (\bibinfo {year}
  {2008}{\natexlab{a}})}\BibitemShut {NoStop}%
\bibitem [{\citenamefont {Schunck}\ and\ \citenamefont
  {Egido}(2008{\natexlab{b}})}]{Schunck2008_PRC78-064305}%
  \BibitemOpen
  \bibfield  {author} {\bibinfo {author} {\bibfnamefont {N.}~\bibnamefont
  {Schunck}}\ and\ \bibinfo {author} {\bibfnamefont {J.~L.}\ \bibnamefont
  {Egido}},\ }\href {\doibase 10.1103/PhysRevC.78.064305} {\bibfield  {journal}
  {\bibinfo  {journal} {Phys. Rev. C}\ }\textbf {\bibinfo {volume} {78}},\
  \bibinfo {pages} {064305} (\bibinfo {year} {2008}{\natexlab{b}})}\BibitemShut
  {NoStop}%
\bibitem [{\citenamefont {Long}\ \emph {et~al.}(2010)\citenamefont {Long},
  \citenamefont {Ring}, \citenamefont {Giai},\ and\ \citenamefont
  {Meng}}]{Long2010_PRC81-024308}%
  \BibitemOpen
  \bibfield  {author} {\bibinfo {author} {\bibfnamefont {W.~H.}\ \bibnamefont
  {Long}}, \bibinfo {author} {\bibfnamefont {P.}~\bibnamefont {Ring}}, \bibinfo
  {author} {\bibfnamefont {N.~V.}\ \bibnamefont {Giai}}, \ and\ \bibinfo
  {author} {\bibfnamefont {J.}~\bibnamefont {Meng}},\ }\href {\doibase
  10.1103/PhysRevC.81.024308} {\bibfield  {journal} {\bibinfo  {journal} {Phys.
  Rev. C}\ }\textbf {\bibinfo {volume} {81}},\ \bibinfo {pages} {024308}
  (\bibinfo {year} {2010})}\BibitemShut {NoStop}%
\bibitem [{\citenamefont {Zhou}\ \emph {et~al.}(2006)\citenamefont {Zhou},
  \citenamefont {Meng},\ and\ \citenamefont {Ring}}]{Zhou2006_AIPCP865-90}%
  \BibitemOpen
  \bibfield  {author} {\bibinfo {author} {\bibfnamefont {S.-G.}\ \bibnamefont
  {Zhou}}, \bibinfo {author} {\bibfnamefont {J.}~\bibnamefont {Meng}}, \ and\
  \bibinfo {author} {\bibfnamefont {P.}~\bibnamefont {Ring}},\ }in\ \href
  {http://link.aip.org/link/?APC/865/90/1} {\emph {\bibinfo {booktitle}
  {Nuclear Physics Trends}}},\ \bibinfo {series} {AIP Conf. Proc.}, Vol.\
  \bibinfo {volume} {865},\ \bibinfo {editor} {edited by\ \bibinfo {editor}
  {\bibfnamefont {Y.-G.}\ \bibnamefont {Ma}}\ and\ \bibinfo {editor}
  {\bibfnamefont {A.}~\bibnamefont {Ozawa}}}\ (\bibinfo  {publisher} {AIP},\
  \bibinfo {year} {2006})\ pp.\ \bibinfo {pages} {90--95}\BibitemShut {NoStop}%
\bibitem [{\citenamefont {Zhou}\ \emph {et~al.}(2008)\citenamefont {Zhou},
  \citenamefont {Meng},\ and\ \citenamefont {Ring}}]{Zhou2008_ISPUN2007}%
  \BibitemOpen
  \bibfield  {author} {\bibinfo {author} {\bibfnamefont {S.-G.}\ \bibnamefont
  {Zhou}}, \bibinfo {author} {\bibfnamefont {J.}~\bibnamefont {Meng}}, \ and\
  \bibinfo {author} {\bibfnamefont {P.}~\bibnamefont {Ring}},\ }in\ \href
  {http://www.worldscibooks.com/physics/6649.html} {\emph {\bibinfo {booktitle}
  {Physics of Unstable Nuclei}}},\ \bibinfo {editor} {edited by\ \bibinfo
  {editor} {\bibfnamefont {D.~T.}\ \bibnamefont {Khoa}}, \bibinfo {editor}
  {\bibfnamefont {P.}~\bibnamefont {Egelhof}}, \bibinfo {editor} {\bibfnamefont
  {S.}~\bibnamefont {Gales}}, \bibinfo {editor} {\bibfnamefont
  {N.}~\bibnamefont {Van~Giai}}, \ and\ \bibinfo {editor} {\bibfnamefont
  {T.}~\bibnamefont {Motobayashi}}}\ (\bibinfo  {publisher} {World
  Scientific},\ \bibinfo {year} {2008})\ pp.\ \bibinfo {pages} {402--408},\
  \bibinfo {note} {proceedings of the International Symposium on Physics of
  Unstable Nuclei, July 3-7, 2007, Hoi An, Vietnam, arXiv: 0803.1376v1
  [nucl-th]}\BibitemShut {NoStop}%
\bibitem [{\citenamefont {Zhou}\ \emph {et~al.}(2010)\citenamefont {Zhou},
  \citenamefont {Meng}, \citenamefont {Ring},\ and\ \citenamefont
  {Zhao}}]{Zhou2010_PRC82-011301R}%
  \BibitemOpen
  \bibfield  {author} {\bibinfo {author} {\bibfnamefont {S.-G.}\ \bibnamefont
  {Zhou}}, \bibinfo {author} {\bibfnamefont {J.}~\bibnamefont {Meng}}, \bibinfo
  {author} {\bibfnamefont {P.}~\bibnamefont {Ring}}, \ and\ \bibinfo {author}
  {\bibfnamefont {E.-G.}\ \bibnamefont {Zhao}},\ }\href {\doibase
  10.1103/PhysRevC.82.011301} {\bibfield  {journal} {\bibinfo  {journal} {Phys.
  Rev. C}\ }\textbf {\bibinfo {volume} {82}},\ \bibinfo {pages} {011301(R)}
  (\bibinfo {year} {2010})},\ \bibinfo {note} {{arXiv:0909.1600v3
  [nucl-th]}}\BibitemShut {NoStop}%
\bibitem [{\citenamefont {Zhou}\ \emph {et~al.}(2011)\citenamefont {Zhou},
  \citenamefont {Meng}, \citenamefont {Ring},\ and\ \citenamefont
  {Zhao}}]{Zhou2011_JPCS312-092067}%
  \BibitemOpen
  \bibfield  {author} {\bibinfo {author} {\bibfnamefont {S.-G.}\ \bibnamefont
  {Zhou}}, \bibinfo {author} {\bibfnamefont {J.}~\bibnamefont {Meng}}, \bibinfo
  {author} {\bibfnamefont {P.}~\bibnamefont {Ring}}, \ and\ \bibinfo {author}
  {\bibfnamefont {E.-G.}\ \bibnamefont {Zhao}},\ }\href {\doibase
  10.1088/1742-6596/312/9/092067} {\bibfield  {journal} {\bibinfo  {journal}
  {J. Phys: Conf. Ser.}\ }\textbf {\bibinfo {volume} {312}},\ \bibinfo {pages}
  {092067} (\bibinfo {year} {2011})},\ \bibinfo {note} {{arXiv:1101.3158v1
  [nucl-th]}}\BibitemShut {NoStop}%
\bibitem [{\citenamefont {Li}\ and\ \citenamefont
  {Heenen}(1996)}]{Li1996_PRC54-1617}%
  \BibitemOpen
  \bibfield  {author} {\bibinfo {author} {\bibfnamefont {X.}~\bibnamefont
  {Li}}\ and\ \bibinfo {author} {\bibfnamefont {P.-H.}\ \bibnamefont
  {Heenen}},\ }\href {\doibase 10.1103/PhysRevC.54.1617} {\bibfield  {journal}
  {\bibinfo  {journal} {Phys. Rev. C}\ }\textbf {\bibinfo {volume} {54}},\
  \bibinfo {pages} {1617} (\bibinfo {year} {1996})}\BibitemShut {NoStop}%
\bibitem [{\citenamefont {Misu}\ \emph {et~al.}(1997)\citenamefont {Misu},
  \citenamefont {Nazarewicz},\ and\ \citenamefont
  {{\AA}berg}}]{Misu1997_NPA614-44}%
  \BibitemOpen
  \bibfield  {author} {\bibinfo {author} {\bibfnamefont {T.}~\bibnamefont
  {Misu}}, \bibinfo {author} {\bibfnamefont {W.}~\bibnamefont {Nazarewicz}}, \
  and\ \bibinfo {author} {\bibfnamefont {S.}~\bibnamefont {{\AA}berg}},\ }\href
  {\doibase 10.1016/S0375-9474(96)00458-7} {\bibfield  {journal} {\bibinfo
  {journal} {Nucl. Phys. A}\ }\textbf {\bibinfo {volume} {614}},\ \bibinfo
  {pages} {44} (\bibinfo {year} {1997})}\BibitemShut {NoStop}%
\bibitem [{\citenamefont {Guo}\ \emph {et~al.}(2003)\citenamefont {Guo},
  \citenamefont {Zhao},\ and\ \citenamefont {Sakata}}]{Guo2003_CTP40-573}%
  \BibitemOpen
  \bibfield  {author} {\bibinfo {author} {\bibfnamefont {L.}~\bibnamefont
  {Guo}}, \bibinfo {author} {\bibfnamefont {E.-G.}\ \bibnamefont {Zhao}}, \
  and\ \bibinfo {author} {\bibfnamefont {F.}~\bibnamefont {Sakata}},\ }\href
  {http://ctp.itp.cn/qikan/Epaper/zhaiyao.asp?bsid=1081} {\bibfield  {journal}
  {\bibinfo  {journal} {Commun. Theor. Phys.}\ }\textbf {\bibinfo {volume}
  {40}},\ \bibinfo {pages} {573} (\bibinfo {year} {2003})}\BibitemShut
  {NoStop}%
\bibitem [{\citenamefont {Meng}\ \emph {et~al.}(2003)\citenamefont {Meng},
  \citenamefont {L\"{u}}, \citenamefont {Zhang},\ and\ \citenamefont
  {Zhou}}]{Meng2003_NPA722-C366}%
  \BibitemOpen
  \bibfield  {author} {\bibinfo {author} {\bibfnamefont {J.}~\bibnamefont
  {Meng}}, \bibinfo {author} {\bibfnamefont {H.~F.}\ \bibnamefont {L\"{u}}},
  \bibinfo {author} {\bibfnamefont {S.~Q.}\ \bibnamefont {Zhang}}, \ and\
  \bibinfo {author} {\bibfnamefont {S.~G.}\ \bibnamefont {Zhou}},\ }\href
  {\doibase 10.1016/S0375-9474(03)01391-5} {\bibfield  {journal} {\bibinfo
  {journal} {Nucl. Phys. A}\ }\textbf {\bibinfo {volume} {722}},\ \bibinfo
  {pages} {366c} (\bibinfo {year} {2003})}\BibitemShut {NoStop}%
\bibitem [{\citenamefont {Nunes}(2005)}]{Nunes2005_NPA757-349}%
  \BibitemOpen
  \bibfield  {author} {\bibinfo {author} {\bibfnamefont {F.~M.}\ \bibnamefont
  {Nunes}},\ }\href {\doibase 10.1016/j.nuclphysa.2005.04.005} {\bibfield
  {journal} {\bibinfo  {journal} {Nucl. Phys. A}\ }\textbf {\bibinfo {volume}
  {757}},\ \bibinfo {pages} {349} (\bibinfo {year} {2005})}\BibitemShut
  {NoStop}%
\bibitem [{\citenamefont {Pei}\ \emph {et~al.}(2006)\citenamefont {Pei},
  \citenamefont {Xu},\ and\ \citenamefont {Stevenson}}]{Pei2006_NPA765-29}%
  \BibitemOpen
  \bibfield  {author} {\bibinfo {author} {\bibfnamefont {J.~C.}\ \bibnamefont
  {Pei}}, \bibinfo {author} {\bibfnamefont {F.~R.}\ \bibnamefont {Xu}}, \ and\
  \bibinfo {author} {\bibfnamefont {P.~D.}\ \bibnamefont {Stevenson}},\ }\href
  {\doibase 10.1016/j.nuclphysa.2005.10.004} {\bibfield  {journal} {\bibinfo
  {journal} {Nucl. Phys. A}\ }\textbf {\bibinfo {volume} {765}},\ \bibinfo
  {pages} {29} (\bibinfo {year} {2006})}\BibitemShut {NoStop}%
\bibitem [{\citenamefont {Nakamura}\ \emph {et~al.}(2009)\citenamefont
  {Nakamura}, \citenamefont {Kobayashi}, \citenamefont {Kondo}, \citenamefont
  {Satou}, \citenamefont {Aoi}, \citenamefont {Baba}, \citenamefont {Deguchi},
  \citenamefont {Fukuda}, \citenamefont {Gibelin}, \citenamefont {Inabe},
  \citenamefont {Ishihara}, \citenamefont {Kameda}, \citenamefont {Kawada},
  \citenamefont {Kubo}, \citenamefont {Kusaka}, \citenamefont {Mengoni},
  \citenamefont {Motobayashi}, \citenamefont {Ohnishi}, \citenamefont {Ohtake},
  \citenamefont {Orr}, \citenamefont {Otsu}, \citenamefont {Otsuka},
  \citenamefont {Saito}, \citenamefont {Sakurai}, \citenamefont {Shimoura},
  \citenamefont {Sumikama},\ and\ \citenamefont
  {Takeda}}]{Nakamura2009_PRL103-262501}%
  \BibitemOpen
  \bibfield  {author} {\bibinfo {author} {\bibfnamefont {T.}~\bibnamefont
  {Nakamura}}, \bibinfo {author} {\bibfnamefont {N.}~\bibnamefont {Kobayashi}},
  \bibinfo {author} {\bibfnamefont {Y.}~\bibnamefont {Kondo}}, \bibinfo
  {author} {\bibfnamefont {Y.}~\bibnamefont {Satou}}, \bibinfo {author}
  {\bibfnamefont {N.}~\bibnamefont {Aoi}}, \bibinfo {author} {\bibfnamefont
  {H.}~\bibnamefont {Baba}}, \bibinfo {author} {\bibfnamefont {S.}~\bibnamefont
  {Deguchi}}, \bibinfo {author} {\bibfnamefont {N.}~\bibnamefont {Fukuda}},
  \bibinfo {author} {\bibfnamefont {J.}~\bibnamefont {Gibelin}}, \bibinfo
  {author} {\bibfnamefont {N.}~\bibnamefont {Inabe}}, \bibinfo {author}
  {\bibfnamefont {M.}~\bibnamefont {Ishihara}}, \bibinfo {author}
  {\bibfnamefont {D.}~\bibnamefont {Kameda}}, \bibinfo {author} {\bibfnamefont
  {Y.}~\bibnamefont {Kawada}}, \bibinfo {author} {\bibfnamefont
  {T.}~\bibnamefont {Kubo}}, \bibinfo {author} {\bibfnamefont {K.}~\bibnamefont
  {Kusaka}}, \bibinfo {author} {\bibfnamefont {A.}~\bibnamefont {Mengoni}},
  \bibinfo {author} {\bibfnamefont {T.}~\bibnamefont {Motobayashi}}, \bibinfo
  {author} {\bibfnamefont {T.}~\bibnamefont {Ohnishi}}, \bibinfo {author}
  {\bibfnamefont {M.}~\bibnamefont {Ohtake}}, \bibinfo {author} {\bibfnamefont
  {N.~A.}\ \bibnamefont {Orr}}, \bibinfo {author} {\bibfnamefont
  {H.}~\bibnamefont {Otsu}}, \bibinfo {author} {\bibfnamefont {T.}~\bibnamefont
  {Otsuka}}, \bibinfo {author} {\bibfnamefont {A.}~\bibnamefont {Saito}},
  \bibinfo {author} {\bibfnamefont {H.}~\bibnamefont {Sakurai}}, \bibinfo
  {author} {\bibfnamefont {S.}~\bibnamefont {Shimoura}}, \bibinfo {author}
  {\bibfnamefont {T.}~\bibnamefont {Sumikama}}, \ and\ \bibinfo {author}
  {\bibfnamefont {H.}~\bibnamefont {Takeda}},\ }\href {\doibase
  10.1103/PhysRevLett.103.262501} {\bibfield  {journal} {\bibinfo  {journal}
  {Phys. Rev. Lett.}\ }\textbf {\bibinfo {volume} {103}},\ \bibinfo {pages}
  {262501} (\bibinfo {year} {2009})}\BibitemShut {NoStop}%
\bibitem [{\citenamefont {Kanungo}\ \emph {et~al.}(2011)\citenamefont
  {Kanungo}, \citenamefont {Prochazka}, \citenamefont {Horiuchi}, \citenamefont
  {Nociforo}, \citenamefont {Aumann}, \citenamefont {Boutin}, \citenamefont
  {Cortina-Gil}, \citenamefont {Davids}, \citenamefont {Diakaki}, \citenamefont
  {Farinon}, \citenamefont {Geissel}, \citenamefont {Gernhaeuser},
  \citenamefont {Gerl}, \citenamefont {Janik}, \citenamefont {Jonson},
  \citenamefont {Kindler}, \citenamefont {Knoebel}, \citenamefont {Kruecken},
  \citenamefont {Lantz}, \citenamefont {Lenske}, \citenamefont {Litvinov},
  \citenamefont {Lommel}, \citenamefont {Mahata}, \citenamefont {Maierbeck},
  \citenamefont {Musumarra}, \citenamefont {Nilsson}, \citenamefont {Perro},
  \citenamefont {Scheidenberger}, \citenamefont {Sitar}, \citenamefont
  {Strmen}, \citenamefont {Sun}, \citenamefont {Suzuki}, \citenamefont
  {Szarka}, \citenamefont {Tanihata}, \citenamefont {Utsuno}, \citenamefont
  {Weick},\ and\ \citenamefont {Winkler}}]{Kanungo2011_PRC83-021302R}%
  \BibitemOpen
  \bibfield  {author} {\bibinfo {author} {\bibfnamefont {R.}~\bibnamefont
  {Kanungo}}, \bibinfo {author} {\bibfnamefont {A.}~\bibnamefont {Prochazka}},
  \bibinfo {author} {\bibfnamefont {W.}~\bibnamefont {Horiuchi}}, \bibinfo
  {author} {\bibfnamefont {C.}~\bibnamefont {Nociforo}}, \bibinfo {author}
  {\bibfnamefont {T.}~\bibnamefont {Aumann}}, \bibinfo {author} {\bibfnamefont
  {D.}~\bibnamefont {Boutin}}, \bibinfo {author} {\bibfnamefont
  {D.}~\bibnamefont {Cortina-Gil}}, \bibinfo {author} {\bibfnamefont
  {B.}~\bibnamefont {Davids}}, \bibinfo {author} {\bibfnamefont
  {M.}~\bibnamefont {Diakaki}}, \bibinfo {author} {\bibfnamefont
  {F.}~\bibnamefont {Farinon}}, \bibinfo {author} {\bibfnamefont
  {H.}~\bibnamefont {Geissel}}, \bibinfo {author} {\bibfnamefont
  {R.}~\bibnamefont {Gernhaeuser}}, \bibinfo {author} {\bibfnamefont
  {J.}~\bibnamefont {Gerl}}, \bibinfo {author} {\bibfnamefont {R.}~\bibnamefont
  {Janik}}, \bibinfo {author} {\bibfnamefont {B.}~\bibnamefont {Jonson}},
  \bibinfo {author} {\bibfnamefont {B.}~\bibnamefont {Kindler}}, \bibinfo
  {author} {\bibfnamefont {R.}~\bibnamefont {Knoebel}}, \bibinfo {author}
  {\bibfnamefont {R.}~\bibnamefont {Kruecken}}, \bibinfo {author}
  {\bibfnamefont {M.}~\bibnamefont {Lantz}}, \bibinfo {author} {\bibfnamefont
  {H.}~\bibnamefont {Lenske}}, \bibinfo {author} {\bibfnamefont
  {Y.}~\bibnamefont {Litvinov}}, \bibinfo {author} {\bibfnamefont
  {B.}~\bibnamefont {Lommel}}, \bibinfo {author} {\bibfnamefont
  {K.}~\bibnamefont {Mahata}}, \bibinfo {author} {\bibfnamefont
  {P.}~\bibnamefont {Maierbeck}}, \bibinfo {author} {\bibfnamefont
  {A.}~\bibnamefont {Musumarra}}, \bibinfo {author} {\bibfnamefont
  {T.}~\bibnamefont {Nilsson}}, \bibinfo {author} {\bibfnamefont
  {C.}~\bibnamefont {Perro}}, \bibinfo {author} {\bibfnamefont
  {C.}~\bibnamefont {Scheidenberger}}, \bibinfo {author} {\bibfnamefont
  {B.}~\bibnamefont {Sitar}}, \bibinfo {author} {\bibfnamefont
  {P.}~\bibnamefont {Strmen}}, \bibinfo {author} {\bibfnamefont
  {B.}~\bibnamefont {Sun}}, \bibinfo {author} {\bibfnamefont {Y.}~\bibnamefont
  {Suzuki}}, \bibinfo {author} {\bibfnamefont {I.}~\bibnamefont {Szarka}},
  \bibinfo {author} {\bibfnamefont {I.}~\bibnamefont {Tanihata}}, \bibinfo
  {author} {\bibfnamefont {Y.}~\bibnamefont {Utsuno}}, \bibinfo {author}
  {\bibfnamefont {H.}~\bibnamefont {Weick}}, \ and\ \bibinfo {author}
  {\bibfnamefont {M.}~\bibnamefont {Winkler}},\ }\href {\doibase
  10.1103/PhysRevC.83.021302} {\bibfield  {journal} {\bibinfo  {journal} {Phys.
  Rev. C}\ }\textbf {\bibinfo {volume} {83}},\ \bibinfo {pages} {021302(R)}
  (\bibinfo {year} {2011})}\BibitemShut {NoStop}%
\bibitem [{\citenamefont {Terasaki}\ \emph {et~al.}(1996)\citenamefont
  {Terasaki}, \citenamefont {Heenen}, \citenamefont {Flocard},\ and\
  \citenamefont {Bonche}}]{Terasaki1996_NPA600-371}%
  \BibitemOpen
  \bibfield  {author} {\bibinfo {author} {\bibfnamefont {J.}~\bibnamefont
  {Terasaki}}, \bibinfo {author} {\bibfnamefont {P.~H.}\ \bibnamefont
  {Heenen}}, \bibinfo {author} {\bibfnamefont {H.}~\bibnamefont {Flocard}}, \
  and\ \bibinfo {author} {\bibfnamefont {P.}~\bibnamefont {Bonche}},\ }\href
  {\doibase 10.1016/0375-9474(96)00036-X} {\bibfield  {journal} {\bibinfo
  {journal} {Nucl. Phys. A}\ }\textbf {\bibinfo {volume} {600}},\ \bibinfo
  {pages} {371} (\bibinfo {year} {1996})}\BibitemShut {NoStop}%
\bibitem [{\citenamefont {Terasaki}\ \emph {et~al.}(1997)\citenamefont
  {Terasaki}, \citenamefont {Flocard}, \citenamefont {Heenen},\ and\
  \citenamefont {Bonche}}]{Terasaki1997_NPA621-706}%
  \BibitemOpen
  \bibfield  {author} {\bibinfo {author} {\bibfnamefont {J.}~\bibnamefont
  {Terasaki}}, \bibinfo {author} {\bibfnamefont {H.}~\bibnamefont {Flocard}},
  \bibinfo {author} {\bibfnamefont {P.~H.}\ \bibnamefont {Heenen}}, \ and\
  \bibinfo {author} {\bibfnamefont {P.}~\bibnamefont {Bonche}},\ }\href
  {\doibase 10.1016/S0375-9474(97)00183-8} {\bibfield  {journal} {\bibinfo
  {journal} {Nucl. Phys. A}\ }\textbf {\bibinfo {volume} {621}},\ \bibinfo
  {pages} {706} (\bibinfo {year} {1997})}\BibitemShut {NoStop}%
\bibitem [{\citenamefont {Yamagami}\ \emph {et~al.}(2001)\citenamefont
  {Yamagami}, \citenamefont {Matsuyanagi},\ and\ \citenamefont
  {Matsuo}}]{Yamagami2001_NPA693-579}%
  \BibitemOpen
  \bibfield  {author} {\bibinfo {author} {\bibfnamefont {M.}~\bibnamefont
  {Yamagami}}, \bibinfo {author} {\bibfnamefont {K.}~\bibnamefont
  {Matsuyanagi}}, \ and\ \bibinfo {author} {\bibfnamefont {M.}~\bibnamefont
  {Matsuo}},\ }\href {\doibase 10.1016/S0375-9474(01)00918-6} {\bibfield
  {journal} {\bibinfo  {journal} {Nucl. Phys. A}\ }\textbf {\bibinfo {volume}
  {693}},\ \bibinfo {pages} {579} (\bibinfo {year} {2001})}\BibitemShut
  {NoStop}%
\bibitem [{\citenamefont {Teran}\ \emph {et~al.}(2003)\citenamefont {Teran},
  \citenamefont {Oberacker},\ and\ \citenamefont
  {Umar}}]{Teran2003_PRC67-064314}%
  \BibitemOpen
  \bibfield  {author} {\bibinfo {author} {\bibfnamefont {E.}~\bibnamefont
  {Teran}}, \bibinfo {author} {\bibfnamefont {V.~E.}\ \bibnamefont
  {Oberacker}}, \ and\ \bibinfo {author} {\bibfnamefont {A.~S.}\ \bibnamefont
  {Umar}},\ }\href {\doibase 10.1103/PhysRevC.67.064314} {\bibfield  {journal}
  {\bibinfo  {journal} {Phys. Rev. C}\ }\textbf {\bibinfo {volume} {67}},\
  \bibinfo {pages} {064314} (\bibinfo {year} {2003})}\BibitemShut {NoStop}%
\bibitem [{\citenamefont {Oberacker}\ \emph {et~al.}(2003)\citenamefont
  {Oberacker}, \citenamefont {Umar}, \citenamefont {Ter\'{a}n},\ and\
  \citenamefont {Blazkiewicz}}]{Oberacker2003_PRC68-064302}%
  \BibitemOpen
  \bibfield  {author} {\bibinfo {author} {\bibfnamefont {V.~E.}\ \bibnamefont
  {Oberacker}}, \bibinfo {author} {\bibfnamefont {A.~S.}\ \bibnamefont {Umar}},
  \bibinfo {author} {\bibfnamefont {E.}~\bibnamefont {Ter\'{a}n}}, \ and\
  \bibinfo {author} {\bibfnamefont {A.}~\bibnamefont {Blazkiewicz}},\ }\href
  {\doibase 10.1103/PhysRevC.68.064302} {\bibfield  {journal} {\bibinfo
  {journal} {Phys. Rev. C}\ }\textbf {\bibinfo {volume} {68}},\ \bibinfo
  {pages} {064302} (\bibinfo {year} {2003})}\BibitemShut {NoStop}%
\bibitem [{\citenamefont {Pei}\ \emph {et~al.}(2008)\citenamefont {Pei},
  \citenamefont {Stoitsov}, \citenamefont {Fann}, \citenamefont {Nazarewicz},
  \citenamefont {Schunck},\ and\ \citenamefont {Xu}}]{Pei2008_PRC78-064306}%
  \BibitemOpen
  \bibfield  {author} {\bibinfo {author} {\bibfnamefont {J.~C.}\ \bibnamefont
  {Pei}}, \bibinfo {author} {\bibfnamefont {M.~V.}\ \bibnamefont {Stoitsov}},
  \bibinfo {author} {\bibfnamefont {G.~I.}\ \bibnamefont {Fann}}, \bibinfo
  {author} {\bibfnamefont {W.}~\bibnamefont {Nazarewicz}}, \bibinfo {author}
  {\bibfnamefont {N.}~\bibnamefont {Schunck}}, \ and\ \bibinfo {author}
  {\bibfnamefont {F.~R.}\ \bibnamefont {Xu}},\ }\href {\doibase
  10.1103/PhysRevC.78.064306} {\bibfield  {journal} {\bibinfo  {journal} {Phys.
  Rev. C}\ }\textbf {\bibinfo {volume} {78}},\ \bibinfo {pages} {064306}
  (\bibinfo {year} {2008})}\BibitemShut {NoStop}%
\bibitem [{\citenamefont {Tajima}(2004)}]{Tajima2004_PRC69-034305}%
  \BibitemOpen
  \bibfield  {author} {\bibinfo {author} {\bibfnamefont {N.}~\bibnamefont
  {Tajima}},\ }\href {\doibase 10.1103/PhysRevC.69.034305} {\bibfield
  {journal} {\bibinfo  {journal} {Phys. Rev. C}\ }\textbf {\bibinfo {volume}
  {69}},\ \bibinfo {pages} {034305} (\bibinfo {year} {2004})}\BibitemShut
  {NoStop}%
\bibitem [{\citenamefont {Reinhard}\ \emph {et~al.}(1997)\citenamefont
  {Reinhard}, \citenamefont {Bender}, \citenamefont {Rutz},\ and\ \citenamefont
  {Maruhn}}]{Reinhard1997_ZPA358-277}%
  \BibitemOpen
  \bibfield  {author} {\bibinfo {author} {\bibfnamefont {P.-G.}\ \bibnamefont
  {Reinhard}}, \bibinfo {author} {\bibfnamefont {M.}~\bibnamefont {Bender}},
  \bibinfo {author} {\bibfnamefont {K.}~\bibnamefont {Rutz}}, \ and\ \bibinfo
  {author} {\bibfnamefont {J.}~\bibnamefont {Maruhn}},\ }\href {\doibase
  10.1007/s002180050328} {\bibfield  {journal} {\bibinfo  {journal} {Z. Phys.
  A}\ }\textbf {\bibinfo {volume} {358}},\ \bibinfo {pages} {277} (\bibinfo
  {year} {1997})}\BibitemShut {NoStop}%
\bibitem [{\citenamefont {Nakada}(2008)}]{Nakada2008_NPA808-47}%
  \BibitemOpen
  \bibfield  {author} {\bibinfo {author} {\bibfnamefont {H.}~\bibnamefont
  {Nakada}},\ }\href {\doibase 10.1016/j.nuclphysa.2008.05.011} {\bibfield
  {journal} {\bibinfo  {journal} {Nucl. Phys. A}\ }\textbf {\bibinfo {volume}
  {808}},\ \bibinfo {pages} {47} (\bibinfo {year} {2008})}\BibitemShut
  {NoStop}%
\bibitem [{\citenamefont {Oba}\ and\ \citenamefont
  {Matsuo}(2009)}]{Oba2009_PRC80-024301}%
  \BibitemOpen
  \bibfield  {author} {\bibinfo {author} {\bibfnamefont {H.}~\bibnamefont
  {Oba}}\ and\ \bibinfo {author} {\bibfnamefont {M.}~\bibnamefont {Matsuo}},\
  }\href {\doibase 10.1103/PhysRevC.80.024301} {\bibfield  {journal} {\bibinfo
  {journal} {Phys. Rev. C}\ }\textbf {\bibinfo {volume} {80}},\ \bibinfo
  {pages} {024301} (\bibinfo {year} {2009})}\BibitemShut {NoStop}%
\bibitem [{\citenamefont {Zhang}\ \emph
  {et~al.}(2011{\natexlab{b}})\citenamefont {Zhang}, \citenamefont {Matsuo},\
  and\ \citenamefont {Meng}}]{Zhang2011_PRC83-054301}%
  \BibitemOpen
  \bibfield  {author} {\bibinfo {author} {\bibfnamefont {Y.}~\bibnamefont
  {Zhang}}, \bibinfo {author} {\bibfnamefont {M.}~\bibnamefont {Matsuo}}, \
  and\ \bibinfo {author} {\bibfnamefont {J.}~\bibnamefont {Meng}},\ }\href
  {\doibase 10.1103/PhysRevC.83.054301} {\bibfield  {journal} {\bibinfo
  {journal} {Phys. Rev. C}\ }\textbf {\bibinfo {volume} {83}},\ \bibinfo
  {pages} {054301} (\bibinfo {year} {2011}{\natexlab{b}})}\BibitemShut
  {NoStop}%
\bibitem [{\citenamefont {Vretenar}\ \emph {et~al.}(1999)\citenamefont
  {Vretenar}, \citenamefont {Lalazissis},\ and\ \citenamefont
  {Ring}}]{Vretenar1999_PRL82-4595}%
  \BibitemOpen
  \bibfield  {author} {\bibinfo {author} {\bibfnamefont {D.}~\bibnamefont
  {Vretenar}}, \bibinfo {author} {\bibfnamefont {G.~A.}\ \bibnamefont
  {Lalazissis}}, \ and\ \bibinfo {author} {\bibfnamefont {P.}~\bibnamefont
  {Ring}},\ }\href {\doibase 10.1103/PhysRevLett.82.4595} {\bibfield  {journal}
  {\bibinfo  {journal} {Phys. Rev. Lett.}\ }\textbf {\bibinfo {volume} {82}},\
  \bibinfo {pages} {4595} (\bibinfo {year} {1999})}\BibitemShut {NoStop}%
\bibitem [{\citenamefont {Lalazissis}\ \emph
  {et~al.}(1999{\natexlab{a}})\citenamefont {Lalazissis}, \citenamefont
  {Vretenar}, \citenamefont {Ring}, \citenamefont {Stoitsov},\ and\
  \citenamefont {Robledo}}]{Lalazissis1999_PRC60-014310}%
  \BibitemOpen
  \bibfield  {author} {\bibinfo {author} {\bibfnamefont {G.~A.}\ \bibnamefont
  {Lalazissis}}, \bibinfo {author} {\bibfnamefont {D.}~\bibnamefont
  {Vretenar}}, \bibinfo {author} {\bibfnamefont {P.}~\bibnamefont {Ring}},
  \bibinfo {author} {\bibfnamefont {M.}~\bibnamefont {Stoitsov}}, \ and\
  \bibinfo {author} {\bibfnamefont {L.~M.}\ \bibnamefont {Robledo}},\ }\href
  {\doibase 10.1103/PhysRevC.60.014310} {\bibfield  {journal} {\bibinfo
  {journal} {Phys. Rev. C}\ }\textbf {\bibinfo {volume} {60}},\ \bibinfo
  {pages} {014310} (\bibinfo {year} {1999}{\natexlab{a}})}\BibitemShut
  {NoStop}%
\bibitem [{\citenamefont {Lalazissis}\ \emph
  {et~al.}(1999{\natexlab{b}})\citenamefont {Lalazissis}, \citenamefont
  {Vretenar},\ and\ \citenamefont {Ring}}]{Lalazissis1999_NPA650-133}%
  \BibitemOpen
  \bibfield  {author} {\bibinfo {author} {\bibfnamefont {G.~A.}\ \bibnamefont
  {Lalazissis}}, \bibinfo {author} {\bibfnamefont {D.}~\bibnamefont
  {Vretenar}}, \ and\ \bibinfo {author} {\bibfnamefont {P.}~\bibnamefont
  {Ring}},\ }\href {\doibase 10.1016/S0375-9474(99)00121-9} {\bibfield
  {journal} {\bibinfo  {journal} {Nucl. Phys. A}\ }\textbf {\bibinfo {volume}
  {650}},\ \bibinfo {pages} {133} (\bibinfo {year}
  {1999}{\natexlab{b}})}\BibitemShut {NoStop}%
\bibitem [{\citenamefont {Niksic}\ \emph {et~al.}(2002)\citenamefont {Niksic},
  \citenamefont {Vretenar}, \citenamefont {Ring},\ and\ \citenamefont
  {Lalazissis}}]{Niksic2002_PRC65-054320}%
  \BibitemOpen
  \bibfield  {author} {\bibinfo {author} {\bibfnamefont {T.}~\bibnamefont
  {Niksic}}, \bibinfo {author} {\bibfnamefont {D.}~\bibnamefont {Vretenar}},
  \bibinfo {author} {\bibfnamefont {P.}~\bibnamefont {Ring}}, \ and\ \bibinfo
  {author} {\bibfnamefont {G.~A.}\ \bibnamefont {Lalazissis}},\ }\href
  {\doibase 10.1103/PhysRevC.65.054320} {\bibfield  {journal} {\bibinfo
  {journal} {Phys. Rev. C}\ }\textbf {\bibinfo {volume} {65}},\ \bibinfo
  {pages} {054320} (\bibinfo {year} {2002})}\BibitemShut {NoStop}%
\bibitem [{\citenamefont {Niksic}\ \emph {et~al.}(2004)\citenamefont {Niksic},
  \citenamefont {Vretenar}, \citenamefont {Lalazissis},\ and\ \citenamefont
  {Ring}}]{Niksic2004_PRC69-047301}%
  \BibitemOpen
  \bibfield  {author} {\bibinfo {author} {\bibfnamefont {T.}~\bibnamefont
  {Niksic}}, \bibinfo {author} {\bibfnamefont {D.}~\bibnamefont {Vretenar}},
  \bibinfo {author} {\bibfnamefont {G.~A.}\ \bibnamefont {Lalazissis}}, \ and\
  \bibinfo {author} {\bibfnamefont {P.}~\bibnamefont {Ring}},\ }\href {\doibase
  10.1103/PhysRevC.69.047301} {\bibfield  {journal} {\bibinfo  {journal} {Phys.
  Rev. C}\ }\textbf {\bibinfo {volume} {69}},\ \bibinfo {pages} {047301}
  (\bibinfo {year} {2004})}\BibitemShut {NoStop}%
\bibitem [{\citenamefont {Zhang}\ and\ \citenamefont
  {Onley}(1988)}]{Zhang1988_PLB209-145}%
  \BibitemOpen
  \bibfield  {author} {\bibinfo {author} {\bibfnamefont {J.-K.}\ \bibnamefont
  {Zhang}}\ and\ \bibinfo {author} {\bibfnamefont {D.~S.}\ \bibnamefont
  {Onley}},\ }\href {\doibase 10.1016/0370-2693(88)90921-5} {\bibfield
  {journal} {\bibinfo  {journal} {Phys. Lett. B}\ }\textbf {\bibinfo {volume}
  {209}},\ \bibinfo {pages} {145} (\bibinfo {year} {1988})}\BibitemShut
  {NoStop}%
\bibitem [{\citenamefont {Zhang}\ and\ \citenamefont
  {Onley}(1991)}]{Zhang1991_NPA526-245}%
  \BibitemOpen
  \bibfield  {author} {\bibinfo {author} {\bibfnamefont {J.-K.}\ \bibnamefont
  {Zhang}}\ and\ \bibinfo {author} {\bibfnamefont {D.~S.}\ \bibnamefont
  {Onley}},\ }\href {\doibase 10.1016/0375-9474(91)90286-F} {\bibfield
  {journal} {\bibinfo  {journal} {Nucl. Phys. A}\ }\textbf {\bibinfo {volume}
  {526}},\ \bibinfo {pages} {245} (\bibinfo {year} {1991})}\BibitemShut
  {NoStop}%
\bibitem [{\citenamefont {Duerr}\ and\ \citenamefont
  {Teller}(1956)}]{Duerr1956_PR0101-494}%
  \BibitemOpen
  \bibfield  {author} {\bibinfo {author} {\bibfnamefont {H.-P.}\ \bibnamefont
  {Duerr}}\ and\ \bibinfo {author} {\bibfnamefont {E.}~\bibnamefont {Teller}},\
  }\href {\doibase 10.1103/PhysRev.101.494} {\bibfield  {journal} {\bibinfo
  {journal} {Phys. Rev.}\ }\textbf {\bibinfo {volume} {101}},\ \bibinfo {pages}
  {494} (\bibinfo {year} {1956})}\BibitemShut {NoStop}%
\bibitem [{\citenamefont {Duerr}(1956)}]{Duerr1956_PR0103-469}%
  \BibitemOpen
  \bibfield  {author} {\bibinfo {author} {\bibfnamefont {H.-P.}\ \bibnamefont
  {Duerr}},\ }\href {\doibase 10.1103/PhysRev.103.469} {\bibfield  {journal}
  {\bibinfo  {journal} {Phys. Rev.}\ }\textbf {\bibinfo {volume} {103}},\
  \bibinfo {pages} {469} (\bibinfo {year} {1956})}\BibitemShut {NoStop}%
\bibitem [{\citenamefont {Walecka}(1974)}]{Walecka1974_APNY83-491}%
  \BibitemOpen
  \bibfield  {author} {\bibinfo {author} {\bibfnamefont {J.~D.}\ \bibnamefont
  {Walecka}},\ }\href {\doibase 10.1016/0003-4916(74)90208-5} {\bibfield
  {journal} {\bibinfo  {journal} {Ann. Phys.}\ }\textbf {\bibinfo {volume}
  {83}},\ \bibinfo {pages} {491} (\bibinfo {year} {1974})}\BibitemShut
  {NoStop}%
\bibitem [{\citenamefont {Serot}\ and\ \citenamefont
  {Walecka}(1986)}]{Serot1986_ANP16-1}%
  \BibitemOpen
  \bibfield  {author} {\bibinfo {author} {\bibfnamefont {B.~D.}\ \bibnamefont
  {Serot}}\ and\ \bibinfo {author} {\bibfnamefont {J.~D.}\ \bibnamefont
  {Walecka}},\ }\href@noop {} {\bibfield  {journal} {\bibinfo  {journal} {Adv.
  Nucl. Phys.}\ }\textbf {\bibinfo {volume} {16}},\ \bibinfo {pages} {1}
  (\bibinfo {year} {1986})}\BibitemShut {NoStop}%
\bibitem [{\citenamefont {Reinhard}(1989)}]{Reinhard1989_RPP52-439}%
  \BibitemOpen
  \bibfield  {author} {\bibinfo {author} {\bibfnamefont {P.~G.}\ \bibnamefont
  {Reinhard}},\ }\href {\doibase 10.1088/0034-4885/52/4/002} {\bibfield
  {journal} {\bibinfo  {journal} {Rep. Prog. Phys.}\ }\textbf {\bibinfo
  {volume} {52}},\ \bibinfo {pages} {439} (\bibinfo {year} {1989})}\BibitemShut
  {NoStop}%
\bibitem [{\citenamefont {Ring}(1996)}]{Ring1996_PPNP37-193}%
  \BibitemOpen
  \bibfield  {author} {\bibinfo {author} {\bibfnamefont {P.}~\bibnamefont
  {Ring}},\ }\href {\doibase 10.1016/0146-6410(96)00054-3} {\bibfield
  {journal} {\bibinfo  {journal} {Prog. Part. Nucl. Phys.}\ }\textbf {\bibinfo
  {volume} {37}},\ \bibinfo {pages} {193} (\bibinfo {year} {1996})}\BibitemShut
  {NoStop}%
\bibitem [{\citenamefont {Ring}(2001)}]{Ring2001_PPNP46-165}%
  \BibitemOpen
  \bibfield  {author} {\bibinfo {author} {\bibfnamefont {P.}~\bibnamefont
  {Ring}},\ }\href {\doibase 10.1016/S0146-6410(01)00120-X} {\bibfield
  {journal} {\bibinfo  {journal} {Prog. Part. Nucl. Phys.}\ }\textbf {\bibinfo
  {volume} {46}},\ \bibinfo {pages} {165} (\bibinfo {year} {2001})}\BibitemShut
  {NoStop}%
\bibitem [{\citenamefont {Vretenar}\ \emph {et~al.}(2005)\citenamefont
  {Vretenar}, \citenamefont {Afanasjev}, \citenamefont {Lalazissis},\ and\
  \citenamefont {Ring}}]{Vretenar2005_PR409-101}%
  \BibitemOpen
  \bibfield  {author} {\bibinfo {author} {\bibfnamefont {D.}~\bibnamefont
  {Vretenar}}, \bibinfo {author} {\bibfnamefont {A.}~\bibnamefont {Afanasjev}},
  \bibinfo {author} {\bibfnamefont {G.}~\bibnamefont {Lalazissis}}, \ and\
  \bibinfo {author} {\bibfnamefont {P.}~\bibnamefont {Ring}},\ }\href {\doibase
  10.1016/j.physrep.2004.10.001} {\bibfield  {journal} {\bibinfo  {journal}
  {Phys. Rep.}\ }\textbf {\bibinfo {volume} {409}},\ \bibinfo {pages} {101}
  (\bibinfo {year} {2005})}\BibitemShut {NoStop}%
\bibitem [{\citenamefont {Meng}\ \emph {et~al.}(2006)\citenamefont {Meng},
  \citenamefont {Toki}, \citenamefont {Zhou}, \citenamefont {Zhang},
  \citenamefont {Long},\ and\ \citenamefont {Geng}}]{Meng2006_PPNP57-470}%
  \BibitemOpen
  \bibfield  {author} {\bibinfo {author} {\bibfnamefont {J.}~\bibnamefont
  {Meng}}, \bibinfo {author} {\bibfnamefont {H.}~\bibnamefont {Toki}}, \bibinfo
  {author} {\bibfnamefont {S.~G.}\ \bibnamefont {Zhou}}, \bibinfo {author}
  {\bibfnamefont {S.~Q.}\ \bibnamefont {Zhang}}, \bibinfo {author}
  {\bibfnamefont {W.~H.}\ \bibnamefont {Long}}, \ and\ \bibinfo {author}
  {\bibfnamefont {L.~S.}\ \bibnamefont {Geng}},\ }\href {\doibase
  10.1016/j.ppnp.2005.06.001} {\bibfield  {journal} {\bibinfo  {journal} {Prog.
  Part. Nucl. Phys.}\ }\textbf {\bibinfo {volume} {57}},\ \bibinfo {pages}
  {470} (\bibinfo {year} {2006})}\BibitemShut {NoStop}%
\bibitem [{\citenamefont {Boguta}\ and\ \citenamefont
  {Bodmer}(1977)}]{Boguta1977_NPA292-413}%
  \BibitemOpen
  \bibfield  {author} {\bibinfo {author} {\bibfnamefont {J.}~\bibnamefont
  {Boguta}}\ and\ \bibinfo {author} {\bibfnamefont {A.~R.}\ \bibnamefont
  {Bodmer}},\ }\href {\doibase 10.1016/0375-9474(77)90626-1} {\bibfield
  {journal} {\bibinfo  {journal} {Nucl. Phys. A}\ }\textbf {\bibinfo {volume}
  {292}},\ \bibinfo {pages} {413} (\bibinfo {year} {1977})}\BibitemShut
  {NoStop}%
\bibitem [{\citenamefont {Kucharek}\ and\ \citenamefont
  {Ring}(1991)}]{Kucharek1991_ZPA339-23}%
  \BibitemOpen
  \bibfield  {author} {\bibinfo {author} {\bibfnamefont {H.}~\bibnamefont
  {Kucharek}}\ and\ \bibinfo {author} {\bibfnamefont {P.}~\bibnamefont
  {Ring}},\ }\href {\doibase 10.1007/BF01282930} {\bibfield  {journal}
  {\bibinfo  {journal} {Z. Phys. A}\ }\textbf {\bibinfo {volume} {339}},\
  \bibinfo {pages} {23} (\bibinfo {year} {1991})}\BibitemShut {NoStop}%
\bibitem [{\citenamefont {Ring}\ and\ \citenamefont {Schuck}(1980)}]{Ring1980}%
  \BibitemOpen
  \bibfield  {author} {\bibinfo {author} {\bibfnamefont {P.}~\bibnamefont
  {Ring}}\ and\ \bibinfo {author} {\bibfnamefont {P.}~\bibnamefont {Schuck}},\
  }\href@noop {} {\emph {\bibinfo {title} {The Nuclear Many-Body Problem}}}\
  (\bibinfo  {publisher} {Springer},\ \bibinfo {year} {1980})\BibitemShut
  {NoStop}%
\bibitem [{\citenamefont {Price}\ and\ \citenamefont
  {Walker}(1987)}]{Price1987_PRC36-354}%
  \BibitemOpen
  \bibfield  {author} {\bibinfo {author} {\bibfnamefont {C.~E.}\ \bibnamefont
  {Price}}\ and\ \bibinfo {author} {\bibfnamefont {G.~E.}\ \bibnamefont
  {Walker}},\ }\href {\doibase 10.1103/PhysRevC.36.354} {\bibfield  {journal}
  {\bibinfo  {journal} {Phys. Rev. C}\ }\textbf {\bibinfo {volume} {36}},\
  \bibinfo {pages} {354} (\bibinfo {year} {1987})}\BibitemShut {NoStop}%
\bibitem [{\citenamefont {Koepf}\ and\ \citenamefont
  {Ring}(1991)}]{Koepf1991_ZPA339-81}%
  \BibitemOpen
  \bibfield  {author} {\bibinfo {author} {\bibfnamefont {W.}~\bibnamefont
  {Koepf}}\ and\ \bibinfo {author} {\bibfnamefont {P.}~\bibnamefont {Ring}},\
  }\href {\doibase 10.1007/BF01282936} {\bibfield  {journal} {\bibinfo
  {journal} {Z. Phys. A}\ }\textbf {\bibinfo {volume} {339}},\ \bibinfo {pages}
  {81} (\bibinfo {year} {1991})}\BibitemShut {NoStop}%
\bibitem [{\citenamefont {Bonche}\ \emph {et~al.}(1985)\citenamefont {Bonche},
  \citenamefont {Flocard}, \citenamefont {Heenen}, \citenamefont {Krieger},\
  and\ \citenamefont {Weiss}}]{Bonche1985_NPA443-39}%
  \BibitemOpen
  \bibfield  {author} {\bibinfo {author} {\bibfnamefont {P.}~\bibnamefont
  {Bonche}}, \bibinfo {author} {\bibfnamefont {H.}~\bibnamefont {Flocard}},
  \bibinfo {author} {\bibfnamefont {P.~H.}\ \bibnamefont {Heenen}}, \bibinfo
  {author} {\bibfnamefont {S.~J.}\ \bibnamefont {Krieger}}, \ and\ \bibinfo
  {author} {\bibfnamefont {M.~S.}\ \bibnamefont {Weiss}},\ }\href {\doibase
  10.1016/0375-9474(85)90320-3} {\bibfield  {journal} {\bibinfo  {journal}
  {Nucl. Phys. A}\ }\textbf {\bibinfo {volume} {443}},\ \bibinfo {pages} {39}
  (\bibinfo {year} {1985})}\BibitemShut {NoStop}%
\bibitem [{\citenamefont {Long}\ \emph {et~al.}(2004)\citenamefont {Long},
  \citenamefont {Meng}, \citenamefont {Giai},\ and\ \citenamefont
  {Zhou}}]{Long2004_PRC69-034319}%
  \BibitemOpen
  \bibfield  {author} {\bibinfo {author} {\bibfnamefont {W.}~\bibnamefont
  {Long}}, \bibinfo {author} {\bibfnamefont {J.}~\bibnamefont {Meng}}, \bibinfo
  {author} {\bibfnamefont {N.~V.}\ \bibnamefont {Giai}}, \ and\ \bibinfo
  {author} {\bibfnamefont {S.-G.}\ \bibnamefont {Zhou}},\ }\href {\doibase
  10.1103/PhysRevC.69.034319} {\bibfield  {journal} {\bibinfo  {journal} {Phys.
  Rev. C}\ }\textbf {\bibinfo {volume} {69}},\ \bibinfo {pages} {034319}
  (\bibinfo {year} {2004})}\BibitemShut {NoStop}%
\bibitem [{\citenamefont {Zhao}\ \emph {et~al.}(2009)\citenamefont {Zhao},
  \citenamefont {Sun},\ and\ \citenamefont {Meng}}]{Zhao2009_CPL26-112102}%
  \BibitemOpen
  \bibfield  {author} {\bibinfo {author} {\bibfnamefont {P.-W.}\ \bibnamefont
  {Zhao}}, \bibinfo {author} {\bibfnamefont {B.-Y.}\ \bibnamefont {Sun}}, \
  and\ \bibinfo {author} {\bibfnamefont {J.}~\bibnamefont {Meng}},\ }\href
  {http://stacks.iop.org/0256-307X/26/i=11/a=112102} {\bibfield  {journal}
  {\bibinfo  {journal} {Chin. Phys. Lett.}\ }\textbf {\bibinfo {volume} {26}},\
  \bibinfo {pages} {112102} (\bibinfo {year} {2009})}\BibitemShut {NoStop}%
\bibitem [{\citenamefont {Lalazissis}\ \emph {et~al.}(1997)\citenamefont
  {Lalazissis}, \citenamefont {Konig},\ and\ \citenamefont
  {Ring}}]{Lalazissis1997_PRC55-540}%
  \BibitemOpen
  \bibfield  {author} {\bibinfo {author} {\bibfnamefont {G.~A.}\ \bibnamefont
  {Lalazissis}}, \bibinfo {author} {\bibfnamefont {J.}~\bibnamefont {Konig}}, \
  and\ \bibinfo {author} {\bibfnamefont {P.}~\bibnamefont {Ring}},\ }\href
  {\doibase 10.1103/PhysRevC.55.540} {\bibfield  {journal} {\bibinfo  {journal}
  {Phys. Rev. C}\ }\textbf {\bibinfo {volume} {55}},\ \bibinfo {pages} {540}
  (\bibinfo {year} {1997})}\BibitemShut {NoStop}%
\bibitem [{\citenamefont {Serra}\ and\ \citenamefont
  {Ring}(2002)}]{Serra2002_PRC65-064324}%
  \BibitemOpen
  \bibfield  {author} {\bibinfo {author} {\bibfnamefont {M.}~\bibnamefont
  {Serra}}\ and\ \bibinfo {author} {\bibfnamefont {P.}~\bibnamefont {Ring}},\
  }\href {\doibase 10.1103/PhysRevC.65.064324} {\bibfield  {journal} {\bibinfo
  {journal} {Phys. Rev. C}\ }\textbf {\bibinfo {volume} {65}},\ \bibinfo
  {pages} {064324} (\bibinfo {year} {2002})}\BibitemShut {NoStop}%
\bibitem [{\citenamefont {Berger}\ \emph {et~al.}(1984)\citenamefont {Berger},
  \citenamefont {Girod},\ and\ \citenamefont {Gogny}}]{Berger1984_NPA428-23}%
  \BibitemOpen
  \bibfield  {author} {\bibinfo {author} {\bibfnamefont {J.~F.}\ \bibnamefont
  {Berger}}, \bibinfo {author} {\bibfnamefont {M.}~\bibnamefont {Girod}}, \
  and\ \bibinfo {author} {\bibfnamefont {D.}~\bibnamefont {Gogny}},\ }\href
  {\doibase 10.1016/0375-9474(84)90240-9} {\bibfield  {journal} {\bibinfo
  {journal} {Nucl. Phys. A}\ }\textbf {\bibinfo {volume} {428}},\ \bibinfo
  {pages} {23} (\bibinfo {year} {1984})}\BibitemShut {NoStop}%
\bibitem [{\citenamefont {Audi}\ \emph {et~al.}(2003)\citenamefont {Audi},
  \citenamefont {Wapstra},\ and\ \citenamefont
  {Thibault}}]{Audi2003_NPA729-337}%
  \BibitemOpen
  \bibfield  {author} {\bibinfo {author} {\bibfnamefont {G.}~\bibnamefont
  {Audi}}, \bibinfo {author} {\bibfnamefont {A.~H.}\ \bibnamefont {Wapstra}}, \
  and\ \bibinfo {author} {\bibfnamefont {C.}~\bibnamefont {Thibault}},\ }\href
  {\doibase 10.1016/j.nuclphysa.2003.11.003} {\bibfield  {journal} {\bibinfo
  {journal} {Nucl. Phys. A}\ }\textbf {\bibinfo {volume} {729}},\ \bibinfo
  {pages} {337} (\bibinfo {year} {2003})}\BibitemShut {NoStop}%
\bibitem [{\citenamefont {Baumann}\ \emph {et~al.}(2007)\citenamefont
  {Baumann}, \citenamefont {Amthor}, \citenamefont {Bazin}, \citenamefont
  {Brown}, \citenamefont {III}, \citenamefont {Gade}, \citenamefont {Ginter},
  \citenamefont {Hausmann}, \citenamefont {Matos}, \citenamefont {Morrissey},
  \citenamefont {Portillo}, \citenamefont {Schiller}, \citenamefont {Sherrill},
  \citenamefont {Stolz}, \citenamefont {Tarasov},\ and\ \citenamefont
  {Thoennessen}}]{Baumann2007_Nature449-1022}%
  \BibitemOpen
  \bibfield  {author} {\bibinfo {author} {\bibfnamefont {T.}~\bibnamefont
  {Baumann}}, \bibinfo {author} {\bibfnamefont {A.~M.}\ \bibnamefont {Amthor}},
  \bibinfo {author} {\bibfnamefont {D.}~\bibnamefont {Bazin}}, \bibinfo
  {author} {\bibfnamefont {B.~A.}\ \bibnamefont {Brown}}, \bibinfo {author}
  {\bibfnamefont {C.~M.~F.}\ \bibnamefont {III}}, \bibinfo {author}
  {\bibfnamefont {A.}~\bibnamefont {Gade}}, \bibinfo {author} {\bibfnamefont
  {T.~N.}\ \bibnamefont {Ginter}}, \bibinfo {author} {\bibfnamefont
  {M.}~\bibnamefont {Hausmann}}, \bibinfo {author} {\bibfnamefont
  {M.}~\bibnamefont {Matos}}, \bibinfo {author} {\bibfnamefont {D.~J.}\
  \bibnamefont {Morrissey}}, \bibinfo {author} {\bibfnamefont {M.}~\bibnamefont
  {Portillo}}, \bibinfo {author} {\bibfnamefont {A.}~\bibnamefont {Schiller}},
  \bibinfo {author} {\bibfnamefont {B.~M.}\ \bibnamefont {Sherrill}}, \bibinfo
  {author} {\bibfnamefont {A.}~\bibnamefont {Stolz}}, \bibinfo {author}
  {\bibfnamefont {O.~B.}\ \bibnamefont {Tarasov}}, \ and\ \bibinfo {author}
  {\bibfnamefont {M.}~\bibnamefont {Thoennessen}},\ }\href {\doibase
  10.1038/nature06213} {\bibfield  {journal} {\bibinfo  {journal} {Nature}\
  }\textbf {\bibinfo {volume} {449}},\ \bibinfo {pages} {1022} (\bibinfo {year}
  {2007})}\BibitemShut {NoStop}%
\bibitem [{\citenamefont {M\"oller}\ \emph {et~al.}(1995)\citenamefont
  {M\"oller}, \citenamefont {Nix}, \citenamefont {Myers},\ and\ \citenamefont
  {Swiatecki}}]{Moller1995_ADNDT59-185}%
  \BibitemOpen
  \bibfield  {author} {\bibinfo {author} {\bibfnamefont {P.}~\bibnamefont
  {M\"oller}}, \bibinfo {author} {\bibfnamefont {J.~R.}\ \bibnamefont {Nix}},
  \bibinfo {author} {\bibfnamefont {W.~D.}\ \bibnamefont {Myers}}, \ and\
  \bibinfo {author} {\bibfnamefont {W.~J.}\ \bibnamefont {Swiatecki}},\ }\href
  {\doibase 10.1006/adnd.1995.1002} {\bibfield  {journal} {\bibinfo  {journal}
  {At. Data Nucl. Data Tables}\ }\textbf {\bibinfo {volume} {59}},\ \bibinfo
  {pages} {185} (\bibinfo {year} {1995})}\BibitemShut {NoStop}%
\bibitem [{\citenamefont {Zhi}\ and\ \citenamefont
  {Ren}(2006)}]{Zhi2006_PLB638-166}%
  \BibitemOpen
  \bibfield  {author} {\bibinfo {author} {\bibfnamefont {Q.}~\bibnamefont
  {Zhi}}\ and\ \bibinfo {author} {\bibfnamefont {Z.}~\bibnamefont {Ren}},\
  }\href {\doibase 10.1016/j.physletb.2006.05.057} {\bibfield  {journal}
  {\bibinfo  {journal} {Phys. Lett. B}\ }\textbf {\bibinfo {volume} {638}},\
  \bibinfo {pages} {166} (\bibinfo {year} {2006})}\BibitemShut {NoStop}%
\bibitem [{\citenamefont {Ren}\ \emph {et~al.}(1996)\citenamefont {Ren},
  \citenamefont {Zhu}, \citenamefont {Cai},\ and\ \citenamefont
  {Xu}}]{Ren1996_PLB380-241}%
  \BibitemOpen
  \bibfield  {author} {\bibinfo {author} {\bibfnamefont {Z.}~\bibnamefont
  {Ren}}, \bibinfo {author} {\bibfnamefont {Z.~Y.}\ \bibnamefont {Zhu}},
  \bibinfo {author} {\bibfnamefont {Y.~H.}\ \bibnamefont {Cai}}, \ and\
  \bibinfo {author} {\bibfnamefont {G.}~\bibnamefont {Xu}},\ }\href {\doibase
  10.1016/0370-2693(96)00462-5} {\bibfield  {journal} {\bibinfo  {journal}
  {Phys. Lett. B}\ }\textbf {\bibinfo {volume} {380}},\ \bibinfo {pages} {241}
  (\bibinfo {year} {1996})}\BibitemShut {NoStop}%
\bibitem [{\citenamefont {Goriely}\ \emph {et~al.}(2010)\citenamefont
  {Goriely}, \citenamefont {Chamel},\ and\ \citenamefont
  {Pearson}}]{Goriely2010_PRC82-035804}%
  \BibitemOpen
  \bibfield  {author} {\bibinfo {author} {\bibfnamefont {S.}~\bibnamefont
  {Goriely}}, \bibinfo {author} {\bibfnamefont {N.}~\bibnamefont {Chamel}}, \
  and\ \bibinfo {author} {\bibfnamefont {J.~M.}\ \bibnamefont {Pearson}},\
  }\href {\doibase 10.1103/PhysRevC.82.035804} {\bibfield  {journal} {\bibinfo
  {journal} {Phys. Rev. C}\ }\textbf {\bibinfo {volume} {82}},\ \bibinfo
  {pages} {035804} (\bibinfo {year} {2010})},\ \bibinfo {note} {and
  \href{http://www.astro.ulb.ac.be/pmwiki/Brusslib/Hfb17}{the HFB21 mass
  table}}\BibitemShut {NoStop}%
\bibitem [{\citenamefont {Raman}\ \emph {et~al.}(2001)\citenamefont {Raman},
  \citenamefont {Nestor},\ and\ \citenamefont
  {Tikkanen}}]{RAMAN2001_ADNDT78-1}%
  \BibitemOpen
  \bibfield  {author} {\bibinfo {author} {\bibfnamefont {S.}~\bibnamefont
  {Raman}}, \bibinfo {author} {\bibfnamefont {C.~W.}\ \bibnamefont {Nestor}}, \
  and\ \bibinfo {author} {\bibfnamefont {P.}~\bibnamefont {Tikkanen}},\ }\href
  {\doibase http://dx.doi.org/10.1006/adnd.2001.0858} {\bibfield  {journal}
  {\bibinfo  {journal} {At. Data Nucl. Data Tables}\ }\textbf {\bibinfo
  {volume} {78}},\ \bibinfo {pages} {1} (\bibinfo {year} {2001})}\BibitemShut
  {NoStop}%
\bibitem [{\citenamefont {Patra}\ and\ \citenamefont
  {Praharaj}(1991)}]{Patra1991_PLB273-13}%
  \BibitemOpen
  \bibfield  {author} {\bibinfo {author} {\bibfnamefont {S.~K.}\ \bibnamefont
  {Patra}}\ and\ \bibinfo {author} {\bibfnamefont {C.~R.}\ \bibnamefont
  {Praharaj}},\ }\href {\doibase 10.1016/0370-2693(91)90545-2} {\bibfield
  {journal} {\bibinfo  {journal} {Phys. Lett. B}\ }\textbf {\bibinfo {volume}
  {273}},\ \bibinfo {pages} {13} (\bibinfo {year} {1991})}\BibitemShut
  {NoStop}%
\bibitem [{\citenamefont {Lalazissis}\ \emph
  {et~al.}(1999{\natexlab{c}})\citenamefont {Lalazissis}, \citenamefont
  {Raman},\ and\ \citenamefont {Ring}}]{Lalazissis1999_ADNDT71-1}%
  \BibitemOpen
  \bibfield  {author} {\bibinfo {author} {\bibfnamefont {G.}~\bibnamefont
  {Lalazissis}}, \bibinfo {author} {\bibfnamefont {S.}~\bibnamefont {Raman}}, \
  and\ \bibinfo {author} {\bibfnamefont {P.}~\bibnamefont {Ring}},\ }\href
  {\doibase 10.1006/adnd.1998.0795} {\bibfield  {journal} {\bibinfo  {journal}
  {At. Data Nucl. Data Tables}\ }\textbf {\bibinfo {volume} {71}},\ \bibinfo
  {pages} {1} (\bibinfo {year} {1999}{\natexlab{c}})}\BibitemShut {NoStop}%
\bibitem [{\citenamefont {Rodriguez-Guzman}\ \emph {et~al.}(2000)\citenamefont
  {Rodriguez-Guzman}, \citenamefont {Egido},\ and\ \citenamefont
  {Robledo}}]{Rodriguez-Guzman2000_PRC62-054319}%
  \BibitemOpen
  \bibfield  {author} {\bibinfo {author} {\bibfnamefont {R.~R.}\ \bibnamefont
  {Rodriguez-Guzman}}, \bibinfo {author} {\bibfnamefont {J.~L.}\ \bibnamefont
  {Egido}}, \ and\ \bibinfo {author} {\bibfnamefont {L.~M.}\ \bibnamefont
  {Robledo}},\ }\href {\doibase 10.1103/PhysRevC.62.054319} {\bibfield
  {journal} {\bibinfo  {journal} {Phys. Rev. C}\ }\textbf {\bibinfo {volume}
  {62}},\ \bibinfo {pages} {054319} (\bibinfo {year} {2000})}\BibitemShut
  {NoStop}%
\bibitem [{\citenamefont {Rodriguez-Guzman}\ \emph {et~al.}(2002)\citenamefont
  {Rodriguez-Guzman}, \citenamefont {Egido},\ and\ \citenamefont
  {Robledo}}]{Rodriguez-Guzman2002_NPA709-201}%
  \BibitemOpen
  \bibfield  {author} {\bibinfo {author} {\bibfnamefont {R.}~\bibnamefont
  {Rodriguez-Guzman}}, \bibinfo {author} {\bibfnamefont {J.}~\bibnamefont
  {Egido}}, \ and\ \bibinfo {author} {\bibfnamefont {L.~M.}\ \bibnamefont
  {Robledo}},\ }\href {\doibase 10.1016/S0375-9474(02)01019-9} {\bibfield
  {journal} {\bibinfo  {journal} {Nucl. Phys. A}\ }\textbf {\bibinfo {volume}
  {709}},\ \bibinfo {pages} {201} (\bibinfo {year} {2002})}\BibitemShut
  {NoStop}%
\bibitem [{\citenamefont {Niksic}\ \emph {et~al.}(2006)\citenamefont {Niksic},
  \citenamefont {Vretenar},\ and\ \citenamefont
  {Ring}}]{Niksic2006_PRC73-034308}%
  \BibitemOpen
  \bibfield  {author} {\bibinfo {author} {\bibfnamefont {T.}~\bibnamefont
  {Niksic}}, \bibinfo {author} {\bibfnamefont {D.}~\bibnamefont {Vretenar}}, \
  and\ \bibinfo {author} {\bibfnamefont {P.}~\bibnamefont {Ring}},\ }\href
  {\doibase 10.1103/PhysRevC.73.034308} {\bibfield  {journal} {\bibinfo
  {journal} {Phys. Rev. C}\ }\textbf {\bibinfo {volume} {73}},\ \bibinfo
  {pages} {034308} (\bibinfo {year} {2006})}\BibitemShut {NoStop}%
\bibitem [{\citenamefont {Yao}\ \emph {et~al.}(2010)\citenamefont {Yao},
  \citenamefont {Meng}, \citenamefont {Ring},\ and\ \citenamefont
  {Vretenar}}]{Yao2010_PRC81-044311}%
  \BibitemOpen
  \bibfield  {author} {\bibinfo {author} {\bibfnamefont {J.~M.}\ \bibnamefont
  {Yao}}, \bibinfo {author} {\bibfnamefont {J.}~\bibnamefont {Meng}}, \bibinfo
  {author} {\bibfnamefont {P.}~\bibnamefont {Ring}}, \ and\ \bibinfo {author}
  {\bibfnamefont {D.}~\bibnamefont {Vretenar}},\ }\href {\doibase
  10.1103/PhysRevC.81.044311} {\bibfield  {journal} {\bibinfo  {journal} {Phys.
  Rev. C}\ }\textbf {\bibinfo {volume} {81}},\ \bibinfo {pages} {044311}
  (\bibinfo {year} {2010})}\BibitemShut {NoStop}%
\bibitem [{\citenamefont {Yao}\ \emph {et~al.}(2011)\citenamefont {Yao},
  \citenamefont {Mei}, \citenamefont {Chen}, \citenamefont {Meng},
  \citenamefont {Ring},\ and\ \citenamefont {Vretenar}}]{Yao2011_PRC83-014308}%
  \BibitemOpen
  \bibfield  {author} {\bibinfo {author} {\bibfnamefont {J.~M.}\ \bibnamefont
  {Yao}}, \bibinfo {author} {\bibfnamefont {H.}~\bibnamefont {Mei}}, \bibinfo
  {author} {\bibfnamefont {H.}~\bibnamefont {Chen}}, \bibinfo {author}
  {\bibfnamefont {J.}~\bibnamefont {Meng}}, \bibinfo {author} {\bibfnamefont
  {P.}~\bibnamefont {Ring}}, \ and\ \bibinfo {author} {\bibfnamefont
  {D.}~\bibnamefont {Vretenar}},\ }\href {\doibase 10.1103/PhysRevC.83.014308}
  {\bibfield  {journal} {\bibinfo  {journal} {Phys. Rev. C}\ }\textbf {\bibinfo
  {volume} {83}},\ \bibinfo {pages} {014308} (\bibinfo {year}
  {2011})}\BibitemShut {NoStop}%
\bibitem [{\citenamefont {Sorlin}\ and\ \citenamefont
  {Porquet}(2008)}]{Sorlin2008_PPNP61-602}%
  \BibitemOpen
  \bibfield  {author} {\bibinfo {author} {\bibfnamefont {O.}~\bibnamefont
  {Sorlin}}\ and\ \bibinfo {author} {\bibfnamefont {M.-G.}\ \bibnamefont
  {Porquet}},\ }\href {\doibase 10.1016/j.ppnp.2008.05.001} {\bibfield
  {journal} {\bibinfo  {journal} {Prog. Part. Nucl. Phys.}\ }\textbf {\bibinfo
  {volume} {61}},\ \bibinfo {pages} {602} (\bibinfo {year} {2008})}\BibitemShut
  {NoStop}%
\bibitem [{\citenamefont {Tripathi}\ \emph {et~al.}(2008)\citenamefont
  {Tripathi}, \citenamefont {Tabor}, \citenamefont {Mantica}, \citenamefont
  {Utsuno}, \citenamefont {Bender}, \citenamefont {Cook}, \citenamefont
  {Hoffman}, \citenamefont {Lee}, \citenamefont {Otsuka}, \citenamefont
  {Pereira}, \citenamefont {Perry}, \citenamefont {Pepper}, \citenamefont
  {Pinter}, \citenamefont {Stoker}, \citenamefont {Volya},\ and\ \citenamefont
  {Weisshaar}}]{Tripathi2008_PRL101-142504}%
  \BibitemOpen
  \bibfield  {author} {\bibinfo {author} {\bibfnamefont {V.}~\bibnamefont
  {Tripathi}}, \bibinfo {author} {\bibfnamefont {S.~L.}\ \bibnamefont {Tabor}},
  \bibinfo {author} {\bibfnamefont {P.~F.}\ \bibnamefont {Mantica}}, \bibinfo
  {author} {\bibfnamefont {Y.}~\bibnamefont {Utsuno}}, \bibinfo {author}
  {\bibfnamefont {P.}~\bibnamefont {Bender}}, \bibinfo {author} {\bibfnamefont
  {J.}~\bibnamefont {Cook}}, \bibinfo {author} {\bibfnamefont {C.~R.}\
  \bibnamefont {Hoffman}}, \bibinfo {author} {\bibfnamefont {S.}~\bibnamefont
  {Lee}}, \bibinfo {author} {\bibfnamefont {T.}~\bibnamefont {Otsuka}},
  \bibinfo {author} {\bibfnamefont {J.}~\bibnamefont {Pereira}}, \bibinfo
  {author} {\bibfnamefont {M.}~\bibnamefont {Perry}}, \bibinfo {author}
  {\bibfnamefont {K.}~\bibnamefont {Pepper}}, \bibinfo {author} {\bibfnamefont
  {J.~S.}\ \bibnamefont {Pinter}}, \bibinfo {author} {\bibfnamefont
  {J.}~\bibnamefont {Stoker}}, \bibinfo {author} {\bibfnamefont
  {A.}~\bibnamefont {Volya}}, \ and\ \bibinfo {author} {\bibfnamefont
  {D.}~\bibnamefont {Weisshaar}},\ }\href {\doibase
  10.1103/PhysRevLett.101.142504} {\bibfield  {journal} {\bibinfo  {journal}
  {Phys. Rev. Lett.}\ }\textbf {\bibinfo {volume} {101}},\ \bibinfo {pages}
  {142504} (\bibinfo {year} {2008})}\BibitemShut {NoStop}%
\bibitem [{\citenamefont {Doornenbal}\ \emph {et~al.}(2009)\citenamefont
  {Doornenbal}, \citenamefont {Scheit}, \citenamefont {Aoi}, \citenamefont
  {Takeuchi}, \citenamefont {Li}, \citenamefont {Takeshita}, \citenamefont
  {Wang}, \citenamefont {Baba}, \citenamefont {Deguchi}, \citenamefont
  {Fukuda}, \citenamefont {Geissel}, \citenamefont {Gernhauser}, \citenamefont
  {Gibelin}, \citenamefont {Hachiuma}, \citenamefont {Hara}, \citenamefont
  {Hinke}, \citenamefont {Inabe}, \citenamefont {Itahashi}, \citenamefont
  {Itoh}, \citenamefont {Kameda}, \citenamefont {Kanno}, \citenamefont
  {Kawada}, \citenamefont {Kobayashi}, \citenamefont {Kondo}, \citenamefont
  {Krucken}, \citenamefont {Kubo}, \citenamefont {Kuboki}, \citenamefont
  {Kusaka}, \citenamefont {Lantz}, \citenamefont {Michimasa}, \citenamefont
  {Motobayashi}, \citenamefont {Nakamura}, \citenamefont {Nakao}, \citenamefont
  {Namihira}, \citenamefont {Nishimura}, \citenamefont {Ohnishi}, \citenamefont
  {Ohtake}, \citenamefont {Orr}, \citenamefont {Otsu}, \citenamefont {Ozeki},
  \citenamefont {Satou}, \citenamefont {Shimoura}, \citenamefont {Sumikama},
  \citenamefont {Takechi}, \citenamefont {Takeda}, \citenamefont {Tanaka},
  \citenamefont {Tanaka}, \citenamefont {Togano}, \citenamefont {Winkler},
  \citenamefont {Yanagisawa}, \citenamefont {Yoneda}, \citenamefont {Yoshida},
  \citenamefont {Yoshida},\ and\ \citenamefont
  {Sakurai}}]{Doornenbal2009_PRL103-032501}%
  \BibitemOpen
  \bibfield  {author} {\bibinfo {author} {\bibfnamefont {P.}~\bibnamefont
  {Doornenbal}}, \bibinfo {author} {\bibfnamefont {H.}~\bibnamefont {Scheit}},
  \bibinfo {author} {\bibfnamefont {N.}~\bibnamefont {Aoi}}, \bibinfo {author}
  {\bibfnamefont {S.}~\bibnamefont {Takeuchi}}, \bibinfo {author}
  {\bibfnamefont {K.}~\bibnamefont {Li}}, \bibinfo {author} {\bibfnamefont
  {E.}~\bibnamefont {Takeshita}}, \bibinfo {author} {\bibfnamefont
  {H.}~\bibnamefont {Wang}}, \bibinfo {author} {\bibfnamefont {H.}~\bibnamefont
  {Baba}}, \bibinfo {author} {\bibfnamefont {S.}~\bibnamefont {Deguchi}},
  \bibinfo {author} {\bibfnamefont {N.}~\bibnamefont {Fukuda}}, \bibinfo
  {author} {\bibfnamefont {H.}~\bibnamefont {Geissel}}, \bibinfo {author}
  {\bibfnamefont {R.}~\bibnamefont {Gernhauser}}, \bibinfo {author}
  {\bibfnamefont {J.}~\bibnamefont {Gibelin}}, \bibinfo {author} {\bibfnamefont
  {I.}~\bibnamefont {Hachiuma}}, \bibinfo {author} {\bibfnamefont
  {Y.}~\bibnamefont {Hara}}, \bibinfo {author} {\bibfnamefont {C.}~\bibnamefont
  {Hinke}}, \bibinfo {author} {\bibfnamefont {N.}~\bibnamefont {Inabe}},
  \bibinfo {author} {\bibfnamefont {K.}~\bibnamefont {Itahashi}}, \bibinfo
  {author} {\bibfnamefont {S.}~\bibnamefont {Itoh}}, \bibinfo {author}
  {\bibfnamefont {D.}~\bibnamefont {Kameda}}, \bibinfo {author} {\bibfnamefont
  {S.}~\bibnamefont {Kanno}}, \bibinfo {author} {\bibfnamefont
  {Y.}~\bibnamefont {Kawada}}, \bibinfo {author} {\bibfnamefont
  {N.}~\bibnamefont {Kobayashi}}, \bibinfo {author} {\bibfnamefont
  {Y.}~\bibnamefont {Kondo}}, \bibinfo {author} {\bibfnamefont
  {R.}~\bibnamefont {Krucken}}, \bibinfo {author} {\bibfnamefont
  {T.}~\bibnamefont {Kubo}}, \bibinfo {author} {\bibfnamefont {T.}~\bibnamefont
  {Kuboki}}, \bibinfo {author} {\bibfnamefont {K.}~\bibnamefont {Kusaka}},
  \bibinfo {author} {\bibfnamefont {M.}~\bibnamefont {Lantz}}, \bibinfo
  {author} {\bibfnamefont {S.}~\bibnamefont {Michimasa}}, \bibinfo {author}
  {\bibfnamefont {T.}~\bibnamefont {Motobayashi}}, \bibinfo {author}
  {\bibfnamefont {T.}~\bibnamefont {Nakamura}}, \bibinfo {author}
  {\bibfnamefont {T.}~\bibnamefont {Nakao}}, \bibinfo {author} {\bibfnamefont
  {K.}~\bibnamefont {Namihira}}, \bibinfo {author} {\bibfnamefont
  {S.}~\bibnamefont {Nishimura}}, \bibinfo {author} {\bibfnamefont
  {T.}~\bibnamefont {Ohnishi}}, \bibinfo {author} {\bibfnamefont
  {M.}~\bibnamefont {Ohtake}}, \bibinfo {author} {\bibfnamefont {N.~A.}\
  \bibnamefont {Orr}}, \bibinfo {author} {\bibfnamefont {H.}~\bibnamefont
  {Otsu}}, \bibinfo {author} {\bibfnamefont {K.}~\bibnamefont {Ozeki}},
  \bibinfo {author} {\bibfnamefont {Y.}~\bibnamefont {Satou}}, \bibinfo
  {author} {\bibfnamefont {S.}~\bibnamefont {Shimoura}}, \bibinfo {author}
  {\bibfnamefont {T.}~\bibnamefont {Sumikama}}, \bibinfo {author}
  {\bibfnamefont {M.}~\bibnamefont {Takechi}}, \bibinfo {author} {\bibfnamefont
  {H.}~\bibnamefont {Takeda}}, \bibinfo {author} {\bibfnamefont {K.~N.}\
  \bibnamefont {Tanaka}}, \bibinfo {author} {\bibfnamefont {K.}~\bibnamefont
  {Tanaka}}, \bibinfo {author} {\bibfnamefont {Y.}~\bibnamefont {Togano}},
  \bibinfo {author} {\bibfnamefont {M.}~\bibnamefont {Winkler}}, \bibinfo
  {author} {\bibfnamefont {Y.}~\bibnamefont {Yanagisawa}}, \bibinfo {author}
  {\bibfnamefont {K.}~\bibnamefont {Yoneda}}, \bibinfo {author} {\bibfnamefont
  {A.}~\bibnamefont {Yoshida}}, \bibinfo {author} {\bibfnamefont
  {K.}~\bibnamefont {Yoshida}}, \ and\ \bibinfo {author} {\bibfnamefont
  {H.}~\bibnamefont {Sakurai}},\ }\href {\doibase
  10.1103/PhysRevLett.103.032501} {\bibfield  {journal} {\bibinfo  {journal}
  {Phys. Rev. Lett.}\ }\textbf {\bibinfo {volume} {103}},\ \bibinfo {pages}
  {032501} (\bibinfo {year} {2009})}\BibitemShut {NoStop}%
\bibitem [{\citenamefont {Wimmer}\ \emph {et~al.}(2010)\citenamefont {Wimmer},
  \citenamefont {Kroell}, \citenamefont {Kruecken}, \citenamefont {Bildstein},
  \citenamefont {Gernhaeuser}, \citenamefont {Bastin}, \citenamefont {Bree},
  \citenamefont {Diriken}, \citenamefont {Van~Duppen}, \citenamefont {Huyse},
  \citenamefont {Patronis}, \citenamefont {Vermaelen}, \citenamefont {Voulot},
  \citenamefont {Van~de Walle}, \citenamefont {Wenander}, \citenamefont
  {Fraile}, \citenamefont {Chapman}, \citenamefont {Hadinia}, \citenamefont
  {Orlandi}, \citenamefont {Smith}, \citenamefont {Lutter}, \citenamefont
  {Thirolf}, \citenamefont {Labiche}, \citenamefont {Blazhev}, \citenamefont
  {Kalkuehler}, \citenamefont {Reiter}, \citenamefont {Seidlitz}, \citenamefont
  {Warr}, \citenamefont {Macchiavelli}, \citenamefont {Jeppesen}, \citenamefont
  {Fiori}, \citenamefont {Georgiev}, \citenamefont {Schrieder}, \citenamefont
  {Das~Gupta}, \citenamefont {Lo~Bianco}, \citenamefont {Nardelli},
  \citenamefont {Butterworth}, \citenamefont {Johansen},\ and\ \citenamefont
  {Riisager}}]{Wimmer2010_PRL105-252501}%
  \BibitemOpen
  \bibfield  {author} {\bibinfo {author} {\bibfnamefont {K.}~\bibnamefont
  {Wimmer}}, \bibinfo {author} {\bibfnamefont {T.}~\bibnamefont {Kroell}},
  \bibinfo {author} {\bibfnamefont {R.}~\bibnamefont {Kruecken}}, \bibinfo
  {author} {\bibfnamefont {V.}~\bibnamefont {Bildstein}}, \bibinfo {author}
  {\bibfnamefont {R.}~\bibnamefont {Gernhaeuser}}, \bibinfo {author}
  {\bibfnamefont {B.}~\bibnamefont {Bastin}}, \bibinfo {author} {\bibfnamefont
  {N.}~\bibnamefont {Bree}}, \bibinfo {author} {\bibfnamefont {J.}~\bibnamefont
  {Diriken}}, \bibinfo {author} {\bibfnamefont {P.}~\bibnamefont {Van~Duppen}},
  \bibinfo {author} {\bibfnamefont {M.}~\bibnamefont {Huyse}}, \bibinfo
  {author} {\bibfnamefont {N.}~\bibnamefont {Patronis}}, \bibinfo {author}
  {\bibfnamefont {P.}~\bibnamefont {Vermaelen}}, \bibinfo {author}
  {\bibfnamefont {D.}~\bibnamefont {Voulot}}, \bibinfo {author} {\bibfnamefont
  {J.}~\bibnamefont {Van~de Walle}}, \bibinfo {author} {\bibfnamefont
  {F.}~\bibnamefont {Wenander}}, \bibinfo {author} {\bibfnamefont {L.~M.}\
  \bibnamefont {Fraile}}, \bibinfo {author} {\bibfnamefont {R.}~\bibnamefont
  {Chapman}}, \bibinfo {author} {\bibfnamefont {B.}~\bibnamefont {Hadinia}},
  \bibinfo {author} {\bibfnamefont {R.}~\bibnamefont {Orlandi}}, \bibinfo
  {author} {\bibfnamefont {J.~F.}\ \bibnamefont {Smith}}, \bibinfo {author}
  {\bibfnamefont {R.}~\bibnamefont {Lutter}}, \bibinfo {author} {\bibfnamefont
  {P.~G.}\ \bibnamefont {Thirolf}}, \bibinfo {author} {\bibfnamefont
  {M.}~\bibnamefont {Labiche}}, \bibinfo {author} {\bibfnamefont
  {A.}~\bibnamefont {Blazhev}}, \bibinfo {author} {\bibfnamefont
  {M.}~\bibnamefont {Kalkuehler}}, \bibinfo {author} {\bibfnamefont
  {P.}~\bibnamefont {Reiter}}, \bibinfo {author} {\bibfnamefont
  {M.}~\bibnamefont {Seidlitz}}, \bibinfo {author} {\bibfnamefont
  {N.}~\bibnamefont {Warr}}, \bibinfo {author} {\bibfnamefont {A.~O.}\
  \bibnamefont {Macchiavelli}}, \bibinfo {author} {\bibfnamefont {H.~B.}\
  \bibnamefont {Jeppesen}}, \bibinfo {author} {\bibfnamefont {E.}~\bibnamefont
  {Fiori}}, \bibinfo {author} {\bibfnamefont {G.}~\bibnamefont {Georgiev}},
  \bibinfo {author} {\bibfnamefont {G.}~\bibnamefont {Schrieder}}, \bibinfo
  {author} {\bibfnamefont {S.}~\bibnamefont {Das~Gupta}}, \bibinfo {author}
  {\bibfnamefont {G.}~\bibnamefont {Lo~Bianco}}, \bibinfo {author}
  {\bibfnamefont {S.}~\bibnamefont {Nardelli}}, \bibinfo {author}
  {\bibfnamefont {J.}~\bibnamefont {Butterworth}}, \bibinfo {author}
  {\bibfnamefont {J.}~\bibnamefont {Johansen}}, \ and\ \bibinfo {author}
  {\bibfnamefont {K.}~\bibnamefont {Riisager}},\ }\href {\doibase
  10.1103/PhysRevLett.105.252501} {\bibfield  {journal} {\bibinfo  {journal}
  {Phys. Rev. Lett.}\ }\textbf {\bibinfo {volume} {105}},\ \bibinfo {pages}
  {252501} (\bibinfo {year} {2010})}\BibitemShut {NoStop}%
\bibitem [{\citenamefont {Church}\ \emph {et~al.}(2005)\citenamefont {Church},
  \citenamefont {Campbell}, \citenamefont {Dinca}, \citenamefont {Enders},
  \citenamefont {Gade}, \citenamefont {Glasmacher}, \citenamefont {Hu},
  \citenamefont {Janssens}, \citenamefont {Mueller}, \citenamefont {Olliver},
  \citenamefont {Perry}, \citenamefont {Riley},\ and\ \citenamefont
  {Yurkewicz}}]{Church2005_PRC72-054320}%
  \BibitemOpen
  \bibfield  {author} {\bibinfo {author} {\bibfnamefont {J.~A.}\ \bibnamefont
  {Church}}, \bibinfo {author} {\bibfnamefont {C.~M.}\ \bibnamefont
  {Campbell}}, \bibinfo {author} {\bibfnamefont {D.-C.}\ \bibnamefont {Dinca}},
  \bibinfo {author} {\bibfnamefont {J.}~\bibnamefont {Enders}}, \bibinfo
  {author} {\bibfnamefont {A.}~\bibnamefont {Gade}}, \bibinfo {author}
  {\bibfnamefont {T.}~\bibnamefont {Glasmacher}}, \bibinfo {author}
  {\bibfnamefont {Z.}~\bibnamefont {Hu}}, \bibinfo {author} {\bibfnamefont
  {R.~V.~F.}\ \bibnamefont {Janssens}}, \bibinfo {author} {\bibfnamefont
  {W.~F.}\ \bibnamefont {Mueller}}, \bibinfo {author} {\bibfnamefont
  {H.}~\bibnamefont {Olliver}}, \bibinfo {author} {\bibfnamefont {B.~C.}\
  \bibnamefont {Perry}}, \bibinfo {author} {\bibfnamefont {L.~A.}\ \bibnamefont
  {Riley}}, \ and\ \bibinfo {author} {\bibfnamefont {K.~L.}\ \bibnamefont
  {Yurkewicz}},\ }\href {http://link.aps.org/abstract/PRC/v72/e054320}
  {\bibfield  {journal} {\bibinfo  {journal} {Phys. Rev. C}\ }\textbf {\bibinfo
  {volume} {72}},\ \bibinfo {pages} {054320} (\bibinfo {year}
  {2005})}\BibitemShut {NoStop}%
\bibitem [{\citenamefont {Heenen}\ \emph {et~al.}(2000)\citenamefont {Heenen},
  \citenamefont {Bonche}, \citenamefont {Cwiok}, \citenamefont {Nazarewicz},\
  and\ \citenamefont {Valor}}]{Heenen2000_RIKENRev26-31}%
  \BibitemOpen
  \bibfield  {author} {\bibinfo {author} {\bibfnamefont {P.-H.}\ \bibnamefont
  {Heenen}}, \bibinfo {author} {\bibfnamefont {P.}~\bibnamefont {Bonche}},
  \bibinfo {author} {\bibfnamefont {S.}~\bibnamefont {Cwiok}}, \bibinfo
  {author} {\bibfnamefont {W.}~\bibnamefont {Nazarewicz}}, \ and\ \bibinfo
  {author} {\bibfnamefont {A.}~\bibnamefont {Valor}},\ }\href
  {http://www.riken.go.jp/lab-www/library/publication/review/conts/conts26.html}
  {\bibfield  {journal} {\bibinfo  {journal} {RIKEN Rev.}\ }\textbf {\bibinfo
  {volume} {26}},\ \bibinfo {pages} {31} (\bibinfo {year} {2000})}\BibitemShut
  {NoStop}%
\bibitem [{\citenamefont {Suzuki}\ \emph {et~al.}(1998)\citenamefont {Suzuki},
  \citenamefont {Geissel}, \citenamefont {Bochkarev}, \citenamefont {Chulkov},
  \citenamefont {Golovkov}, \citenamefont {Fukunishi}, \citenamefont {Hirata},
  \citenamefont {Irnich}, \citenamefont {Janas}, \citenamefont {Keller},
  \citenamefont {Kobayashi}, \citenamefont {Kraus}, \citenamefont
  {M\"{u}zenberg}, \citenamefont {Neumaier}, \citenamefont {Nickel},
  \citenamefont {Ozawa}, \citenamefont {Piechaczeck}, \citenamefont {Roeckl},
  \citenamefont {Schwab}, \citenamefont {S\"{u}merer}, \citenamefont
  {Yoshida},\ and\ \citenamefont {Tanihata}}]{Suzuki1998_NPA630-661}%
  \BibitemOpen
  \bibfield  {author} {\bibinfo {author} {\bibfnamefont {T.}~\bibnamefont
  {Suzuki}}, \bibinfo {author} {\bibfnamefont {H.}~\bibnamefont {Geissel}},
  \bibinfo {author} {\bibfnamefont {O.}~\bibnamefont {Bochkarev}}, \bibinfo
  {author} {\bibfnamefont {L.}~\bibnamefont {Chulkov}}, \bibinfo {author}
  {\bibfnamefont {M.}~\bibnamefont {Golovkov}}, \bibinfo {author}
  {\bibfnamefont {N.}~\bibnamefont {Fukunishi}}, \bibinfo {author}
  {\bibfnamefont {D.}~\bibnamefont {Hirata}}, \bibinfo {author} {\bibfnamefont
  {H.}~\bibnamefont {Irnich}}, \bibinfo {author} {\bibfnamefont
  {Z.}~\bibnamefont {Janas}}, \bibinfo {author} {\bibfnamefont
  {H.}~\bibnamefont {Keller}}, \bibinfo {author} {\bibfnamefont
  {T.}~\bibnamefont {Kobayashi}}, \bibinfo {author} {\bibfnamefont
  {G.}~\bibnamefont {Kraus}}, \bibinfo {author} {\bibfnamefont
  {G.}~\bibnamefont {M\"{u}zenberg}}, \bibinfo {author} {\bibfnamefont
  {S.}~\bibnamefont {Neumaier}}, \bibinfo {author} {\bibfnamefont
  {F.}~\bibnamefont {Nickel}}, \bibinfo {author} {\bibfnamefont
  {A.}~\bibnamefont {Ozawa}}, \bibinfo {author} {\bibfnamefont
  {A.}~\bibnamefont {Piechaczeck}}, \bibinfo {author} {\bibfnamefont
  {E.}~\bibnamefont {Roeckl}}, \bibinfo {author} {\bibfnamefont
  {W.}~\bibnamefont {Schwab}}, \bibinfo {author} {\bibfnamefont
  {K.}~\bibnamefont {S\"{u}merer}}, \bibinfo {author} {\bibfnamefont
  {K.}~\bibnamefont {Yoshida}}, \ and\ \bibinfo {author} {\bibfnamefont
  {I.}~\bibnamefont {Tanihata}},\ }\href {\doibase
  10.1016/S0375-9474(98)00799-4} {\bibfield  {journal} {\bibinfo  {journal}
  {Nucl. Phys. A}\ }\textbf {\bibinfo {volume} {630}},\ \bibinfo {pages} {661}
  (\bibinfo {year} {1998})}\BibitemShut {NoStop}%
\bibitem [{\citenamefont {Lu}\ \emph {et~al.}(2011)\citenamefont {Lu},
  \citenamefont {Zhao},\ and\ \citenamefont {Zhou}}]{Lu2011_PRC84-014328}%
  \BibitemOpen
  \bibfield  {author} {\bibinfo {author} {\bibfnamefont {B.-N.}\ \bibnamefont
  {Lu}}, \bibinfo {author} {\bibfnamefont {E.-G.}\ \bibnamefont {Zhao}}, \ and\
  \bibinfo {author} {\bibfnamefont {S.-G.}\ \bibnamefont {Zhou}},\ }\href
  {\doibase 10.1103/PhysRevC.84.014328} {\bibfield  {journal} {\bibinfo
  {journal} {Phys. Rev. C}\ }\textbf {\bibinfo {volume} {84}},\ \bibinfo
  {pages} {014328} (\bibinfo {year} {2011})},\ \bibinfo {note}
  {arXiv:1104.4638v1 [nucl-th]}\BibitemShut {NoStop}%
\bibitem [{\citenamefont {Lu}\ \emph {et~al.}(2012)\citenamefont {Lu},
  \citenamefont {Zhao},\ and\ \citenamefont {Zhou}}]{Lu2012_PRC85-011301R}%
  \BibitemOpen
  \bibfield  {author} {\bibinfo {author} {\bibfnamefont {B.-N.}\ \bibnamefont
  {Lu}}, \bibinfo {author} {\bibfnamefont {E.-G.}\ \bibnamefont {Zhao}}, \ and\
  \bibinfo {author} {\bibfnamefont {S.-G.}\ \bibnamefont {Zhou}},\ }\href
  {\doibase 10.1103/PhysRevC.85.011301} {\bibfield  {journal} {\bibinfo
  {journal} {Phys. Rev. C}\ }\textbf {\bibinfo {volume} {85}},\ \bibinfo
  {pages} {011301(R)} (\bibinfo {year} {2012})},\ \bibinfo {note} {{arXiv:
  1110.6769v2 [nucl-th]}}\BibitemShut {NoStop}%
\bibitem [{\citenamefont {Meng}\ \emph {et~al.}(1997)\citenamefont {Meng},
  \citenamefont {P\"oschl},\ and\ \citenamefont {Ring}}]{Meng1997_ZPA358-123}%
  \BibitemOpen
  \bibfield  {author} {\bibinfo {author} {\bibfnamefont {J.}~\bibnamefont
  {Meng}}, \bibinfo {author} {\bibfnamefont {W.}~\bibnamefont {P\"oschl}}, \
  and\ \bibinfo {author} {\bibfnamefont {P.}~\bibnamefont {Ring}},\ }\href
  {\doibase 10.1007/s002180050285} {\bibfield  {journal} {\bibinfo  {journal}
  {Z. Phys. A}\ }\textbf {\bibinfo {volume} {358}},\ \bibinfo {pages} {123}
  (\bibinfo {year} {1997})}\BibitemShut {NoStop}%
\bibitem [{\citenamefont {Meng}\ and\ \citenamefont
  {Ring}(1998)}]{Meng1998_PRL80-460}%
  \BibitemOpen
  \bibfield  {author} {\bibinfo {author} {\bibfnamefont {J.}~\bibnamefont
  {Meng}}\ and\ \bibinfo {author} {\bibfnamefont {P.}~\bibnamefont {Ring}},\
  }\href {\doibase 10.1103/PhysRevLett.80.460} {\bibfield  {journal} {\bibinfo
  {journal} {Phys. Rev. Lett.}\ }\textbf {\bibinfo {volume} {80}},\ \bibinfo
  {pages} {460} (\bibinfo {year} {1998})}\BibitemShut {NoStop}%
\bibitem [{\citenamefont {Edmonds}(1957)}]{Edmonds1957}%
  \BibitemOpen
  \bibfield  {author} {\bibinfo {author} {\bibfnamefont {A.~R.}\ \bibnamefont
  {Edmonds}},\ }\href@noop {} {\emph {\bibinfo {title} {Angular Momentum in
  Quantum Mechanics}}}\ (\bibinfo  {publisher} {University Press, Princeton},\
  \bibinfo {year} {1957})\BibitemShut {NoStop}%
\bibitem [{\citenamefont {Peierls}\ and\ \citenamefont
  {Yoccoz}(1957)}]{Peierls1957_ProcPhysSocA70-381}%
  \BibitemOpen
  \bibfield  {author} {\bibinfo {author} {\bibfnamefont {R.~E.}\ \bibnamefont
  {Peierls}}\ and\ \bibinfo {author} {\bibfnamefont {J.}~\bibnamefont
  {Yoccoz}},\ }\href {\doibase 10.1088/0370-1298/70/5/309} {\bibfield
  {journal} {\bibinfo  {journal} {Proc. Phys. Soc. A}\ }\textbf {\bibinfo
  {volume} {70}},\ \bibinfo {pages} {381} (\bibinfo {year} {1957})}\BibitemShut
  {NoStop}%
\bibitem [{\citenamefont {Bender}\ \emph {et~al.}(2000)\citenamefont {Bender},
  \citenamefont {Rutz}, \citenamefont {Reinhard},\ and\ \citenamefont
  {Maruhn}}]{Bender2000_EPJA7-467}%
  \BibitemOpen
  \bibfield  {author} {\bibinfo {author} {\bibfnamefont {M.}~\bibnamefont
  {Bender}}, \bibinfo {author} {\bibfnamefont {K.}~\bibnamefont {Rutz}},
  \bibinfo {author} {\bibfnamefont {P.-G.}\ \bibnamefont {Reinhard}}, \ and\
  \bibinfo {author} {\bibfnamefont {J.~A.}\ \bibnamefont {Maruhn}},\ }\href
  {\doibase 10.1007/PL00013645} {\bibfield  {journal} {\bibinfo  {journal}
  {Eur. Phys. J. A}\ }\textbf {\bibinfo {volume} {7}},\ \bibinfo {pages} {467}
  (\bibinfo {year} {2000})}\BibitemShut {NoStop}%
\end{thebibliography}

%

\appendix

\section{\label{appendix:pre}Spherical spinors in coordinate space}

In this work we use three different representations of the wave functions. 
The starting point is the coordinate space representation $x=(\br{s} p)$, 
where $s$ is the spin coordinate and $p$ describes large (or upper) 
($p=1$ or $p=+$) and small (or lower) ($p=2$ or $p=-$) components.
The second basis is a discrete basis of spherical Dirac spinors 
$|n\kappa m\rangle$ which is obtained by the diagonalization of 
the spherical Dirac Hamiltonian with fields of Woods-Saxon shape. 
This basis is called the Woods-Saxon basis in the following.
In this basis the RHB equations is solved and the solutions form 
a basis of quasi-particle states labeled by $|k\rangle$. 
The Dirac spinors of the Woods-Saxon basis are represented in coordinate space as
\begin{equation}
 \langle \br{s} p|n\kappa m\rangle 
 = \phi_{n\kappa m}(\br{s} p)
 = i^p \frac{ R_{n\kappa}(r,p)}{r} Y^{l(p)}_{\kappa m}(\Omega,{s}),
\end{equation}
The orbital angular momenta of these components are 
$l(p=1)=j+\frac{1}{2}{\rm sign}(\kappa)$ and
$l(p=2)=j-\frac{1}{2}{\rm sign}(\kappa)$.
$R_{n\kappa}(r,1) = G_{n\kappa}(r)$, 
$R_{n\kappa}(r,2) = F_{n\kappa}(r)$ are the radial wave functions,
and $Y^{l}_{\kappa m}$ are the spinor spherical harmonics
\begin{eqnarray}
 Y^{l}_{\kappa m}(\Omega,{s}) 
 & = & 
 \sum_{m_l,m_s} C(\frac{1}{2}m_s lm_l|jm)Y_{lm_l}(\Omega)\chi_{\frac{1}{2}m_s}
\end{eqnarray}
The time reversal state reads
\begin{eqnarray}
 \bar\phi_{n\kappa m}(\br{s} p)
 & = &
 (-1)^{p+l(p)+j-m}
 \phi_{n\kappa-m}(\br{s} p).
\end{eqnarray}

These basis functions are obtained from the solution of a Dirac equation with 
spherical potentials of Woods-Saxon-shape~\cite{Koepf1991_ZPA339-81}
\begin{equation}
 h^{(0)}_D = \balp \cdot \bp + \beta\left[ M + S^{(0)}(r) \right] + V^{(0)}(r),
\label{eq:hD0}
\end{equation}
on a mesh in $r$-space using the Runge-Kutta method.
For each $\kappa$ we have eigenstates with positive and negative eigenvalues 
$\epsilon_{n\kappa}$ and for completeness of the basis the sum over $n\kappa$ 
has to include states with positive eigenvalues and those with negative 
eigenvalues~\cite{Zhou2003_PRC68-034323}. This has nothing to do with the no-sea approximation
which is applied in the final quasiparticle basis where the sums over $k$ 
in Eq. (\ref{eq:mesonsource}) runs only over solutions with positive single particle energies.

Since the RHB equation (\ref{eq:RHB1})
has to be solved in this basis one has to evaluate matrix elements of the form
\begin{equation}
 \langle n\kappa m |h_D   |n'\kappa' m\rangle  ~~~~~~{\rm and}~~~~~~
 \langle n\kappa m |\Delta|n'\kappa' m\rangle .
\label{eq:A4}
\end{equation}
In order to simplify the calculations, the integrations over the angles are 
carried out analytically using well known
angular momentum coupling techniques and only the radial integrals are calculated numerically.
For local potentials we need the following products of basis wave-functions
\begin{equation}
 \sum_{s} \phi^{}_{n\kappa m}(\bm{r}sp)
          \phi_{n'\kappa' m}^{\ast}(\bm{r}sp) .
\label{eq:A5}
\end{equation}
Following Eq.~(\ref{eq:expansion}) they are expanded in terms of Legendre polynomials. 
For the coefficient of rank $\lambda$ depending only on the radius $r$ we find
\begin{equation}
 \left[ \sum_{s} \phi^{}_{n\kappa m} \phi_{n'\kappa' m}^{\ast} \right]_\lambda
 =
 \frac{R_{n\kappa}(r,p)}{r} \frac{R_{n^{\prime}\kappa^{\prime}}(r,p)}{r}
 \langle \kappa m | P_{\lambda} | \kappa^{\prime} m \rangle.
\label{eq:A6}
\end{equation}
The angular matrix elements $\langle\kappa m|P_{\lambda}^{{}}|\kappa^{\prime}m\rangle$
can be derived with the help of the Wigner-Eckart theorem~\cite{Edmonds1957}.
For even values of $l+\lambda+l'$ we find
\begin{equation}
 \langle \kappa m | P_{\lambda}^{{}} | \kappa^{\prime} m \rangle
 = (-)^{m-\frac{1}{2}} \hat{\jmath}\hat{\jmath}'
 \left( {j \atop -m } {\lambda \atop 0 }\, {j' \atop m } \right)
 \left( {j \atop -\frac12 } {\lambda \atop 0 }\, {j' \atop \frac12 } \right) ,
\end{equation}
where $\hat{\jmath}=\sqrt{2j+1}$. For odd values of $l+\lambda+l'$
these matrix elements vanish.

\section{\label{appendix:matrix}Matrix elements of the DRHB Hamiltonian}

The Dirac Hartree-Bogoliubov equations~\cite{Kucharek1991_ZPA339-23}
read in coordinate space
\begin{widetext}
\begin{eqnarray}
 \sum_{{s}'p'} \int d^3 \br'
 \left(
  \begin{array}{cc}
   h_D(\br{s} p,\br'{s}'p') - \lambda &
   \Delta(\br{s} p,\br'{s}' p') \\
  -\Delta^*(\br{s} p,\br'{s}' p')
   & -h_D(\br{s} p,\br'{s}'p') + \lambda \\
  \end{array}
 \right)
 \left(
  { U_{k}(\br'{s}' p') \atop V_{k}(\br'{s}' p') }
 \right)
 & = &
 E_{k}
  \left(
   { U_{k}(\br{s} p) \atop V_{k}(\br{s} p) }
  \right) ,
\end{eqnarray}
\end{widetext}
where $E_{k}$ is the quasiparticle energy and $\lambda$ the chemical potential. On the
Hartree level the Dirac Hamiltonian is local
\begin{equation}
h_D(\br{s} p,\br'{s}'p') = h_D(\br,{s} p,{s}'p') \delta(\br-\br') .
\end{equation}
For the zero range pairing force in Eq.~(\ref{eq:pairing_force})
which projects onto the $S=0$ part of the pairing density, the pairing field 
is local too and does not depend on the spin variables
\begin{equation}
 \Delta(\br p,\br'p') = \delta_{pp'}\Delta(\br p) \delta(\br-\br') .
\end{equation}
In this work we restrict ourselves on pairing fields diagonal in the quantum
number $p$ (see Appendix \ref{appendix:reldel}).
These equations of motions are
solved by expanding the spinors $U_k$ and $V_k$ in terms of a Woods-Saxon basis
of Dirac spinors $\varphi_{n\kappa m}(\bm{r}{s})$ in Eq.~(\ref{eq:SRHspinor})
with positive and negative single particle energies $\epsilon_{n\kappa}$.

For the self-consistent solution of the Dirac equation (\ref{eq:Dirac0}) 
with deformed potentials of axial symmetry, we expand the potentials 
$S(\br)$ and $V(\br)$ in terms of the Legendre polynomials as in Eq. (\ref{eq:expansion}).
The deformed Dirac Hamiltonian $h_D$ is divided into two parts, 
the spherical Woods-Saxon Hamiltonian $h^{(0)}_D$ of Eq.~(\ref{eq:hD0})
and the deformed rest
\begin{eqnarray}
 h_D & = & h^{(0)}_D + \sum_\lambda \left[ \beta S'_\lambda(r) + V'_\lambda(r) \right]
                                    P_\lambda(\Omega) ,
\end{eqnarray}
with $S'_0=S^{}_0-S^{(0)}$, $V'_0=V^{}_0-V^{(0)}$, and  
$S'_\lambda=S_\lambda$, and $V'_\lambda=V_\lambda$ for $\lambda>0$.
Using Eq. (\ref{eq:A6}) the matrix elements of the Dirac Hamiltonian read,
\begin{widetext}
\begin{eqnarray}
 \langle n\kappa| h_D |n'\kappa' \rangle  
 & = &
 \epsilon_{n\kappa} \delta_{nn'}\delta_{\kappa\kappa'} 
 + \sum_\lambda \langle\kappa m|P_{\lambda}|\kappa^{\prime}m\rangle%
    \int dr \left[ G_{n\kappa}(r) (V'_\lambda(r) + S'_\lambda(r))G_{n'\kappa'}(r)
            \right.
 \nonumber \\ 
 &   & \mbox{\hspace{60mm}} 
            \left.
                 + F_{n\kappa}(r)  (V'_\lambda(r) - S'_\lambda(r))F_{n'\kappa'}(r)\right]
 \ .
\end{eqnarray}
\end{widetext}

The integral in the pairing matrix element $\langle n\kappa m |\Delta|n'\kappa' m\rangle$
contains the time reversal basis function. Since the pairing interaction Eq.~(\ref{eq:pairing_force})
projects onto the $S=0$ we have to couple the product 
$\phi^{}_{n\kappa m}(\bm{r}s) \bar{\phi}_{n'\kappa' m}(\bm{r}s)$ to spin $S=0$ and find
\begin{equation}
 \sum_{s} (-)^{\frac{1}{2}-s} \phi^{}_{n\kappa m}(s) \bar{\phi}_{n'\kappa' m}(-s)
 =
 \sum_{s} \phi^{}_{n\kappa m}(s) \phi_{n'\kappa' m}^\ast(s).
\label{eq:A10}
\end{equation}
Using again Eq.~(\ref{eq:A6}) one finds
\begin{equation}
 \langle n\kappa|\Delta^{++}|n'\kappa' \rangle  
 = 
 \sum_\lambda \langle \kappa m | P_{\lambda} | \kappa^{\prime} m \rangle%
 \int dr G_{n\kappa} \Delta_\lambda(r) G_{n'\kappa'}
\end{equation}
and
\begin{equation}
 \langle n \kappa | \Delta^{--} | n' \kappa' \rangle  
 = 
 \sum_\lambda \langle \kappa m | P_{\lambda} | \kappa^{\prime} m \rangle%
 \int dr F_{n\kappa} \Delta_\lambda(r) F_{n'\kappa'}
\end{equation}
where the potentials $\Delta_\lambda(r)$ will be given in Appendix~\ref{appendix:pair}.

\section{\label{appendix:den}Calculation of the densities}

In order to determine the self-consistent fields in the next step
of the iteration we first have to determine the densities. Starting
from the expansion coefficients $u^{(m)}_{k,(n\kappa)}$ and
$v^{(m)}_{k,(n\kappa)}$ obtained through the
diagonalization of the RHB matrix (\ref{eq:RHB1})
we find the density matrix in the Woods-Saxon basis
\begin{equation}
\rho^{(m)}_{(n\kappa)(n'\kappa')}
  = \sum_{k>0}v^{(m)*}_{k,(n\kappa)} v^{(m)}_{k,(n'\kappa')}.
\label{eq:rhows}
\end{equation}
Next we transform these densities to coordinate space and find for the
local part
\begin{eqnarray}
 \rho(\bm{r}p) 
 & = &
 2 \sum_{m>0} \sum_{n\kappa}^{n'\kappa'} \sum_{s} 
   \phi^{}_{n\kappa{m}}(\bm{r}sp) 
   \rho^{(m)}_{(n\kappa),(n'\kappa')}
   \phi_{n'\kappa'{m}}^{\ast}(\bm{r}sp)
\nonumber\\
 & = &
 \sum_{\lambda} \rho_{\lambda}(r,p) P_{\lambda}(\Omega).
\label{eq:density}
\end{eqnarray}
Using Eq.~(\ref{eq:A6}) we finally obtain  the various local densities
\begin{widetext}
\begin{eqnarray}
 \rho^s_{\lambda}(r)
 & = & 
 2 \frac{2\lambda+1} {4\pi r^2} 
 \sum_{m>0} \sum_{n\kappa}^{n'\kappa'}
  \rho^{(m)}_{(n\kappa)(n'\kappa')} \left[ G_{n\kappa}(r) G_{n'\kappa'}(r)
                                         - F_{n\kappa}(r) F_{n'\kappa'}(r) \right]
  \langle \kappa m | P_{\lambda} | \kappa^{\prime} m \rangle,
\\
 \rho^v_{\lambda}(r)
 & = &
 2 \frac{2\lambda+1} {4\pi r^2}
 \sum_{m>0} \sum_{n\kappa}^{n'\kappa'}
  \rho^{(m)}_{(n\kappa)(n'\kappa')} \left[ G_{n\kappa}(r) G_{n'\kappa'}(r)
                                         + F_{n\kappa}(r) F_{n'\kappa'}(r) \right]
  \langle \kappa m | P_{\lambda} | \kappa^{\prime} m \rangle
  ,
\end{eqnarray}
\end{widetext}
and similar equations for the isovector density $\rho^3_\lambda(r)$ and for the charge
density $\rho^c_\lambda(r)$.

\section{\label{appendix:KG}Solution of the Klein-Gordon equation}

The various densities are the sources of the meson fields in the
Klein Gordon equations (\ref{eq:mesonmotion}).
These equations are solved by integrating the densities
over the static Green functions in spherical coordinates.
For simplicity we give here the details only for the $\sigma$ meson
\begin{eqnarray}
 D(r,\theta,r',\theta',m_\sigma)
 & = &
 - m_\sigma \sum_{\lambda}
   j_\lambda(i m_\sigma r_<) h^{(1)}_\lambda(i m_\sigma r_>)
 \nonumber \\ 
 &   &
 \times (2\lambda+1) P_\lambda(\cos\theta) P_\lambda(\cos\theta'),
\end{eqnarray}
and the photon
\begin{eqnarray}
 D(r,\theta,r',\theta')  
 & = &
 \sum_{\lambda}
  \frac{r^\lambda_<}{r^{\lambda+1}_>} P_\lambda(\cos\theta) P_\lambda(\cos\theta').
\end{eqnarray}
Here $r_>={\rm max}(r,r')$ and $r_<={\rm min}(r,r')$.
The solution for the $\sigma$ field is
\begin{equation}
 \sigma(\br)= \sum_\lambda \sigma_{\lambda}(r) P_\lambda(\cos\theta) ,
\end{equation}
with
\begin{eqnarray}
 \sigma_{\lambda}(r)
 & = &
  -4\pi g_{\sigma} m_\sigma
  \left( h_\lambda(i m_\sigma r)
  \int_0^r      dr' j_\lambda(i m_\sigma r') \rho^s_{\lambda}(r')
  \right. \nonumber \\ &&
  + \left. j_\lambda(i m_\sigma r)
  \int_r^\infty dr' h_\lambda(i m_\sigma r') \rho^s_{\lambda}(r')
  \right) ,
\end{eqnarray}
where $j_\lambda$ and $h_\lambda$ are the spherical Bessel and Hankel functions. 
Similarly we find for the Coulomb field
\begin{eqnarray}
 A^0_\lambda(r)
 & = &
 \frac{1}{r^{\lambda+1}} \int_0^r dr' r'^\lambda \rho^{c}_{\lambda}(r')
 +r^\lambda \int_r^\infty dr' \frac{1}{r'^{\lambda+1}} \rho^{c}_{\lambda}(r')
 . \nonumber \\
\end{eqnarray}
From the $\lambda$ components of the meson fields $\sigma_\lambda(r)$, 
$\omega^0_\lambda(r)$ $\rho^0_\lambda(r)$ and $A^0_\lambda(r)$
we find immediately the corresponding components of the scalar and 
the vector potential given in Eqs. (\ref{eq:vaspot}) and (\ref{eq:vavpot}).

\section{\label{appendix:pair}Pairing fields and tensors}

As in the case of the normal density we first calculate the pairing tensor 
$\kappa$ in the Woods Saxon basis
\begin{equation}
\kappa^{(m)}_{(n\kappa)(n'\kappa')}
  = \sum_{k>0}v^{(m)*}_{k,(n\kappa)} u^{(m)}_{k,(n'\kappa')}.
\label{eq.kappaws}
\end{equation}
Next we transform it to coordinate space and obtain $\kappa(\br sp,\br' s'p')$.
This is a 2$\times$2 matrix in spin space and therefore it can be expressed 
in terms of the unity and the Pauli matrices
\begin{equation}
 \kappa(\bm{r}sp, \br's'p') = 
 \kappa(\br p,\br'p') + \bm{\kappa}(\br p, \br'p') \cdot \bm{\sigma},
\end{equation}
where $\kappa(\br p,\br'p')$ is the $S=0$ part and $\bm{\kappa}(\br p,\br'p')$
is a vector, the $S=1$ part of the pairing tensor. We realize that the special form of the
pairing interaction in Eq.~(\ref{eq:pairing_force}) guarantees that we do not need the full matrix
$\kappa(\br sp,\br' s'p')$. As a consequence of the zero range we need only the local
part of this matrix, and since the force acts in the $S=0$ channel, only the
spin scalar part of $\kappa$ contributes. It is obtained by coupling
to $S=0$:
\begin{equation}
 \kappa(\bm{r},p,p') = \sum_{s}(-)^{s+\frac12} \kappa(\br sp,\br{\rm-}sp')
 .
\end{equation}
As mentioned above, in this work we take into account only pairing fields which are diagonal in
the quantum number $p$.

Because of the symplectic structure of the RHB equations
the pairing tensor $\kappa$ connects basis states $|n\kappa m\rangle$ with the
time reversal states $|\overline{n'\kappa' m}\rangle$. Using the same arguments 
as in Eq. (\ref{eq:A10}) we obtain
for the local and scalar part of the pairing density
\begin{eqnarray}
 \kappa(\bm{r}p)
 & = &
 2 \sum_{m>0} \sum_{n\kappa}^{n'\kappa'} \sum_{s}
   \phi_{k^{}}(\bm{r}sp)
   \kappa^{(m)}_{(n\kappa),(n'\kappa')}
   \phi_{k^{\prime}}^{\ast}(\bm{r}sp)
 \nonumber\\
 & = &
 \sum_{\lambda} \kappa_{\lambda}(r,p) P_{\lambda}(\Omega).
\end{eqnarray}
Finally with the help of Eq.~(\ref{eq:A6}) we obtain the pairing densities 
in various $\lambda$-channels
\begin{eqnarray}
 \kappa^{++}_{\lambda}(r)
 & = & 
 2 \frac{2\lambda+1} {4\pi r^2}
 \sum_{m>0} \sum_{n\kappa}^{n'\kappa'}
  G_{n\kappa} \kappa^{(m)}_{(n\kappa),(n'\kappa')} G_{n'\kappa'}  \nonumber\\
 &   &~~~~~~~~~~~~~~~~~~~~~~~~~~~
 \times \langle \kappa m|P_{\lambda}|\kappa^{\prime}m \rangle,
\label{eq:E5} \\
 \kappa^{--}_{\lambda}(r)
 & = & 
 2 \frac{2\lambda+1} {4\pi r^2}
 \sum_{m>0} \sum_{n\kappa}^{n'\kappa'}
  F_{n\kappa} \kappa^{(m)}_{(n\kappa),(n'\kappa')} F_{n'\kappa'}  \nonumber\\
 &   &~~~~~~~~~~~~~~~~~~~~~~~~~~~
 \times \langle \kappa m|P_{\lambda}|\kappa^{\prime}m \rangle.
\label{eq:E6}
\end{eqnarray}

As a consequence of these simplifications the gap equation~(\ref{eq:gap12}) has the local form
\begin{equation}
 \Delta(\br,p) = V_0 f(\br)\kappa(\br,p)
\end{equation}
with $f(\br)=(1-\rho(\br)/\rho_{\rm sat})$. The decomposition of this equation 
into spherical harmonics yields
\begin{equation}
 \Delta_\lambda(r) = (2\lambda+1) V_0 \sum_{\lambda',\lambda''} 
                     f_{\lambda'}(r)\kappa_{\lambda''}(r)
                     \left({\lambda \atop 0}{\lambda' \atop 0}{\lambda'' \atop 0}\right)^2
 \ ,
\label{eq:gap}
\end{equation}
and $f_\lambda(r)=(\delta_{\lambda 0}-\rho(r)_\lambda/\rho_{\rm sat})$.

\section{\label{appendix:reldel}Relativistic structure of the pairing field}

So far we have neglected parts of the pairing field which connect large
and small components, i.e., we have assumed that
\begin{equation}
 \Delta^{+-}(\br) = \kappa^{+-}(\br) = 0 .
\end{equation}
Since the density $\rho(\br)$ and the density-dependent function $f(\rho(\br))$ 
of the pairing interaction does not mix these components, the structure of 
Eq. (\ref{eq:gap}) shows also very clearly that the pairing tensor $\kappa(\br)$ 
and the pairing field $\Delta(\br)$ are in this respect completely connected. 
If $\kappa(\br)$ mixes these components, so does $\Delta$.

Considering the structure of Eqs. (\ref{eq:E5}) and (\ref{eq:E6}) 
we find that a non-vanishing term $\kappa^{+-}(\br)$ would have the form
\begin{eqnarray}
 \kappa^{+-}_{\lambda}(r)
 & = & 
 2 \frac{2\lambda+1} {4\pi r^2}
 \sum_{m>0} \sum_{n\kappa}^{n'\kappa'}
  G_{n\kappa} \kappa^{(m)}_{(n\kappa),(n'\kappa')} F_{n'\kappa'}  \nonumber  \\
 &   &~~~~~~~~~~~~~~~~~~~~~~~~~~~
 \times \langle \kappa m|P_{\lambda}|\kappa^{\prime}m \rangle .
\end{eqnarray}
Since large and small components have different parity, non-vanishing values of this function
would require odd values of $l+l'$ and because of the parity selection rule in
$\langle\kappa m|P_{\lambda}|\kappa^{\prime}m\rangle$ also odd values of $\lambda$. This means
the parts of $\kappa^{+-}(\br)$ and $\Delta^{+-}(\br)$ can only be expanded in 
components with odd $\lambda$ values.
Of course, this fact is rather trivial. It does not violate parity, 
because even the simple Dirac equation with
parity conserving fields have large and small components with different parity: 
$h_D^{+-}=\bm{\sigma}\cdot\bm{p}$
has also a negative parity.

We can conclude that in the spherical case, where $\lambda=0$ and even, 
the field  $\Delta^{+-}(\br)$ has to vanish. 
In the deformed case this is not necessarily true. On the other side, these considerations
depend on the interaction, as for instance on the fact that the pairing force we have used here
excludes $S=1$. In particular we did not take into account odd $\lambda$-values in the pairing
field and therefore $\Delta^{+-}$ fields are excluded from the beginning. If we would allow
for $S=1$ pairs, spin-vector components of the form $\bm{\Delta}\cdot\bm{\alpha}$ mixing large
and small components are not excluded, even in the spherical case, because in this case $L=1$ and
$S=1$ can couple to $J=0$. Of course this depends on the interaction. 
In Ref.~\cite{Serra2002_PRC65-064324} the $S=1$ part of the zero range pairing force 
was not excluded and non-vanishing pairing fields $\Delta^{+-}$ were taken into account. 
However, they turned out to be an order of magnitude smaller than the diagonal
matrix elements $\Delta^{++}$. In particular they are very small as compared to 
the term $\bm{\sigma}\cdot\bm{p}$ which mixes large and small components in 
the Dirac Hamiltonian. Therefore they can be neglected as a very good approximation.

\section{\label{appendix:com}Microscopic center of mass correction}

The center of mass correction in Eq. (\ref{Ecm})
which is widely used in the literature~\cite{Ring1980} can be derived as
a first order correction to a projection after variation~\cite{Peierls1957_ProcPhysSocA70-381}
onto good linear momentum. In Ref.~\cite{Bender2000_EPJA7-467} this term has been derived
in the framework of the BCS approximation as
\begin{eqnarray}
\frac{\langle \hat{\mathbf{P}}^2 \rangle}{2Am}
   &=&
   - \frac{\hbar^2}{Am} \left[ \sum_{{i} > 0} v_{i}^2   \Delta_{{i}{i}}~+\right. \\
   &+&\left.
   \sum_{{i}, {i}'>0} \left[
   v_{i} v_{{i}'} (v_{i} v_{{i}'} + u_{i} u_{{i}'})
   \left( | \nabla_{{i} {i}'} |^2 + | \nabla_{ {i} \bar{{i}'} } |^2  \right)
   \right] \right].
   \nonumber
\label{eq:cm-corr}
\end{eqnarray}
In the following we show, how this formula can be applied in the framework of
relativistic Hartree-Bogoliubov theory. In a first step we use the fact that
any Hartree-Bogoliubov wavefunction can be expressed in the form of a BCS-state
in the canonical basis~\cite{Ring1980}. This basis is obtained by the diagonalization
of the density matrix in $\rho=V^*V^\intercal$ in the Woods-Saxon basis~(\ref{eq:SRHspinor})
\begin{equation}
 \sum_{n'\kappa'} \rho^{m}_{n\kappa,n'\kappa'} c^{i}_{n'\kappa'} = v^2_{i} c^{i}_{n\kappa}
 .
\label{eq:rodiag}
\end{equation}
The eigenvalues $v^2_{i}$ are the BCS-occupation probabilities and the eigenvectors are
the expansion coefficients of the canonical wave functions in the spherical spinors
of the Woods-Saxon basis
\begin{eqnarray}
 \Phi_{i}  ( {\br} {s})
 = \sum_{n\kappa} c^{i}_{n\kappa} \varphi_{n\kappa m} ( \br {s} )
 .
\end{eqnarray}
Here, ${i} = (n m \pi)$ where $m$ is the third component of the angular momentum $j$
and $\pi = \pm$ is the parity.

Of course the eigenvalues of Eq.~(\ref{eq:rodiag}) provide us only the absolute values
of the occupation amplitudes $v_i$ and $u_i=\sqrt{1-v_i^2}$. In Eq.~(\ref{eq:cm-corr}) we
also need the sign of $u_iv_i$. It is determined by the diagonal elements of the pairing
tensor in the canonical basis,
\begin{eqnarray}
 u_iv_i~= \sum_{n\kappa}^{n'\kappa'} 
           c^{i}_{n\kappa} c^{i}_{n'\kappa'} \kappa_{(n\kappa)(n'\kappa')}
 \ .
\end{eqnarray}

For the direct term we need the diagonal matrix elements of the Laplacian
\begin{widetext}
\begin{eqnarray}
 \Delta_{{i} {i}} 
 & = & 
 -\sum_{n n'\kappa }
   c^{i}_{n\kappa} c^{i}_{n'\kappa} \int dr
   \left\{
    \frac{\partial G_{n\kappa} (r)}{dr} \frac{ \partial  G_{n'\kappa} (r) }{ \partial r}
    + l(l+1) \frac{ G_{n\kappa} (r) G_{n'\kappa} (r) }{r^2}  
   \right.
 \nonumber \\
 &   & \mbox{} \ \ \ \ \ \ \ \ \ \ \ \ \ \ \ \ \ \ \ \ \ \ \ \ \ \
   \left.
    + \frac{\partial F_{n\kappa} (r)}{dr}\frac{ \partial F_{n'\kappa} (r) }{ \partial r }
    + \tilde{l}(\tilde{l}+1) \frac{ F_{n\kappa} (r) F_{n'\kappa} (r) }{r^2}
   \right\} .
\end{eqnarray}
For the exchange term, we have 
$|\nabla_{{i} {i}'} |^2 = \sum_\mu (-)^\mu (\nabla_\mu)_{{i} {i}'}(\nabla_{-\mu})_{{i}{i}'}$
and according to the Wigner-Eckart theorem~\cite{Edmonds1957} we find 
for the spherical coordinate $\mu$ of the gradient operator
\begin{equation}
 (\nabla_\mu)_{{i}{i}'} = 
  \sum_{n\kappa}^{n'\kappa'} c^{i}_{n\kappa} c^{{i}'}_{n'\kappa'}
   (-)^{j-m} \left( \begin{matrix}j & 1 & j'\\-m& \mu& m'\\\end{matrix}\right)
   \langle \phi_{n\kappa}||\nabla||\phi_{n'\kappa'}\rangle ,
\end{equation}
with the reduced matrix element
\begin{eqnarray}
\langle \phi_{n\kappa}||\nabla||\phi_{n'\kappa'}\rangle
  & = &
  (-)^{j-1/2} \hat{\jmath} \hat{\jmath}'
  \left[ (-)^{l'}
    \left\{ {j' \atop l} {j \atop l'} {1 \atop \frac{1}{2}} \right\}
    \int dr G_{n\kappa}(r)\langle l||\nabla||l'\rangle G_{n'\kappa'}(r)
     \right. \nonumber \\ && \left.
    + (-)^{ \tilde{l}' }
    \left\{ {j' \atop \tilde{l}} {j \atop \tilde{l}'} {1 \atop \frac{1}{2}} \right\}
    \int dr F_{n\kappa}(r)\langle\tilde{l}||\nabla||\tilde{l'}\rangle F_{n'\kappa'}(r)
  \right] ,
\end{eqnarray}
where $\hat{\jmath}=\sqrt{2j+1}$ and the expression $\langle l||\nabla||l'\rangle$ 
is the reduced matrix element of $\nabla$ with respect to the integration over the angles. 
Of course, it still contains derivatives with respect to the radial coordinate. 
Following Sect.~5.7 of Ref.~\cite{Edmonds1957}, we obtain
\begin{equation}
 \langle l||\nabla||l'\rangle = 
   \delta_{l,l'+1} \sqrt{l } \, \left[\frac{d}{dr}-\frac{l }{r}\right]
 - \delta_{l,l'-1} \sqrt{l'} \, \left[\frac{d}{dr}+\frac{l'}{r}\right] .
\end{equation}

\end{widetext}

\end{document}